\pdfoutput=1
\documentclass[11pt,a4paper]{article}
\usepackage{jheppub}

\usepackage{empheq}
\usepackage{color}
\usepackage{hyperref}

\usepackage{float}
\restylefloat{table}
\restylefloat{figure}

\usepackage{multirow}
\usepackage{graphicx}
\usepackage{epstopdf}
\usepackage{array}
\usepackage{hhline}
\usepackage{microtype}
\usepackage[usenames,dvipsnames]{xcolor}
\usepackage{mathtools}
\usepackage{todonotes}

\usepackage{verbatim}

\usepackage[hang]{footmisc}
\usepackage{tikz}
\usepackage{tikz-cd}
\usepackage{tikz-3dplot}
\usetikzlibrary{decorations.markings,arrows}
\usetikzlibrary{positioning}
\usetikzlibrary{intersections} 

\tikzset{
    arrowMe/.style={
        postaction=decorate,
        decoration={
            markings,
            mark=at position .5 with {\arrow[thick]{#1}}
        }
    }
}

\newlength{\PicScale}
    \def\CF{{\cal F}}
    \def\cC{\CF}
     \def\Tr{\text{Tr}}

\newcommand{\nn}{\nonumber}

\newcommand{\nc}{\newcommand}
\nc{\beq}{\begin{equation}}
\nc{\eeq}{\end{equation}}
\nc{\bea}{\begin{eqnarray}}
\nc{\eea}{\end{eqnarray}}

\newcommand{\bi}{\begin{itemize}}
\newcommand{\ei}{\end{itemize}}
\newcommand{\ben}{\begin{enumerate}}
\newcommand{\een}{\end{enumerate}}
\newcommand{\be}{\begin{equation}}
\newcommand{\ee}{\end{equation}}
\newcommand{\ba}{\begin{align}}
\newcommand{\ea}{\end{align}}

\newcommand{\comments}[1]{}
\def\nn{\nonumber}

\def\LVS{{\scriptscriptstyle \rm LVS}}

\newcommand\vo{{\mathcal{V}}}

\newcommand{\mc}{\mathcal}

\newcommand{\beqa}{\begin{eqnarray}}
\newcommand{\eeqa}{\end{eqnarray}}

\allowdisplaybreaks[2]
\numberwithin{equation}{section}

\newcommand{\Sp}{U\!Sp}

\newcommand{\ccC}{\mathcal{C}}

\newcommand{\cO}{\mathcal{O}}

\newcommand{\cD}{\mathcal{D}}

\newcommand{\cN}{\mathcal{N}}

\newcommand{\cF}{\mathcal{F}}

\newcommand{\cV}{\mathcal{V}}

\newcommand{\cM}{\mathcal M}

\newcommand{\I}{\text{i}}

\usepackage{cancel}
\usepackage[normalem]{ulem}

\usepackage{amsthm}

\newcommand{\kom}{\, ,\quad }
\newcommand*{\raw}{\rightarrow}

\newcommand*{\gs}{g_{\text{s}}}

\newcommand{\ch}{\text{ch}}
\newcommand{\rk}{\text{rk}}

\newcommand{\tr}{\mathrm{Tr}}

\usepackage{tikz-cd}
\usepackage{tikz}
\usepackage{dsfont}
\usepgflibrary{arrows} 
\usetikzlibrary{arrows,shapes,positioning} 
\usetikzlibrary{decorations.markings}
\usetikzlibrary{shapes.geometric}
\usetikzlibrary{arrows.meta,arrows}

\tikzstyle arrowstyle=[scale=1]
\tikzstyle directed=[postaction={decorate,decoration={markings,
    mark=at position .5 with {\pgftransformscale{2}\arrow[arrowstyle]{stealth}}}}]
\tikzstyle directedThree=[postaction={decorate,decoration={markings,
    mark=at position .5 with {\pgftransformscale{2}\arrow[arrowstyle]{stealth}},mark=at position .4 with {\pgftransformscale{2}\arrow[arrowstyle]{stealth}},mark=at position .6 with {\pgftransformscale{2}\arrow[arrowstyle]{stealth}}}}]
    
    \tikzstyle directedTwo=[postaction={decorate,decoration={markings,
    mark=at position .5 with {\pgftransformscale{2}\arrow[arrowstyle]{stealth}},mark=at position .6 with {\pgftransformscale{2}\arrow[arrowstyle]{stealth}}}}]

\tikzstyle directedBF=[postaction={decorate,decoration={markings,
    mark=at position .9 with {\pgftransformscale{2}\arrow[arrowstyle]{stealth}},mark=at position .1 with {\pgftransformscale{2}\arrowreversed[arrowstyle]{stealth}}}}]

\tikzstyle directedThreeBF=[postaction={decorate,decoration={markings,
    mark=at position .9 with {\pgftransformscale{2}\arrow[arrowstyle]{stealth}},mark=at position .8 with {\pgftransformscale{2}\arrow[arrowstyle]{stealth}},mark=at position .7 with {\pgftransformscale{2}\arrow[arrowstyle]{stealth}},mark=at position .1 with {\pgftransformscale{2}\arrowreversed[arrowstyle]{stealth}},,mark=at position .2 with {\pgftransformscale{2}\arrowreversed[arrowstyle]{stealth}},mark=at position .3 with {\pgftransformscale{2}\arrowreversed[arrowstyle]{stealth}}}}]
    
\tikzstyle directedThreeFF=[postaction={decorate,decoration={markings,
    mark=at position .9 with {\pgftransformscale{2}\arrow[arrowstyle]{stealth}},mark=at position .8 with {\pgftransformscale{2}\arrow[arrowstyle]{stealth}},mark=at position .7 with {\pgftransformscale{2}\arrow[arrowstyle]{stealth}},mark=at position .1 with {\pgftransformscale{2}\arrow[arrowstyle]{stealth}},,mark=at position .2 with {\pgftransformscale{2}\arrow[arrowstyle]{stealth}},mark=at position .3 with {\pgftransformscale{2}\arrow[arrowstyle]{stealth}}}}]

\tikzstyle reverse directed=[postaction={decorate,decoration={markings,
    mark=at position .5 with {\pgftransformscale{2}\arrowreversed[arrowstyle]{stealth};}}}]

\allowdisplaybreaks[2]
\numberwithin{equation}{section}

\title{The Standard Model Quiver in  de Sitter \\ String Compactifications}

\author[1,2]{\small M. Cicoli,}
\author[3]{\small I. Garc{\'i}a Etxebarria,}
\author[4]{\small F. Quevedo,}
\author[4]{\small A. Schachner,}
\author[5]{\small P. Shukla,}
\author[6,7]{\small R. Valandro}

\affiliation[1]{\small Dipartimento di Fisica e Astronomia, Universit\`a di Bologna, via Irnerio 46, 40126 Bologna, Italy}
\affiliation[2]{\small INFN, Sezione di Bologna, viale Berti Pichat 6/2, 40127 Bologna, Italy}
\affiliation[3]{\small Department of Mathematical Sciences, Durham University, Durham, DH1 3LE, UK}
\affiliation[4]{\small DAMTP, Centre for Mathematical Sciences, Wilberforce Road, Cambridge, CB3 0WA, UK}
\affiliation[5]{\small ICTP, Strada Costiera 11, Trieste 34151, Italy}
\affiliation[6]{\small Dipartimento di Fisica, Universit\'a di Trieste, Strada Costiera 11, 34151 Trieste, Italy}
\affiliation[7]{\small INFN, Sezione di Trieste, Via Valerio 2, 34127 Trieste, Italy}

\emailAdd{michele.cicoli at unibo.it}
\emailAdd{inaki.garcia-etxebarria at durham.ac.uk}
\emailAdd{fq201 at damtp.cam.ac.uk}
\emailAdd{as2673 at maths.cam.ac.uk}
\emailAdd{pramodmaths at gmail.com}
\emailAdd{roberto.valandro at ts.infn.it}

\abstract{We argue that the Standard Model quiver can be embedded into compact Calabi-Yau geometries through orientifolded D3-branes at del Pezzo singularities $\mathrm{dP}_n$ with $n\geq 5$ in a framework including moduli stabilisation. To illustrate our approach, we explicitly construct a local $\mathrm{dP}_{5}$ model via a combination of Higgsing and orientifolding. This procedure reduces the original $\mathrm{dP}_{5}$ quiver gauge theory to the Left-Right symmetric model with three families of quarks and leptons as well as a Higgs sector to further break the symmetries to the Standard Model gauge group. We  embed this local model in a globally consistent Calabi-Yau flux compactification with tadpole and Freed-Witten anomaly cancellations. The model features closed string moduli stabilisation with a de Sitter minimum from T-branes, supersymmetry  broken by the K\"ahler moduli, and the MSSM as the low energy spectrum. We further discuss phenomenological and cosmological implications of this construction.}

\begin{document}

\makeatletter
\let\old@fpheader\@fpheader
\renewcommand{\@fpheader}{\old@fpheader\hfill

\hfill }
\makeatother

\maketitle

\bigskip

\section{Introduction}

The main argument for the study of string theory remains its potential to explain all natural phenomena, including gravity, within a consistent quantum framework. In this sense, it is usually stated that string theory provides the UV completion of the Standard Model. However, despite many efforts for the past 30 years, it is fair to say that there is not yet a consistent string construction that includes the Standard Model and does not have some unrealistic features. 

The search for realistic string models seems hopeless due to the huge degeneracy of string compactifications. This is usually compared with the proverbial search for a needle in a haystack. Similar  to using a magnet to find the needle we can follow a bottom-up modular approach in the search for a realistic string model \cite{Aldazabal:2000sa}. We may split the search into at least three independent challenges:
\begin{itemize}
\item Search for  local string constructions in terms of intersecting branes \cite{Aldazabal:2000sa, Blumenhagen:2000wh, Aldazabal:2000dg, Blumenhagen:2001te, Cvetic:2001tj,Conlon:2008wa,Blumenhagen:2008zz} or local F-theory models \cite{Donagi:2008ca,Beasley:2008dc,Beasley:2008kw,Donagi:2008kj} that includes the chiral matter of the Standard Model in which gravity is decoupled.

\item Search for string mechanisms in which global issues such as moduli stabilisation, supersymmetry breaking, inflation or alternatives are addressed, ignoring the potential realisation of the Standard Model particles and interactions beyond gravity.

\item Once a successful framework for each of the two challenges above has been found, combine both constructions to incorporate the Standard Model in a fully-fledged string compactification.

\end{itemize}
Each of these steps is a major challenge by itself but the approach is much simpler and systematic than direct top-down searches for realistic string models.

The last decade has brought enormous advances in our ability to
construct semi-realistic vacua in the framework of type IIB
compactifications. Much of the focus has been on F-theory model
building (starting with
\cite{Donagi:2008ca,Beasley:2008dc,Beasley:2008kw,Donagi:2008kj}), and
has resulted in a rich set of models, with a number of promising
features when it comes to model building (see for instance \cite{Cvetic:2019gnh} and references therein).

One of the defining characteristics of F-theory model building is the
description of sectors where the string coupling constant $g_s$ becomes
large. It is possible to understand such regimes using duality with
M-theory. However, since our knowledge of the behaviour of M-theory on
highly curved manifolds is rather limited, our understanding of
F-theory models is generically limited to features that can be
continued in a supersymmetric way to weakly curved backgrounds.
Furthermore, moduli stabilisation is not included in these constructions limiting their potential contact with the real world.

In this paper we focus instead on the complementary regime of type IIB
models which contain highly curved --- singular, in fact ---
regions.
More concretely, we will explore the case of D3-branes at del
Pezzo singularities in global type IIB Calabi-Yau (CY)
compactifications. Branes probing singularities lead to interesting
low energy dynamics, which can be understood at sufficiently small
string coupling $g_s$.  The visible sector arising from the modes at
the singularity is described by a collection of fractional branes
which are conveniently represented as nodes in $2$-dimensional graphs
referred to as \emph{quiver diagrams}.  Open strings stretched between
stacks of fractional branes give rise to a massless spectrum of matter
fields in bi-fundamental representations joining the various quiver
nodes via directed lines.

Many of the required tools for studying the singular regions are
familiar from various previous analyses of branes at IIB singularities
with orientifolds; our contribution is the construction of explicit
global models that include singularities relevant to realistic model
building, and the detailed analysis of their features.
The constructions provided in this paper reproduce the MSSM exactly in stark contrast to previous local \cite{Franco:2006es,Franco:2012mm,Franco:2013ana,Bianchi:2013gka,Bianchi:2020fuk}
and global \cite{Cicoli:2012vw,Cicoli:2013mpa,Cicoli:2013zha,Cicoli:2013cha,Cicoli:2017shd} investigations. We will find that reproducing the SM spectrum does not require flavour D7-branes.
In spite of the local nature of the model,
it does not imply that one should expect to be able
to simply ``glue" the local physics to any arbitrary compact manifold since satisfying the
multiple phenomenological constraints on the model proves to be very stringent. In fact,
after extensive searches for candidates, we only found a few models which possess realistic
features both at the local and global level.

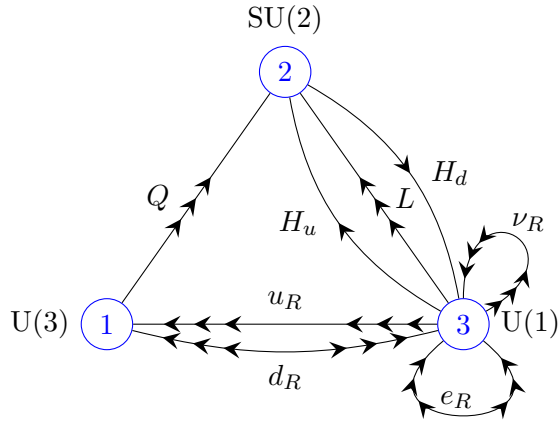
\begin{figure}
\centering
\begin{tikzpicture}[scale=1.]
\draw (0,9) node(1)[anchor=south,circle,draw,blue]{$2$}node[above=20pt] {$\mathrm{SU}(2)$};
\draw (-2,6) node(8)[anchor=east,circle,draw,blue]{1}node[left=20pt] {$\mathrm{U}(3)$};
\draw (2,6) node(4)[anchor=west,circle,draw,blue]{3}node[right=20pt] {$\mathrm{U}(1)$};
\draw (3.2,7.35) node(5) {$\nu_{R}$};
\draw (2.25,5.) node(5) {$e_{R}$};
\draw[directed,black] (1) to[bend left=25]node[right=4pt] {$H_{d}$} (4);
\draw[directed,black] (4) to[bend left=25]node[left=4pt] {$H_{u}$} (1);
\draw[directedThree,black] (4) --node[right=4pt] {$L$} (1);
\draw[directedThreeBF,black] (4) to[bend left=14]node[below=1pt]{$d_{R}$} (8);
\draw[black,directedThreeBF] (4) to [out=320,in=220,looseness=10] (4);
\draw[black,directedThreeFF] (4) to [out=30,in=90,looseness=12] (4);
\draw[directedThreeFF,black] (4) --node[above=2pt] {$u_{R}$} (8);
\draw[directedThree,black] (8) --node[left=6pt] {$Q$} (1);
\end{tikzpicture}
\caption{Standard Model quiver containing an additional $\mathrm{U}(1)_{B}$.
The right-handed neutrinos transform in the (trivial) adjoint of $\mathrm{U}(1)$ and are thus uncharged under all gauge group factors.
}\label{fig:SMQuiver} 
\end{figure}

In order to find a suitable local realisation of the Standard Model,
we have been guided by the seminal work by Wijnholt
\cite{Wijnholt:2007vn}, who provided two scenarios to obtaining the
Standard Model quiver from a single D3-brane at a $\mathrm{dP}_{5}$
singularity. Essential features of both of these two scenarios are the
presence of appropriate orientifold involutions\footnote{This seems to be a general theme: minimal quiver extensions of the supersymmetric Standard Model themselves are often unoriented \cite{Ibanez:2001nd, Anastasopoulos:2006da,Berenstein:2006pk}.} and intricate
Higgsing operations. The resulting quiver in these scenarios is of
form given in fig.~\ref{fig:SMQuiver}.  It consists of the Standard
Model degrees of freedom together with right-handed neutrinos and an
extra $\mathrm{U}(1)_{B}$.  The $\mathrm{U}(1)$ on the bottom right is
obtained either from identifying two $\mathrm{U}(1)$'s in a covering
quiver or from a larger quiver via Higgsing
$\mathrm{SU}(2)\times \mathrm{U}(1)\raw \mathrm{U}(1)$.  In fact, the
latter scenario corresponds to having a supersymmetric version of the
Minimal Left-Right Symmetric Model
\cite{Pati:1974yy,Mohapatra:1974hk,Senjanovic:1975rk} as an
intermediate step.  The local models studied in this paper will be of
this form. It should be noted at this point, though, that there
remains a rich structure of largely unexplored but phenomenologically
promising models from fractional D3-branes at orientifolded
$\mathrm{dP}_{n}$ singularities with $n\geq 5$. In this paper we will
restrict ourselves to the dP$_5$ case, but the analysis of cases with
higher $n$ would certainly be interesting.

In this paper we will go further than a purely local analysis, and
provide a seemingly phenomenologically viable global embedding of the
orientifolded quiver gauge theory, so that our gauge dynamics is
coupled to gravity.
The presence of a $\mathrm{dP}_{5}$ singularity is ensured by having a
\emph{diagonal} divisor of $\mathrm{dP}_{5}$ topology inside the CY.
The diagonality condition allows to take the singular limit for the
dP$_5$ divisor by taking a single linear combination of $2$-cycle
volumes to zero without shrinking any additional divisors, see for
instance \cite{Cicoli:2011it}.

A noteworthy complication in carrying out this program is the absence
of diagonal $\mathrm{dP}_{n}$ divisors with $1\leq n\leq 5$ at Hodge
numbers $h^{1,1}\leq 40$ in the Kreuzer-Skarke (KS) database
\cite{Kreuzer:2000xy}.\footnote{We observed this empirically, but we
  do not know why it is so.} We proceed by instead constructing CY threefolds
$X$ as complete intersections of two equations in $5$-dimensional
toric spaces.
We show that the global orientifold action on $X$ can be made
consistent with the orientifold involution of the local model.

Once we have achieved this global embedding of the local physics, we
will devote the rest of the paper to checking that our background is
workable and phenomenologically promising. These checks will be
developed in detail in the bulk of the paper, but we give a brief
summary here. First, since O7/O3-planes carry non-trivial RR-charges,
tadpole cancellation requires the presence of further ingredients
in the compact space. Freed-Witten anomaly cancellation
\cite{Freed:1999vc} then demands suitable flux backgrounds for various
anti-symmetric tensors. These fluxes, in turn, affect the $4$D
effective theory through non-trivial $F$- and $D$-terms which
stabilise a subset of geometric moduli.  Moreover, they can be chosen
to give rise to a T-brane background
\cite{Donagi:2003hh,Cecotti:2010bp} which leads to a controlled uplift
to $4$D de Sitter minima \cite{Cicoli:2015ylx}.  

K\"ahler moduli
stabilisation necessitates additional non-perturbative effects which
arise from wrapping Euclidean D3-branes on internal $4$-cycles
\cite{Witten:1996bn}.
At the minimum, supersymmetry is broken spontaneously in the hidden
sector by non-vanishing F-terms for bulk K\"ahler moduli
\cite{Balasubramanian:2005zx}.
This breaking is mediated to the
visible sector through gravitational interactions.
In the absence of sequestering effects for orientifolded quivers \cite{Conlon:2010ji}, soft terms are of order the gravitino mass which can be either around $10^{10}$ GeV or at the TeV-scale, depending on the tuning allowed on the flux superpotential. Interestingly, if the gravitino mass is at intermediate scales, our models have all the required features to provide a viable description of the cosmological evolution of our universe, from inflation \cite{Conlon:2005jm} to the post-reheating epoch \cite{Cicoli:2010ha, Cicoli:2010yj} involving non-thermal dark matter \cite{Allahverdi:2020uax}, Affleck-Dine baryogenesis \cite{Allahverdi:2016yws} and axionic dark radiation \cite{Cicoli:2012aq, Higaki:2012ar, Cicoli:2015bpq}.

\medskip

This paper is organised as follows.
Section~\ref{sec:SMQuiver} concerns the local model construction from a D3-brane at a $\mathrm{dP}_{5}$ singularity.
Subsequently, we highlight several obstacles that appear in obtaining suitable CY threefolds from polytope triangulations in the KS database in section~\ref{sec:CYThreefoldScan}.
Afterwards, we specialise to a specific complete intersection CY threefold exhibiting the required $\mathrm{dP}_{5}$ singularity.
We show that the local model is consistently embedded into the compact CY orientifold background.
Further, we provide a fully explicit construction of the D-brane configuration featuring a T-brane background.
In section~\ref{ModStab}, we confirm that closed string moduli can indeed be stabilised in Minkowski or slightly de Sitter minima.
We discuss phenomenological implications in section~\ref{sec:PhenoImp} and summarise our conclusions in section~\ref{sec:Conclusions}.

\vspace{0.5cm}

\medskip

\section{The Standard Model Quiver and Orientifolded dP$_n$ Singularities }\label{sec:SMQuiver} 

In this section we will describe in detail the local model giving rise
to the Standard Model sector.

\subsection{Calabi-Yau singularities and del Pezzo surfaces}

Let $X$ be a CY threefold. The moduli space
$\cM_{\text{K}}(X)$ of K\"ahler classes is characterised by the K\"ahler
cone. Upon approaching the boundary wall of this cone, some parts of
$X$ shrink to zero size typically giving rise to singular geometries.
This shrinking can happen in various distinct ways as outlined in \cite{CYBU4}, see also \cite{1995alg.geom..7016H,Aspinwall:1995mh,Bershadsky:1995sp,Katz:1996ht}.

Here, we are particularly interested in the scenario where a single $4$-cycle shrinks to a point which was
first extensively studied from the string theory point of view in
\cite{Chiang:1995hi,Morrison:1996pp}. The associated singularities are referred to as ``isolated canonical singularities
with a crepant blow up''.  We are interested in the case  when the $4$-cycle to be shrunk to a point is a special type of complex algebraic surface known as
(\emph{generalised}\footnote{Generalised refers to the fact that, in
  principle, these surfaces can have singularities themselves, see,
  e.g., \cite{Douglas:1996xp} for definitions.})  \emph{del Pezzo
  surface}\footnote{More generally, Del Pezzo
surfaces  are complex $2$-dimensional Fano surfaces, i.e., projective
algebraic surfaces with ample anti-canonical divisor class $-K$ so
that $-K\cdot C>0$ for each curve $C$.
Among the Fano surfaces we find also 
$\mathbb{P}^{1}\times \mathbb{P}^{1}$ (sometimes also known as the
Hirzebruch surface $\mathbb{F}_0$), with $h^{(1,1)}=2$.} $\mathrm{dP}_{n}$ \cite{CYBU45,Cordova:2009fg}.  
They are
obtained by blowing up $\mathbb{P}^{2}$ at $0\leq n\leq 8$ points. The
non-vanishing Hodge numbers of these surfaces are
\begin{align}
&h^{(0,0)}=h^{(2,2)}=1\kom h^{(1,1)}=n+1 \, .\end{align}
The generators of $H^{(1,1)}(\mathrm{dP}_{n})$ are given by a hyperplane class $H$ from $\mathbb{P}^{2}$ and exceptional divisors $E_i$ ($i=1,...,n$) of the individual blow ups of $n$ points.

\subsection{Large volume perspective of D3-branes at singularities}\label{sec:SEC}

In the following, we will denote the del Pezzo surface as $Y$, and
consider the case in which it is embedded within a compact CY
threefold $X$. When $Y$ collapses to zero size, this leads to a
singular point in $X$. We now want to understand what happens when
D-branes in IIB string theory probe this singular point.

\begin{figure}[t!]
\centering
 \includegraphics[scale=0.3]{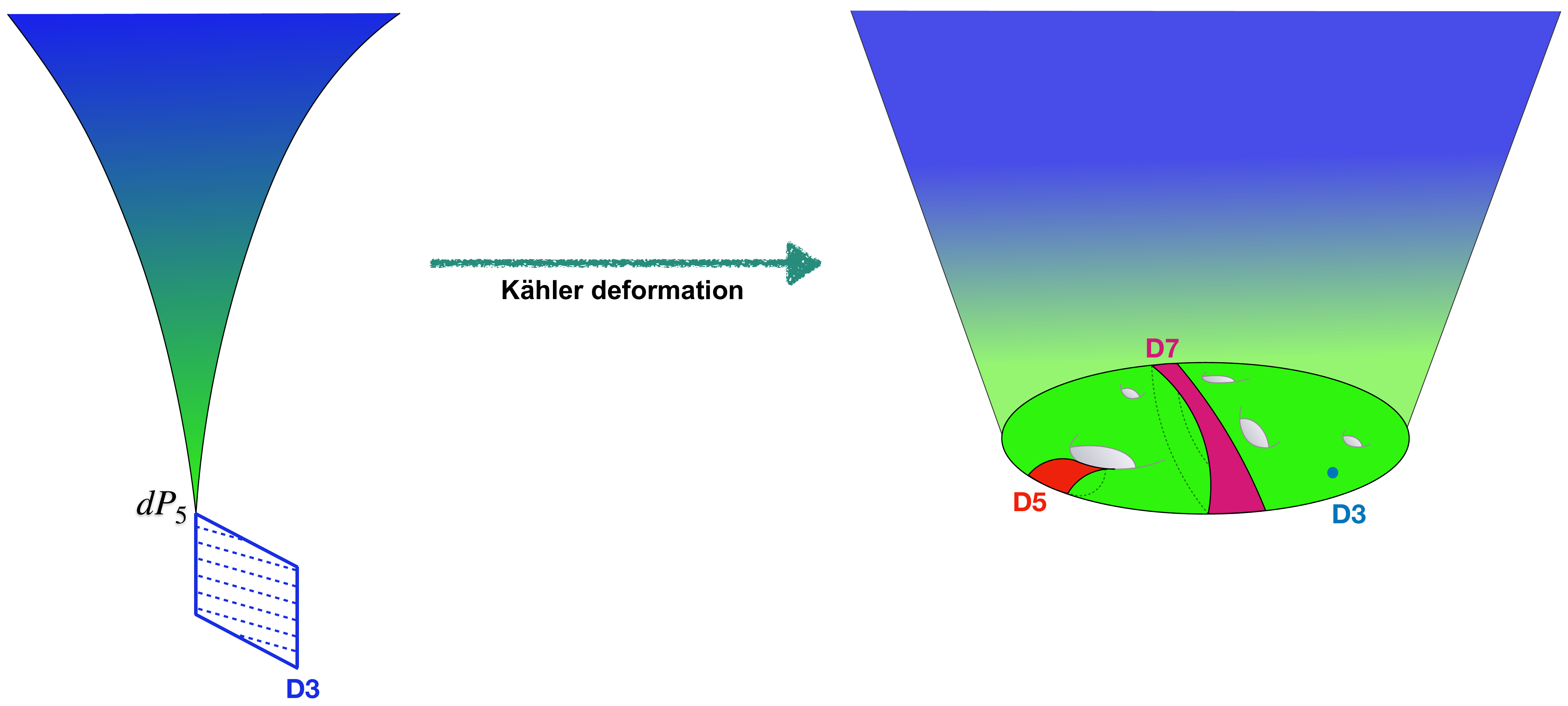}
\caption{The splitting of a D3-brane at a del Pezzo singularity into fractional branes by considering the large volume limit of the corresponding $4$-cycle.}\label{fig:LVP}
\end{figure}

In fact, we can obtain quite a bit of information by going to the
B-model: the result of computing quantities such as the chiral
spectrum or the superpotential in the B-model agree with the results
in the full string theory \cite{Sharpe:1999qz,Douglas:2000gi} (see
\cite{Aspinwall:2004jr} for a review). The computations in the
B-model are insensitive to K\"ahler moduli deformations, so we can
compute these protected quantities using classical geometry: we can
resolve the singularity by deforming the K{\"a}hler structure to large
cycle volumes and thereby small curvature. This gives rise to the
\emph{large volume perspective} \cite{Douglas:2000ah,Douglas:2000qw,Cachazo:2001sg,Wijnholt:2002qz,Aspinwall:2004jr,Herzog:2004qw,Aspinwall:2004vm,Herzog:2005sy,Hanany:2006nm} of a the
D3-brane at the singularity as depicted in~Fig.~\ref{fig:LVP}.

In doing this, it is important that the D3-brane can be understood as
a bound state of so-called \emph{fractional branes} $F_{i}$ supported
on blow-up cycles (whether the D3 or the fractional branes are
realised in any specific physical configuration depends on the K\"ahler
data).  These fractional branes are described by complexes of
\emph{sheaves}, which are intuitively vector bundles supported only on
submanifolds.\footnote{More precisely, the B-model branes should be
  understood as elements of the derived category of coherent sheaves of
  $X$. We refer the reader to \cite{Aspinwall:2004jr} for an
  introduction to this formalism.} For instance, a D3-brane on a
smooth point $p$ corresponds to a \emph{sky-scraper sheaf} $\cO_{p}$:
heuristically a ``vector bundle'' which is non-trivial at $p$ and
trivial everywhere else.

It is convenient to represent the world-volume gauge theory as a
$2$-dimensional graph known as a quiver diagram.  The total gauge
group receives an additional factor $\mathrm{U}(N_{i})$ for each stack
of $N_{i}$ fractional brane $F_{i}$.  In the quiver, we then draw a
node with an assigned gauge multiplet of $\mathrm{U}(N_{i})$.  Open
strings stretched between two fractional branes $F_{i}$ and $F_{j}$
give rise to a massless spectrum of chiral multiplets in
bi-fundamental representations of
$\mathrm{U}(N_{i})\times \mathrm{U}(N_{j})$ which join pairs of nodes
via directed lines. In the large volume description, these modes arise
from elements of the groups $\mathrm{Ext}^{k}(F_{i},F_{j})$, which if
$F_i$ and $F_j$ are ordinary vector bundles reduces to Dolbeault
cohomology of the tensor bundle. See~\cite{Aspinwall:2004jr} for
details of the general case.

We need to describe how to choose an appropriate basis of fractional
branes.\footnote{The choice is not unique \cite{1998IzMat..62..429K}.
Different choices leading to the same IR physics are related Seiberg
dualities \cite{Wijnholt:2002qz,Herzog:2003zc}.} We will do so by
focusing on \emph{exceptional collections}, constructed as follows. A
sheaf $F$ is called \emph{exceptional} if
$\mathrm{dim}(\text{Hom}(F,F))=1$ as well as $\text{Ext}^{k}(F,F)=0$
for all $k>0$. Similarly, a collection $\lbrace F_{i}\rbrace$ of
sheaves is called \emph{exceptional} if all $F_{i}$ are exceptional
and, in addition, there exists an ordering such that
$\text{Ext}^{k}(F_{i},F_{j})=0$ for $i>j$ for any $k$, while
$\text{Ext}^{k}(F_{i},F_{j})\neq 0$ for one $k$ if $j>i$.
There are
systematic techniques to construct exceptional collections in the case
of del Pezzo singularities (and more generally), we refer the reader
to
\cite{1998IzMat..62..429K,Wijnholt:2002qz,Herzog:2003zc,Hanany:2006nm}
for the details.

The sheaves within an exceptional collection have the right property
to describe the fractional branes having support on the internal
cycles in the large volume perspective
\cite{Wijnholt:2002qz,Herzog:2003zc,Herzog:2003dj}. The property
$\text{Ext}^{k}(F_{i},F_{i})=0$ for all $k>0$
ensures that besides the gauge multiplet there is no adjoint matter in the world-volume gauge theory, i.e., there are no arrows beginning and ending at the same node.
Moreover, the matter fields between two nodes have only one chirality due to the imposed ordering, that is, there is only a single direction for each arrow between any pairs of nodes.

In terms of string theory, any given fractional brane $F_{i}$ itself
corresponds to a bound state of D7-, D5- and D3-brane states,
cf.~Fig.~\ref{fig:LVP}. In general some of these branes will be
anti-branes, and at large volume the configuration badly breaks
supersymmetry, but the fractional branes become mutually
supersymmetric at the singular point (for appropriate choices of
$B$-fields) due to $\alpha'$ corrections
\cite{Douglas:2000gi,Aspinwall:2004jr}. Its RR-charges are combined
into a charge vector \cite{Verlinde:2005jr,Buican:2006sn}
\begin{equation}
\ch(F_{i})=\left (\rk(F_{i}),c_{1}(F_{i}),\ch_{2}(F_{i})\right )\, .
\end{equation}
corresponding to the D7-, D5- and D3-charge of $F_{i}$ respectively.
The individual components are interpreted as follows:
\begin{itemize}
\item the D7-charge corresponding to the wrapping number of a D7-brane around the del Pezzo surface $Y$ is just given by the rank of $F_{i}$;
\item the D5-charge is specified by the first Chern class $c_{1}(F_{i})$ of $F_{i}$.
That is, the D5 wrapping number $p^{A}_{i}$ around an integral basis of $2$-cycles $\alpha_{A}\in H_{2}(Y)$ is given by
\begin{equation}\label{eq:chD5} 
p_{i}^{A}=\int_{\alpha_{A}}\, c_{1}(F_{i})\, ;
\end{equation}
\item the D3-brane charge represents a non-trivial instanton number which is obtained from the $2$nd Chern character $\ch_{2}(F_{i})$.
\end{itemize}

\subsubsection*{Matter spectra and exceptional collections}

As mentioned above, the massless matter spectrum of the gauge theory
is encoded in $\text{Ext}^{k}(F_{i},F_{j})$.  For our purposes, it is
sufficient to know the number of arrows between any two nodes.  It is
obtained from the relative Euler character
\begin{equation}
\chi(F_{i},F_{j})=\sum_{k}\, (-1)^{k}\,\dim(\mathrm{Ext}^{k}(F_{i},F_{j})) \:,
\end{equation}
which, after using the Riemann-Roch formula \cite{Verlinde:2005jr}, becomes
\begin{equation}\label{eq:EulerChar}
\chi(F_{i},F_{j})=\int_{Y}\, \mathrm{ch}(F_{i}^{*})\mathrm{ch}(F_{j})\mathrm{Td}(Y) \:.
\end{equation}
Here, $
\mathrm{ch}(F_{i})=\rk(F_{i})+c_{1}(F_{i})+\ch_{2}(F_{i})$ and $\mathrm{Td}(Y)=1-\dfrac{1}{2}K+H^{2}
$
are the Chern character of $F_{i}$ and the Todd class of $Y$ respectively ($-K$ is the anti-canonical divisor class).
The equation \eqref{eq:EulerChar} can then be expressed as
\begin{align}\label{eq:MatterContentQuiverGB} 
\chi(F_{i},F_{j})&=\rk(F_{i})\rk(F_{j})+\rk(F_{i})\ch_{2}(F_{j})+\ch_{2}(F_{i})\rk(F_{j})\nn\\
&\quad -c_{1}(F_{i})\cdot c_{1}(F_{j})+\dfrac{1}{2}\left (\text{rk}(F_{j})\text{deg}(F_{i})-\text{rk}(F_{i})\text{deg}(F_{j})\right )
\end{align}
with $\text{deg}(F_{i})$ being the degree of $F_{i}$ defined as
\begin{equation}
\text{deg}(F_{i})=-c_{1}(F_{i})\cdot K
\end{equation}
in terms of the canonical class $K$ of $Y$. Note that by \eqref{eq:chD5} $\text{deg}(F_{i})$ corresponds to the intersection number between the del Pezzo surface $Y$ and the D5-component of $F_{i}$. Furthermore, $\chi(F_{i},F_{i})=1$ corresponds to the presence of the gauge multiplet of the corresponding node. If this is true for all $F_{i}$ such as in exceptional collections, then this implies the absence of any adjoint matter.

One can show that \cite{Wijnholt:2002qz}
\begin{equation}\label{eq:NonAbGAnomGen} 
\sum_{j}\, \chi(F_{j},F_{i})\, N_{j}=0\, .
\end{equation}
For the gauge theory, this implies that the total number of ingoing and outgoing lines are equal and hence the quiver gauge theory is free of non-abelian gauge anomalies.
There can still be gauge anomalies associated to the $\mathrm{U}(1)$-factors since there is mixing with closed string modes \cite{Ibanez:1998qp}.
For an exceptional collection, \eqref{eq:MatterContentQuiverGB} results in an upper triangular matrix with only $1$'s on the diagonal.
The spectrum of chiral fields is completely contained in the anti-symmetrised expression of \eqref{eq:MatterContentQuiverGB}:
\begin{equation}\label{eq:AsymEulerChar} 
\chi_{-}(F_{i},F_{j})=\chi(F_{i},F_{j})-\chi(F_{j},F_{i})=\text{rk}(F_{j})\text{deg}(F_{i})-\text{rk}(F_{i})\text{deg}(F_{j})\, .
\end{equation}
The direction of arrows is determined by the sign of $\chi_{-}(F_{i},F_{j})$.
We can rewrite \eqref{eq:NonAbGAnomGen} for an exceptional collection as \cite{Verlinde:2005jr}
\begin{equation}\label{eq:NonAbGAnom} 
\sum_{i}\, N_{i}\, \chi_{-}(F_{i},F_{j})=0\, .
\end{equation}
This is again equivalent to the absence of non-abelian gauge anomalies.

\subsection{Why branes at singularities?}

Del Pezzo surfaces are ubiquitous in CY manifolds.
Their purpose for string model building is twofold.
On the one hand, the possible rigidity of these $4$-cycles makes them prime candidates to support non-perturbative effects contributing to the $4$d superpotential.
On the other hand, the worldvolume theories of D3-branes at \emph{del Pezzo singularities} obtained in the limit of vanishing $4$-cycle volumes host interesting particle phenomenology.
In particular:
\begin{enumerate}
\item Chiral matter fields in the worldvolume gauge theory of branes at singularities arise from intersections of $2$- and $4$-cycles.
Every CY threefold singularity will have some $2$- or $4$-cycle volumes shrinking to zero size.
The presence of chiral states necessitates the existence of $4$-cycles collapsing to zero size.
By our previous reasoning, $\mathrm{dP}_{n}$-singularities are the simplest examples associated with the vanishing volume limit of a single $4$-cycle.
\item D-brane constructions of the SM from unoriented quivers come with one local anomalous $\mathrm{U}(1)$ corresponding to baryon number $\mathrm{U}(1)_{B}$. In the oriented covering quiver, we therefore must find two anomalous $\mathrm{U}(1)$'s.
Oriented del Pezzo quivers naturally come with precisely two anomalous $\mathrm{U}(1)$'s.
Geometrically speaking, this is due to two compact cycles in the non-compact CY geometry of a complex cone over del Pezzos. They can be identified with the canonical class and the del Pezzo surface itself \cite{Buican:2006sn,Kennaway:2007tq}.\footnote{In fact,
a combination of $\mathrm{U}(1)$'s is non-anomalous precisely when the associated collection of fractional brane has zero D7-charge (i.e. vanishing rank) as well as no intersection of the D5-component with the canonical class of the del Pezzo \cite{Malyshev:2007zz}.
According to \eqref{eq:AsymEulerChar},
this ensures the absence of chiral matter at the intersection of fractional branes and thus the absence of mixed anomalies.
Branes wrapping the two compact Poincar\'e dual cycles, however, have chiral states inducing mixed anomalies in the $\mathrm{U}(1)$ worldvolume gauge theory.}
From a field theory perspective, we would expect that for the $n+3$ fractional branes of a $\mathrm{dP}_{n}$-singularity, there exist $n+3$ gauge couplings and $n+3$ Fayet-Iliopoulos (FI) parameters. However, two of the latter are not freely tunable and associated with the two anomalous $\mathrm{U}(1)$'s \cite{Intriligator:2005aw}.
Although this is not specific to del Pezzo singularities {\it per se},
it is a favourable criterion for the models in our construction.
\item Fractional branes at the singularity are sufficient to generate a large variety of gauge groups together with the required matter spectrum.
In particular, there is no necessity for the existence of flavour D7-branes, cf. Sect.~5 in~\cite{Cicoli:2012vw}.
Furthermore, turning on VEVs for bi-fundamental fields not only relates various $\mathrm{dP}_{n}$ models \cite{Wijnholt:2002qz}, but also has the potential of generating additional matter fields \cite{Verlinde:2005jr,Wijnholt:2007vn}.
\end{enumerate}

Early local constructions based on oriented quivers at $\mathrm{dP}_{8}$-singularities \cite{Verlinde:2005jr} generated a gauge group containing the SM group.\footnote{However, an embedding into a CY threefold retaining only the SM subgroup would require $h^{(1,1)}\geq 9$.
This is because at least five $2$-cycles of the local $\mathrm{dP}_{8}$ need to be non-trivial in the full CY to guarantee that the associated $\mathrm{U}(1)$ gauge fields become massive, see Sect.~5 in~\cite{Cicoli:2012vw} for a discussion.}
For toric $\mathrm{dP}_{n}$-singularities (i.e. $n\leq 3$), the authors of \cite{Krippendorf:2010hj} showed that there are at most three families.\footnote{We refer to \cite{Aldazabal:2000sa} for a related result for orbifold singularities.}
Further, they argued that hierarchies in the quark masses can indeed be realised for $n\geq 1$.
As discussed in \cite{Dolan:2011qu}, D3-branes at $\mathrm{dP}_{n}$-singularities with $n>2$ are favourable to realise the hierarchical mixing angles in the CKM matrix.
A brief discussion of higher $\mathrm{dP}_{n}$-singularities together with constraints for global embeddings can be found in \cite{Cicoli:2012vw}.
Further works found realistic extensions of the SM in global orientifolded models \cite{Cicoli:2013mpa,Cicoli:2013zha,Cicoli:2013cha} together with inflation \cite{Cicoli:2017shd}.

\subsection{The $\text{dP}_{5}$-quiver and its involutions}

In the rest of this section, we will use the ideas reviewed above to
construct the (MS)SM as a local model on a system of fractional branes
at a $\mathrm{dP}_{n}$-singularity.  Let us start by identifying the
minimal value of $n$ to realise the \emph{Minimal Quiver Standard
Model} (MQSM) \cite{Anastasopoulos:2006da,Berenstein:2006pk}.  The
MQSM requires at least $3$ nodes for both oriented and unoriented
quivers \cite{Heckman:2007zp}.
Non-supersymmetric unoriented versions of the MQSM were proposed in
\cite{Anastasopoulos:2006da,Berenstein:2006pk} with the minimally
required number of $3$ nodes.  The seminal work \cite{Wijnholt:2007vn}
found analogous supersymmetric constructions of the MQSM from
D3-branes at $\mathrm{dP}_{5}$-singularities.  In fact, the
$\mathrm{dP}_{5}$ quiver is the minimal del Pezzo quiver to contain
the MQSM.  For this reason, we are particularly interested in the two
unoriented models based on dP$_5$ implemented in
\cite{Wijnholt:2007vn}.

\begin{figure}[t!]
\centering
\begin{tikzpicture}[scale=0.9]
\draw (0,6) node(1)[anchor=south,circle,draw,blue]{1}node[above=20pt] {$\mathrm{U}(N_{1})$};
\draw (1.5,6) node(2)[anchor=south,circle,draw,blue]{2}node[above=20pt] {$\mathrm{U}(N_{2})$};
\draw (-1.5,4.5) node(8)[anchor=east,circle,draw,blue]{8}node[left=20pt] {$\mathrm{U}(N_{8})$};
\draw (-1.5,3) node(7)[anchor=east,circle,draw,blue]{7}node[left=20pt] {$\mathrm{U}(N_{7})$};
\draw (3,4.5) node(3)[anchor=west,circle,draw,blue]{3}node[right=20pt] {$\mathrm{U}(N_{3})$};
\draw (3,3) node(4)[anchor=west,circle,draw,blue]{4}node[right=20pt] {$\mathrm{U}(N_{4})$};
\draw (0,1.5) node(6)[anchor=north,circle,draw,blue]{6}node[below=20pt] {$\mathrm{U}(N_{6})$};
\draw (1.5,1.5) node(5)[anchor=north,circle,draw,blue]{5}node[below=20pt] {$\mathrm{U}(N_{5})$};
\draw[directed,black] (1) -- (3);
\draw[black,directed] (1) -- (4);
\draw[directed,black] (2) -- (3);
\draw[directed,black] (2) -- (4);
\draw[directed,black] (3) -- (5);
\draw[directed,black] (3) -- (6);
\draw[directed,black] (4) -- (5);
\draw[directed,black] (4) -- (6);
\draw[directed,black] (5) -- (7);
\draw[directed,black] (5) -- (8);
\draw[directed,black] (6) -- (7);
\draw[directed,black] (6) -- (8);
\draw[directed,black] (7) -- (1);
\draw[directed,black] (7) -- (2);
\draw[directed,black] (8) -- (1);
\draw[directed,black] (8) -- (2);
\end{tikzpicture}
\caption{Quiver diagram for a D3-brane at a $\mathrm{dP}_{5}$-singularity}\label{fig:QuivdP5} 
\end{figure}
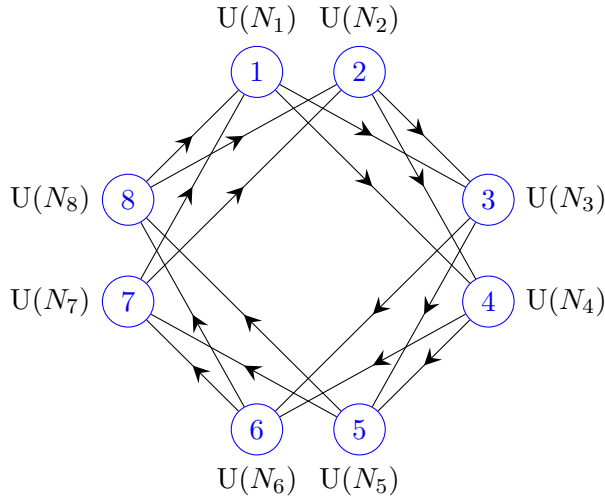

We begin by analysing the quiver gauge theory for a D3-brane at a
$\text{dP}_{5}$-singularity \cite{Wijnholt:2002qz,Wijnholt:2007vn}.
The singularity itself has a well-known toric limit corresponding to
the $\mathbb{Z}_{2}\times\mathbb{Z}_{2}$ orbifold of the conifold.
$\mathbb{P}^{2}$ itself has three independent holomorphic cycles,
i.e., the class of points, the class of the hyperplane and the class
of the $4$-cycle. With the additional $5$ blown up points, we expect
to find $8$ nodes in the $\mathrm{dP}_{5}$ quiver diagram.  It is
easily constructed from intersection numbers of fractional branes.

For del Pezzo surfaces, we choose a basis of generators of $H_{2}(\mathrm{dP}_{n},\mathbb{Z})$ such that
\begin{equation}
H\cdot H=1\kom E_{i}\cdot E_{j}=-\delta_{ij}\kom H\cdot E_{i}=0\, .
\end{equation}
Here, $H$ is the hyperplane class, while the $E_{i}$ are $n$ exceptional curves from blowing up points in $\mathbb{P}^{2}$. The canonical class of a del Pezzo surface is given by
\begin{equation}
K(\mathrm{dP}_{n})=-3H+\sum_{i=1}^{n}\, E_{i}\:.
\end{equation}
The exceptional collection in \cite{Wijnholt:2007vn} is characterised by the charge vectors
\begin{align}
\mathrm{ch}(F_{1})&=\left (1,H-E_{1},0\right ) &\mathrm{deg}(F_{1})&=2\, ,\nn\\
\mathrm{ch}(F_{2})&=\left (1,H-E_{2},0\right ) &\mathrm{deg}(F_{2})&=2\, ,\nn\\
\mathrm{ch}(F_{3})&=-\left (1,2H-E_{1}-E_{2}-E_{4},\dfrac{1}{2}\right ) &\mathrm{deg}(F_{3})&=-3\, ,\nn\\
\mathrm{ch}(F_{4})&=-\left (1,2H-E_{1}-E_{2}-E_{5},\dfrac{1}{2}\right ) &\mathrm{deg}(F_{4})&=-3\, ,\nn\\
\mathrm{ch}(F_{5})&=-\left (1,H-E_{3},0\right )\kom &\mathrm{deg}(F_{5})&=-2\, ,\nn\\
\mathrm{ch}(F_{6})&=-\left (1,E_{4}+E_{5},-1\right ) &\mathrm{deg}(F_{6})&=-2\, ,\nn\\
\mathrm{ch}(F_{7})&=\left (1,H,\dfrac{1}{2}\right ) &\mathrm{deg}(F_{7})&=3\, ,\nn\\
\mathrm{ch}(F_{8})&=\left (1,2H-E_{1}-E_{2}-E_{3},\dfrac{1}{2}\right )&\mathrm{deg}(F_{8})&=3\, .
\end{align}
Since this is an exceptional collection, we
use~\eqref{eq:AsymEulerChar} to compute the number of fields
between each node. We find
\begin{equation}
\chi_{-}(F_{i},F_{j})=\left (\begin{array}{cccccccc}
0 & 0 & 1 & 1 & 0 & 0 & -1 & -1 \\ 
0 & 0 & 1 & 1 & 0 & 0 & -1 & -1 \\ 
-1 & -1 & 0 & 0 & 1 & 1 & 0 & 0 \\ 
-1 & -1 & 0 & 0 & 1 & 1 & 0 & 0 \\ 
0 & 0 & -1 & -1 & 0 & 0 & 1 & 1 \\ 
0 & 0 & -1 & -1 & 0 & 0 & 1 & 1 \\ 
1 & 1 & 0 & 0 & -1 & -1 & 0 & 0 \\ 
1 & 1 & 0 & 0 & -1 & -1 & 0 & 0
\end{array} \right ) \:.
\end{equation}
The corresponding quiver diagram is shown in Fig.~\ref{fig:QuivdP5}. We observe that there is at most one line between each node. The direction of the arrow between node $i$ and $j$ is determined by the sign of $\chi_{-}(F_{i},F_{j})$.

Several key aspects of the worldvolume theory deserve further scrutiny, namely
gauge anomaly cancellation, anomalous $\mathrm{U}(1)$'s and orientifold involutions.

\subsection*{Gauge anomalies}

Gauge anomalies in quiver gauge theories are easily computed through counting incoming and outgoing arrows at the various nodes, see e.g. Sect.~2~in~\cite{Yamazaki:2008bt}. Let us denote with $I_{ij}$ an edge between nodes $i$ and $j$. Then we define
\begin{equation}
\sigma(I_{ij},i)=\begin{cases}
+1 & i\text{ endpoint}\\
-1 & i\text{ startingpoint}\\
0 & \nexists I_{ij}=\text{ no edge}\, .
\end{cases}
\end{equation}
For each node $i$, anomaly cancellation is ensured if
\begin{equation}
\sum_{j}\, \sigma(I_{ij},i)\, N_{j}=0\, .
\end{equation}
This is equivalent to the statement in eq.~\eqref{eq:NonAbGAnom}. For the $\mathrm{dP}_{5}$-quiver in Fig.~\ref{fig:QuivdP5}, we obtain the two conditions
\begin{align}
\label{eq:AnomalydP5C1} N_{1}+N_{2}&=N_{5}+N_{6}\, ,\\
\label{eq:AnomalydP5C2} N_{3}+N_{4}&=N_{7}+N_{8}\, .
\end{align}
The models to be discussed below fall into two classes satisfying the above constraints through the choices
\begin{itemize}
\item model I:
\begin{equation}\label{eq:AnCancCMOne} 
N_{3}=N_{8}\kom N_{4}=N_{7}\kom N_{1}=N_{5}\kom N_{2}=N_{6}\, ;
\end{equation}
\item model II:
\begin{equation}\label{eq:AnCancCMTwo} 
N_{3}=N_{8}\kom N_{4}=N_{7}\kom N_{5}=N_{6}\kom N_{1}+N_{2}=2N_{5}\, .
\end{equation}
\end{itemize}
Below, we argue that the individual choices require distinct types of orientifold involutions.

\subsection*{$\mathrm{U}(1)$ charges}

Each of the $8$ nodes of the quiver in Fig.~\ref{fig:QuivdP5} comes with an $\mathrm{U}(1)$ factor.
The $\mathrm{U}(1)$ charges $q_{i}$ of the corresponding fields can be determined by associating to each ingoing arrow $+1$, to each outgoing arrow $-1$ and to no arrow $0$. Thus, we find for each field $X_{ab}$ the charge vectors $\mathbf{q}^{(ab)}$ with entries $q^{(ab)}_{i}$, $i=1,\ldots ,8$, and
\begin{align}
{q}^{(ab)}_{i}=\delta^{a}_{i}-\delta^{b}_{i}
\end{align}
so that for instance
\begin{equation}
\mathbf{q}^{(13)}=\left (1,0,-1,0,0,0,0,0\right )\, .
\end{equation}
Out of these 8 $\mathrm{U}(1)$'s two have to be anomalous according to our previous reasoning.

In order to find the anomaly-free and anomalous $\mathrm{U}(1)$'s we can look for combinations of the type
\be
Q_i=\sum_{j=1}^8 C_{ij}\frac{q_j}{N_j} \:.
\ee
Normalising $q_i\cdot q_j=N_i N_j\delta_{ij}$ we find an orthogonal basis for the 6 non-anomalous $\mathrm{U}(1)$'s as follows:
\bea\label{eq:AnNonAnUOChargesdP5} 
Q_1 =  \sum_{j=1}^8 \frac{q_j}{N_j} & \qquad\qquad &
 Q_2  =    \frac{q_1}{N_1}-\frac{q_2}{N_2}  \\
   Q_3   =  \frac{q_3}{N_3}-\frac{q_4}{N_4} & \qquad\qquad &
 Q_4  =    \frac{q_5}{N_5}-\frac{q_6}{N_6} \nonumber \\
Q_5   =  \frac{q_7}{N_7}-\frac{q_8}{N_8} &   &
Q_6  =   \frac{q_1}{N_1}+\frac{q_2}{N_2}- \frac{q_3}{N_3}-\frac{q_4}{N_4}+\frac{q_5}{N_5}+\frac{q_6}{N_6}- \frac{q_7}{N_7}-\frac{q_8}{N_8}\nonumber
\eea
 The 2 anomalous\footnote{It turns out that all rank zero and degree zero combinations of fractional branes are free of anomalies because the chiral spectrum from \eqref{eq:AsymEulerChar} vanishes \cite{Buican:2006sn}. Indeed, we find that only $Q_{7}$ and $Q_{8}$ have non-vanishing ranks and degrees corresponding to the expected two anomalous $\mathrm{U}(1)$'s.} $\mathrm{U}(1)$'s are:
 \be\label{eq:AnAnUOChargesdP5} 
 Q_7  =    \frac{q_1}{N_1}+\frac{q_2}{N_2} -  \frac{q_5}{N_5}-\frac{q_6}{N_6} \,,
 \qquad\qquad Q_8  =  \frac{q_3}{N_3}+\frac{q_4}{N_4}-\frac{q_7}{N_7}-\frac{q_8}{N_8}\:.
\ee
One can check that these $\mathrm{U}(1)$ charges are othorgonal so that
\begin{equation}
Q_{i}Q_{j}=\mathrm{diag}\left (8,2,2,2,2,8,4,4\right )\, .
\end{equation}
We summarised the $\mathrm{U}(1)$ charges of the bi-fundamentals $X_{ij}$ in Tab.~\ref{tab:table1}.

\begin{table}[t!]
  \centering
  
  \caption{List of $\mathrm{U}(1)$ charges for the bi-fundamentals appearing in the quiver diagram~\ref{fig:QuivdP5}.}
  \label{tab:table1}

  \begin{tabular}{|c||c|c|c|c|c|c||c|c|}
 \hline
  &&&&&&&&\\[-1.2em]
 Field & $Q_1$ & $Q_2$ & $Q_3$ & $Q_4$ & $Q_5$ & $Q_6$ & $Q_7$ & $Q_8$ \\[0.2em]
 \hline\hline
 &&&&&&&&\\[-1.1em]
  $X_{13}$ & $\frac{1}{N_1}-\frac{1}{N_3}$ & $\frac{1}{N_1}$ & $-\frac{1}{N_3}$ & 0 & 0 &$ \frac{1}{N_1}+\frac{1}{N_3}$ & $ \phantom{x} \frac{1}{N_1} \phantom{x} $ & $ -\frac{1}{N_3}$\\[0.4em] \hline
   &&&&&&&&\\[-1.1em]
   $X_{14}$ & $\frac{1}{N_1}-\frac{1}{N_4}$ & $\frac{1}{N_1}$  & $\frac{1}{N_4}$  & 0 & 0 & $\frac{1}{N_1}+\frac{1}{N_4}$  & $\frac{1}{N_1}$ & $- \frac{1}{N_4}$\\[0.4em] \hline
    &&&&&&&&\\[-1.1em]
   $X_{23}$ & $\frac{1}{N_2}-\frac{1}{N_3}$ & $-\frac{1}{N_2}$ & $-\frac{1}{N_3}$ & 0 & 0 & $\frac{1}{N_2}+\frac{1}{N_3}$ & $\frac{1}{N_2}$ & $-\frac{1}{N_3}$ \\[0.4em] \hline
    &&&&&&&&\\[-1.1em]
   $X_{24}$ & $\frac{1}{N_2}-\frac{1}{N_4}$ & $-\frac{1}{N_2}$ & $\frac{1}{N_4}$ & 0 & 0 & $\frac{1}{N_2}+\frac{1}{N_4}$ & $\frac{1}{N_2}$ & $-\frac{1}{N_4}$ \\ [0.4em]\hline
    &&&&&&&&\\[-1.1em]
   $X_{35}$ & $\frac{1}{N_3}-\frac{1}{N_5}$ & 0 & $\frac{1}{N_3}$ & $-\frac{1}{N_5}$ & 0 & $-\frac{1}{N_3}-\frac{1}{N_5}$ & $\frac{1}{N_5}$ & $\frac{1}{N_3}$ \\[0.4em] \hline
    &&&&&&&&\\[-1.1em]
   $X_{36}$ & $\frac{1}{N_3}-\frac{1}{N_6}$ & 0 & $\frac{1}{N_3}$ & $\frac{1}{N_6}$ & 0 & $-\frac{1}{N_3}-\frac{1}{N_6}$ & $\frac{1}{N_6}$ & $\frac{1}{N_3}$ \\[0.4em] \hline
    &&&&&&&&\\[-1.1em]
   $X_{45}$ & $\frac{1}{N_4}-\frac{1}{N_5}$ & 0 & $-\frac{1}{N_4}$ & $-\frac{1}{N_5}$ & 0 & $-\frac{1}{N_4}-\frac{1}{N_5}$ & $\frac{1}{N_5}$ & $\frac{1}{N_4}$ \\[0.4em] \hline
    &&&&&&&&\\[-1.1em]
   $X_{46}$ & $\frac{1}{N_4}-\frac{1}{N_6}$ & 0 & $-\frac{1}{N_4}$ & $\frac{1}{N_6}$ & 0 & $-\frac{1}{N_4}-\frac{1}{N_6}$ & $\frac{1}{N_6}$ & $\frac{1}{N_4}$ \\[0.4em] \hline
    &&&&&&&&\\[-1.1em]
   $X_{57}$ & $\frac{1}{N_5}-\frac{1}{N_7}$ & 0 & 0 & $\frac{1}{N_5}$ &  $-\frac{1}{N_7}$ & $\frac{1}{N_5}+\frac{1}{N_7}$  & $-\frac{1}{N_5}$ & $\frac{1}{N_7}$ \\[0.4em] \hline
    &&&&&&&&\\[-1.1em]
   $X_{58}$ & $\frac{1}{N_5}-\frac{1}{N_8}$ & 0 & 0 & $\frac{1}{N_5}$ &  $\frac{1}{N_8}$ & $\frac{1}{N_5}+\frac{1}{N_8}$  & $-\frac{1}{N_5}$ & $\frac{1}{N_8}$ \\[0.4em] \hline
    &&&&&&&&\\[-1.1em]
   $X_{67}$ & $\frac{1}{N_6}-\frac{1}{N_7}$ & 0 & 0 & $-\frac{1}{N_6}$ &  $-\frac{1}{N_7}$ & $\frac{1}{N_6}+\frac{1}{N_7}$  & $-\frac{1}{N_6}$ & $\frac{1}{N_7}$ \\[0.4em] \hline
    &&&&&&&&\\[-1.1em]
   $X_{68}$ & $\frac{1}{N_6}-\frac{1}{N_8}$ & 0 & 0 & $-\frac{1}{N_6}$ &  $\frac{1}{N_8}$ & $\frac{1}{N_6}+\frac{1}{N_8}$  & $-\frac{1}{N_6}$ & $\frac{1}{N_8}$ \\[0.4em] \hline
    &&&&&&&&\\[-1.1em]
   $X_{71}$ & $\frac{1}{N_7}-\frac{1}{N_1}$ &  $-\frac{1}{N_1}$ & 0 & 0 &  $\frac{1}{N_7}$ & $-\frac{1}{N_7}-\frac{1}{N_1}$  & $-\frac{1}{N_1}$ & $-\frac{1}{N_7}$ \\[0.4em] \hline
    &&&&&&&&\\[-1.1em]
   $X_{72}$ & $\frac{1}{N_7}-\frac{1}{N_2}$ &  $\frac{1}{N_2}$ & 0 & 0 &  $\frac{1}{N_7}$ & $-\frac{1}{N_7}-\frac{1}{N_2}$  & $-\frac{1}{N_2}$ & $-\frac{1}{N_7}$ \\[0.4em] \hline
    &&&&&&&&\\[-1.1em]
   $X_{81}$ & $\frac{1}{N_8}-\frac{1}{N_1}$ &  $-\frac{1}{N_1}$ & 0 & 0 &  $-\frac{1}{N_8}$ & $-\frac{1}{N_8}-\frac{1}{N_1}$  & $-\frac{1}{N_1}$ & $-\frac{1}{N_8}$ \\[0.4em] \hline
    &&&&&&&&\\[-1.1em]
   $X_{82}$ & $\frac{1}{N_8}-\frac{1}{N_2}$ &  $\frac{1}{N_2}$ & 0 & 0 &  $-\frac{1}{N_8}$ & $-\frac{1}{N_8}-\frac{1}{N_2}$  & $-\frac{1}{N_2}$ & $-\frac{1}{N_8}$ \\[0.4em] \hline
  \end{tabular}
\end{table}

\subsection*{Orientifold involutions -- Models I and II}

In this section, we define two unoriented models based on different orientifold actions. In general, we require that the orientifold involution leaves the superpotential invariant.
For simplicity, we restrict to $\mathrm{USp}$-projections.
In the toric limit, the superpotential is obtained from a dimer diagram and given by \cite{Wijnholt:2007vn}
\begin{align}\label{eq:dPFiveToricW}
W_{Q}^{\text{toric}}&=X_{13}X_{35}X_{58}X_{81}-X_{14}X_{46}X_{68}X_{81}+X_{14}X_{45}X_{57}X_{71}-X_{13}X_{36}X_{67}X_{71}\nn\\
&\quad+X_{24}X_{46}X_{67}X_{72}-X_{23}X_{35}X_{57}X_{72}+X_{23}X_{36}X_{68}X_{82}-X_{24}X_{45}X_{58}X_{82}\, .
\end{align}

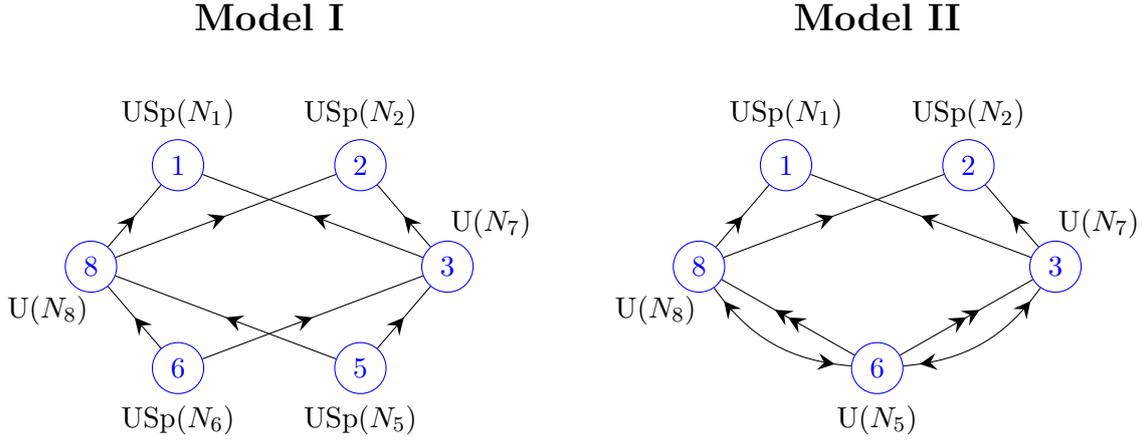
\begin{figure}
\centering
\begin{tikzpicture}[scale=0.8]
\draw (1.5,5.75) node[above=0pt] {\Large\bf Model I};
\draw (11.5,5.75) node[above=0pt] {\Large\bf Model II};
\draw (0,3.25) node(1)[anchor=south,circle,draw,blue]{1}node[above=20pt] {$\mathrm{USp}(N_{1})$};
\draw (3,3.25) node(2)[anchor=south,circle,draw,blue]{2}node[above=20pt] {$\mathrm{USp}(N_{2})$};
\draw (-1.,2) node(8)[anchor=east,circle,draw,blue]{8}node[below left=10pt] {$\mathrm{U}(N_{8})$};
\draw (4.,2) node(3)[anchor=west,circle,draw,blue]{3}node[above right=10pt] {$\mathrm{U}(N_{7})$};
\draw (0,0.75) node(6)[anchor=north,circle,draw,blue]{6}node[below=20pt] {$\mathrm{USp}(N_{6})$};
\draw (3,0.75) node(5)[anchor=north,circle,draw,blue]{5}node[below=20pt] {$\mathrm{USp}(N_{5})$};
\draw[directed,black] (3) -- (1);
\draw[directed,black] (3) -- (2);
\draw[directed,black] (5) -- (3);
\draw[directed,black] (6) -- (3);
\draw[directed,black] (5) -- (8);
\draw[directed,black] (6) -- (8);
\draw[directed,black] (8) -- (1);
\draw[directed,black] (8) -- (2);
\draw (10.,3.25) node(1)[anchor=south,circle,draw,blue]{1}node[above=20pt] {$\mathrm{USp}(N_{1})$};
\draw (13.,3.25) node(2)[anchor=south,circle,draw,blue]{2}node[above=20pt] {$\mathrm{USp}(N_{2})$};
\draw (9.,2) node(8)[anchor=east,circle,draw,blue]{8}node[below left=10pt] {$\mathrm{U}(N_{8})$};
\draw (14.,2) node(3)[anchor=west,circle,draw,blue]{3}node[above right=10pt] {$\mathrm{U}(N_{7})$};
\draw (11.5,0.75) node(5)[anchor=north,circle,draw,blue]{6}node[below=20pt] {$\mathrm{U}(N_{5})$};
\draw[directed,black] (3) -- (1);
\draw[directed,black] (3) -- (2);
\draw[directedTwo,black] (5) -- (3);
\draw[directedTwo,black] (5) -- (8);
\draw[directedBF,black] (5) to[bend right=25] (3);
\draw[directedBF,black] (5) to[bend left=25] (8);
\draw[directed,black] (8) -- (1);
\draw[directed,black] (8) -- (2);
\end{tikzpicture}
\caption{Unoriented quiver diagrams for model I (left) and model II (right)}\label{fig:QuivdP5OIModel1And2} 
\end{figure}

The first model referred to as \emph{model I} in the following in analogy to \cite{Wijnholt:2007vn} corresponds to the orientifold action
\begin{equation}\label{eq:OAModel1} 
1\leftrightarrow 1^{*}\kom 2\leftrightarrow 2^{*}\kom 3\leftrightarrow 8^{*}\kom 4\leftrightarrow 7^{*}\kom 5\leftrightarrow 5^{*}\kom 6\leftrightarrow 6^{*}\, .
\end{equation}
This is consistent with \eqref{eq:AnCancCMOne} and implies that
\begin{align}\label{eq:OAModel1VEVs} 
X_{81}=X_{13}^{T}\gamma_{1}\kom X_{71}=a^{-1}X_{14}^{T}\gamma_{1}\kom X_{82}=b^{-1}X_{23}^{T}\gamma_{2}\kom X_{72}=X_{24}^{T}\gamma_{2}	,,\nn\\
X_{68}=\gamma_{6}^{-1}X_{36}^{T}\kom X_{58}=\gamma_{5}^{-1}X_{35}^{T}\kom X_{67}=\gamma_{6}^{-1} X_{46}^{T}\kom X_{57}=\gamma_{5}^{-1}X_{45}^{T}	\,,
\end{align}
in terms of some phases $a$ and $b$ and $\gamma_{i}^{T}=-\gamma_{i}$ for $\mathrm{USp}$-projections. The resulting quiver diagram is depicted on the left side of Fig.~\ref{fig:QuivdP5OIModel1And2}. For the toric superpotential \eqref{eq:dPFiveToricW}, the projection on node $1$ and $5$ as well as $2$ and $6$ must always be identical.

There is a second class of models called \emph{model II} in \cite{Wijnholt:2007vn} satisfying \eqref{eq:AnCancCMOne}.
The orientifold action reads
\begin{equation}\label{eq:OAModel2} 
1\leftrightarrow 1^{*}\kom 2\leftrightarrow 2^{*}\kom 3\leftrightarrow 8^{*}\kom 4\leftrightarrow 7^{*}\kom 5\leftrightarrow 6^{*}\, .
\end{equation}
This implies that the fields are fixed as
\begin{align}
&X_{81}=X_{13}^{T}\gamma_{1}\kom X_{71}=X_{14}^{T}\gamma_{2}\kom X_{82}=X_{23}^{T}\gamma_{2}\kom X_{72}=X_{24}^{T}\gamma_{2}\, ,\nn\\
&X_{68}=X_{35}^{T}\kom X_{58}=X_{36}^{T}\kom X_{67}=X_{45}^{T}\kom X_{57}=X_{46}^{T}\, .
\end{align}
The quiver diagram is shown on the right of Fig.~\ref{fig:QuivdP5OIModel1And2}.

An important question concerns gauge anomalies in the orientifolded quivers.
This becomes an issue whenever the projection acts differently on the positive and negative contributions to an anomaly \cite{Wijnholt:2007vn,Yamazaki:2008bt}.
It is typically the case for orientifolded gauge theories containing (anti-)symmetric tensor representations which arise for fixed points on edges. These representations contribute $N\pm 4$ to the anomaly coefficient rather than $N$ as for the bi-fundamentals in the parent theory, see e.g. footnote 2 of \cite{Heckman:2007zp}.
In both of our models, there are no fixed points on edges and consequently orientifolding does not change the anomaly cancellation conditions which is consistent with the findings of \cite{Argurio:2020dko}.

\subsection{Higgsing the $\mathrm{dP}_{5}$ quiver gauge theory -- Model I}\label{sec:HiggsingModel1} 

The two unoriented quivers in Fig.~\ref{fig:QuivdP5OIModel1And2} contain already the chiral spectrum of the MSSM, albeit only one family of each quarks and leptons.
The necessary number of chiral families together with the right amount of non-chiral matter are obtained from a higher rank gauge theory by turning on suitable Vacuum Expectation Values (VEVs) for bi-fundamental matter fields.
This section is concerned with this Higgsing procedure for model I of the $\mathrm{dP}_{5}$ quiver.
A similar analysis for model II is summarised in appendix~\ref{sec:HiggsinMod2}.
The Higgsing works essentially in a two step procedure by first obtaining a version of a left-right symmetric model\footnote{Recall that there exists an accidental isomorphism $\mathrm{USp}(2)\cong \mathrm{SU}(2)$ with both being used interchangeably in what follows.} $\mathrm{U}(3)\times \mathrm{USp}(2)_{L}\times \mathrm{USp}(2)_{R}\times \mathrm{U}(1)$ which subsequently needs to be reduced to the (MS)SM gauge group via conventional Higgsing $\mathrm{USp}(2)_{R}\times\mathrm{U}(1)\raw \mathrm{U}(1)$ \cite{Wijnholt:2007vn,Heckman:2007zp}.

\subsection*{Breaking patterns for bi-fundamental VEVs}

We begin our analysis with outlining the breaking pattern for fields in bi-fundamental representations of $\mathrm{U}(N_{1})\times \mathrm{U}(N_{2})$. We choose a basis of generators $T_{j}\in \lbrace h_{0},h_{i},E_{pq}^{+},E_{pq}^{-}\rbrace$, $j=1,\ldots ,N^{2}$, for the Lie algebra $\mathfrak{u}(N)$ by defining
\begin{align}\label{eq:BasisUN} 
h_{i}&=\I\left (e_{i,i}-e_{i+1,i+1}\right )\kom i=1,2,\ldots ,N-1\\
E_{pq}^{-}&=e_{pq}-e_{qp}\kom p,q=1,2,\ldots , N\kom p<q\\
E_{pq}^{+}&=\I(e_{pq}+e_{qp})\kom p,q=1,2,\ldots , N\kom p<q\\
h_{0}&=\I\mathds{1}_{N}
\end{align}
where $e_{pq}$ are the $N\times N$-matrices $(e_{pq})_{ik}=\delta_{ip}\delta_{kq}$.
In the remainder of this section, we look at the gauge-invariant kinetic terms for different bi-fundamental fields $(\mathbf{N}_{1},\overline{\mathbf{N}}_{2})$ analysing the breaking patterns for distinct choices of VEVs through the resulting mass matrices for gauge bosons.

As a warm-up, we consider a bi-fundamental field $X$ in $(\mathbf{N},\overline{\mathbf{N}})$ of $\mathrm{U}(N)^{2}$.
This is a simple setup for a matter field between two nodes of $N$ fractional branes.
We may choose VEVs of the form
\begin{equation}
\langle X\rangle=\mathrm{diag}(a_{1},\ldots,a_{N})\, .
\end{equation}
This choice is in unitary gauge where all Goldstone modes are being absorbed by massive gauge bosons to become the longitudinal degree of freedom.
The group $\mathrm{U}(N)^{2}$ is broken to a subgroup $H$ depending on the choice of constants $a_{i}$.
Here, we distinguish three scenarios.
First, if all the $a_{i}$ are inequivalent, then $H=\mathrm{U}(1)^{N}$ so that the number of Goldstone modes is $2N^{2}-N$.
The original field $X$ encodes $2N^{2}$ real degrees of freedom so that after Higgsing $N$ potentially massive fields (the Higgs bosons) remain.
The second scenario considers $a_{i}=a$ for all $i$ which implies $H=\mathrm{U}(N)$.
The number of Goldstone modes is $2N^{2}-N^{2}=N^{2}$ so that $N^{2}$ real degrees of freedom remain as Higgs fields in the adjoint of $\mathrm{U}(N)$.
Finally, the third configuration consists of block-like structures
\begin{equation}
\langle X\rangle=\mathrm{diag}(\underbrace{a_{1},\ldots,a_{1}}_{n_{1}},\underbrace{a_{2},\ldots ,a_{2}}_{n_{2}},\ldots,\underbrace{a_{M},\ldots,a_{M}}_{n_{M}})\kom \sum_{k=1}^{M}\, n_{k}=N
\end{equation}
so that $H=\mathrm{U}(n_{1})\times\ldots\times \mathrm{U}(n_{M})$.
In this case, we find $2N^{2}-\sum_{i=1}^{M}\, n_{i}^{2}$ Goldstone modes and $n_{i}^{2}$ Higgses survive in the adjoint of $\mathrm{U}(n_{i})$ for all $i=1,\ldots,M$.

The next stage is a matter field $X$ in the bi-fundamental $(\mathbf{N+M},\overline{\mathbf{N}})$ of $\mathrm{U}(N+M)\times \mathrm{U}(N)$.
A VEV of the form
\begin{equation}
\langle X\rangle=\left (\begin{array}{c}
\mathrm{diag}(a_{1},\ldots,a_{N}) \\ 
0_{M\times N}
\end{array} \right )
\end{equation}
breaks the group to $H=\mathrm{U}(M)\times \mathrm{U}(N)$ for $a_{i}=a$ for all $i$.
The number of Goldstone modes is given by $(N+M)^{2}+N^{2}-M^{2}-N^{2}=2MN+N^{2}$, while the number of surviving matter fields is $2(N+M)N-(2MN+N^{2})=N^{2}$ in the adjoint of $\mathrm{U}(N)$.

\begin{table}
\centering
\begin{tabular}{|c|c|c|}
\hline 
$\langle X\rangle$ & groups &  choices \\ 
\hline 
\hline 
 &  &    \\ [-0.6em]
\multirow{3}{5.5cm}{$\left (\begin{array}{ccc}
a_{1}\mathds{1}_{N_{1}\times N_{1}} & 0_{N_{1}\times N_{2}} &0_{N_{1}\times M}\\ 
0_{N_{2}\times N_{1}}& a_{2}\mathds{1}_{N_{2}\times N_{2}}  &0_{N_{2}\times M}\\
0_{N_{2}\times N_{1}}& a_{3}\mathds{1}_{N_{2}\times N_{2}} &0_{N_{2}\times M}
\end{array} \right )$} & $G=\mathrm{U}(N_{1}+2N_{2})\times \mathrm{U}(N_{1}+N_{2}+M)$ &   $a_{2},a_{3}\neq 0$ \\ [0.4em]
 & &  $a_{2}=0$  \\ [0.4em]
  & $H=\mathrm{U}(N_{1})\times \mathrm{U}(N_{2})^{2}\times \mathrm{U}(M)$ & $a_{3}=0$  \\ [0.4em]
\hline 
 &  &    \\ [-0.6em]
\multirow{2}{5.5cm}{$\left (\begin{array}{ccc}
a_{1}\mathds{1}_{N_{1}\times N_{1}} & a_{2}\mathds{1}_{N_{1}\times N_{1}} &0_{N_{1}\times M}\\ 
0_{N_{2}\times N_{1}}&0_{N_{2}\times N_{1}}  &0_{N_{2}\times M}\\
\end{array} \right ) $ }& $G=\mathrm{U}(N_{1}+N_{2})\times \mathrm{U}(2N_{1}+M)$  & $a_{2}=0$  \\ [1.em]
 & $H=\mathrm{U}(N_{1})^{2}\times \mathrm{U}(N_{2}) \times \mathrm{U}(M)$ & $a_{2}\neq 0$\\ [0.4em]
\hline 
&  &    \\ [-0.6em]
\multirow{3}{5.5cm}{$\left (\begin{array}{ccc}
a_{1}\mathds{1}_{N_{1}\times N_{1}} & 0_{N_{1}\times N_{2}} &0_{N_{1}\times M}\\ 
a_{2}\mathds{1}_{N_{1}\times N_{1}}& 0_{N_{1}\times N_{2}}  &0_{N_{1}\times M}\\
0_{N_{2}\times N_{1}}& a_{3}\mathds{1}_{N_{2}\times N_{2}} &0_{N_{2}\times M}
\end{array} \right )$} & $G=\mathrm{U}(2N_{1}+N_{2})\times \mathrm{U}(N_{1}+N_{2}+M)$ &   $a_{1},a_{2}\neq 0$ \\ [0.4em]
 &  &  $a_{1}=0$  \\ [0.4em]
  & $H=\mathrm{U}(N_{1})^{2}\times \mathrm{U}(N_{2})\times \mathrm{U}(M)$ &  $a_{2}=0$  \\ [0.4em]
\hline 
\end{tabular} 
\caption{Breaking patterns of bi-fundamental matter fields $X$ for convenient block-like structures.}\label{tab:BPBifFund} 
\end{table}

More generally, we also make heavy use of the breaking patterns summarised in Tab.~\ref{tab:BPBifFund}.
We indicate various possible choices of constants in the third column that do not change the breaking pattern, but modify superpotential couplings.
Below, we require several different versions of these VEVs to break the original $\mathrm{dP}_{5}$ quiver gauge theory in Fig.~\ref{fig:QuivdP5} to a covering quiver of the (MS)SM.

\subsection*{Higgsing -- building bound states of fractional branes}

The quiver gauge theory obtained at the $\mathrm{dP}_{5}$ singularity contains $8$ factors of unitary groups together with $16$ bi-fundamentals.
The breaking pattern is thus significantly more involved.
To gain some intuition,
we begin by outlining the process of condensing several nodes in the original quiver to a single node.
We follow the ideas of \cite{Wijnholt:2007vn} by building \emph{bound states of fractional branes}.
In the Higgsed quiver, the new node is associated with $n$ copies of a new (bound state) fractional brane with associated gauge group $\mathrm{U}(n)$.

\begin{figure}[t!]
\centering
\begin{tikzpicture}[node distance=0.5cm]
\draw (0.5,6.5) node[above=0pt] {\large\bf Bound state $2F_{b}$};
\draw (0,5) node(1)[anchor=south,circle,draw,blue]{1}node[above=20pt] {$\mathrm{U}(2)$};
\draw (-1,4) node(8)[anchor=east,circle,draw,blue]{8}node[above left=15pt] {$\mathrm{U}(2)$};
\draw (-1,3) node(7)[anchor=east,circle,draw,blue]{7}node[below left=15pt] {$\mathrm{U}(2)$};
\draw (2,4) node(3)[anchor=west,circle,draw,blue]{3}node[above right=15pt] {$\mathrm{U}(2)$};
\draw (2,3) node(4)[anchor=west,circle,draw,blue]{4}node[below right=15pt] {$\mathrm{U}(2)$};
\draw (0,2) node(6)[anchor=north,circle,draw,blue]{6}node[below=20pt] {$\mathrm{U}(4)$};
\draw (1,2) node(5)[anchor=north,circle,draw,blue]{5}node[below=20pt] {$\mathrm{U}(4)$};
\draw[directed,black] (1) -- (3);
\draw[directed,black] (1) -- (4);
\draw[directed,black] (3) -- (5);
\draw[directed,black] (3) -- (6);
\draw[directed,black] (4) -- (5);
\draw[directed,black] (4) -- (6);
\draw[directed,black] (5) -- (7);
\draw[directed,black] (5) -- (8);
\draw[directed,black] (6) -- (7);
\draw[directed,black] (6) -- (8);
\draw[directed,black] (7) -- (1);
\draw[directed,black] (8) -- (1);
\end{tikzpicture}
\hspace*{0.5cm}
\begin{tikzpicture}[node distance=0.5cm]
\draw (0.5,6.5) node[above=0pt] {\large\bf Bound state $2F_{a}$};
\draw (0,5) node(1)[anchor=south,circle,draw,blue]{1}node[above=20pt] {$\mathrm{U}(4)$};
\draw (1,5) node(2)[anchor=south,circle,draw,blue]{2}node[above=20pt] {$\mathrm{U}(4)$};
\draw (-1,4) node(8)[anchor=east,circle,draw,blue]{8}node[above left=15pt] {$\mathrm{U}(2)$};
\draw (-1,3) node(7)[anchor=east,circle,draw,blue]{7}node[below left=15pt] {$\mathrm{U}(2)$};
\draw (2,4) node(3)[anchor=west,circle,draw,blue]{3}node[above right=15pt] {$\mathrm{U}(2)$};
\draw (2,3) node(4)[anchor=west,circle,draw,blue]{4}node[below right=15pt] {$\mathrm{U}(2)$};
\draw (1,2) node(5)[anchor=north,circle,draw,blue]{5}node[below=20pt] {$\mathrm{U}(2)$};
\draw[directed,black] (1) -- (3);
\draw[directed,black] (1) -- (4);
\draw[directed,black] (2) -- (3);
\draw[directed,black] (2) -- (4);
\draw[directed,black] (3) -- (5);
\draw[directed,black] (4) -- (5);
\draw[directed,black] (5) -- (7);
\draw[directed,black] (5) -- (8);
\draw[directed,black] (7) -- (1);
\draw[directed,black] (8) -- (1);
\draw[directed,black] (7) -- (2);
\draw[directed,black] (8) -- (2);
\end{tikzpicture}
\caption{Partial quiver for the bound states $2F_{b}$ (left) and $2F_{a}$ (right) of model I.}\label{fig:QuivdP5Mod1SmallFB} 
\end{figure}
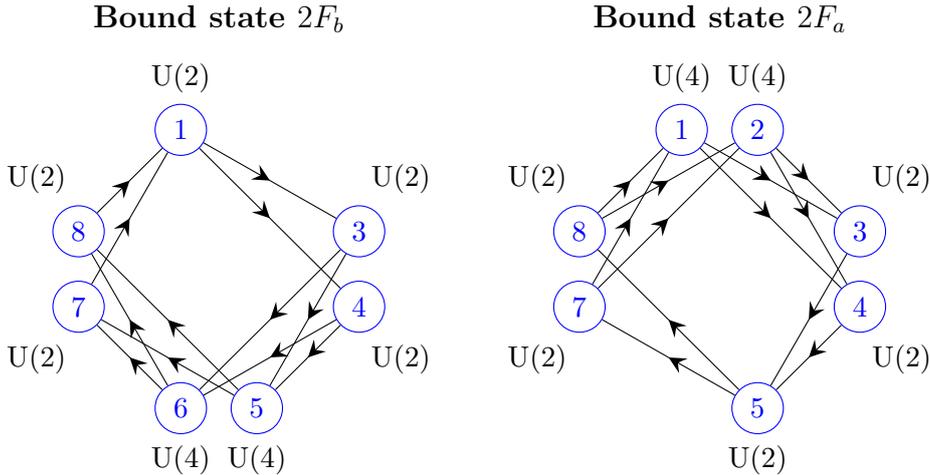

In the remainder of this section, we focus on model I.
Here, the basic working plan is to collapse nodes $1,2,5,6$ in Fig.~\ref{fig:QuivdP5} (together with a sub-sector at nodes $3,4,7,8$) to two individual $\mathrm{U}(2)$'s.
Having achieved this, we only need to ensure that the unbroken symmetry groups at nodes $3,4,7,8$ are $\mathrm{U}(3)$ and $\mathrm{U}(1)$.
Thus, after orientifolding, we end up with the left-right symmetric model.

All in all, these arguments necessitate the construction of two bound states associated with the two $\mathrm{U}(2)$'s.
Practically, bound states are built by combining the charge vectors of fractional branes in the exceptional collection $\lbrace F_{1},\ldots ,F_{8}\rbrace$.
To find suitable linear combinations, we recall that the intersection numbers $\chi_{-}(F_{i},F_{j})$ of the new basis of $6$ fractional branes count additional chiral families.
Looking at \eqref{eq:AsymEulerChar}, we obtain exactly three chiral families if we replace four of our original fractional branes $\lbrace F_{1},F_{2},F_{5},F_{6}\rbrace$ by two rank $\pm 3$ and degree $\mp 6$ objects $\lbrace F_{a},F_{b}\rbrace$.
More explicitly, this implies that for instance
\begin{equation}
\chi_{-}(F_{1},F_{3})=1\quad\raw\quad \chi_{-}(F_{a},F_{3})=3 \kom \chi_{-}(F_{b},F_{3})=-3\, ,
\end{equation}
and similarly for all other nodes.
We stress that the chiral intersection matrix \eqref{eq:AsymEulerChar} only depends on the rank and degree of fractional branes.
At this level of discussion, the non-chiral spectrum remains undetermined and must be analysed separately below.

The first linear combination to satisfy our constraints is defined as
\begin{equation}
\text{ch}(F_{b})=\sum_{i=1}^{8}\, n^{(b)}_{i}\text{ch}(F_{i})=-(3,2H-E_{2}-E_{3}+E_{4}+E_{5},-2)\kom \mathrm{deg}(F_{b})=-6
\end{equation}
in terms of the $8$-vector
\begin{equation}\label{eq:CHFB} 
\mathbf{n}^{(b)}=\left (1,0,1,1,2,2,1,1\right )\, .
\end{equation}
The bound state $2F_{b}$ is realised by turning on VEVs
\begin{align}\label{eq:VEVs2FB} 
&\langle X_{13}\rangle=\left (\begin{array}{c}
a_{1} 
\end{array} \right )\kom \langle X_{14}\rangle=\left (\begin{array}{c}
b_{1} 
\end{array} \right )\kom \langle X_{36}\rangle=\left (\begin{array}{cc}
y_{1} & 0 
\end{array} \right )\, ,\nn\\
&\langle X_{46}\rangle=\left (\begin{array}{cc}
0 & w_{2} 
\end{array} \right )\kom \langle X_{35}\rangle=\left (\begin{array}{cc}
x_{1} & 0 
\end{array} \right )\kom \langle X_{45}\rangle=\left (\begin{array}{cc}
0 & z_{2}
\end{array} \right )
\end{align}
where each entry corresponds to a $2\times 2$-matrix. We also fix the $\gamma$-matrices
\begin{equation}\label{eq:GamMatFB} 
\gamma_{1}=\I\sigma_{2}=\left (\begin{array}{cc}
0 & 1 \\ 
-1 & 0
\end{array} \right )\kom\gamma_{5}=\gamma_{6}=\left (\begin{array}{cc}
\I\sigma_{2} & 0\\ 
0 & \I\sigma_{2}
\end{array} \right )
\end{equation}
so that under orientifold projection the remaining VEVs are fixed through \eqref{eq:OAModel1VEVs} with $a=1$. 
Notice that $\gamma_{i}^{T}=-\gamma_{i}$ as required for a $USp$-projection. We consider the partial quiver diagram for this bound state on the left hand side in Fig.~\ref{fig:QuivdP5Mod1SmallFB} and find that indeed the above VEVs imply the breaking pattern
\begin{equation}
\mathrm{U}(2)^{5}\times \mathrm{U}(4)^{2}\raw \mathrm{U}(2)\xrightarrow{\text{Orientifolding }} \mathrm{USp}(2)\, .
\end{equation}
In the oriented model, we have $48$ Goldstone modes which leaves us with $64$ real scalars in adjoints of $\mathrm{U}(2)$.

The second bound state can be written as
\begin{equation}
\text{ch}(F_{a})=\sum_{i=1}^{8}\, n^{(a)}_{i}\text{ch}(F_{i})=(3,2H-E_{1}-E_{2}+E_{4}+E_{5},0)\kom \mathrm{deg}(F_{a})=6
\end{equation}
where
\begin{equation}\label{eq:CHFA} 
\mathbf{n}^{(a)}=\left (2,2,1,1,1,0,1,1\right )\, .
\end{equation}
We claim that the bound state $2F_{a}$ can be achieved by turning on VEVs
\begin{align}\label{eq:Model1BoundStateFaVEVs} 
&\langle X_{13}\rangle=\left (\begin{array}{c}
a_{2} \\
0 
\end{array} \right )\kom \langle X_{14}\rangle=\left (\begin{array}{c}
0 \\
b_{3}
\end{array} \right )\kom \langle X_{23}\rangle=\left (\begin{array}{c}
c_{1} \\
 0 
\end{array} \right )\, ,\nn\\
&\langle X_{24}\rangle=\left (\begin{array}{c}
0 \\
d_{2} 
\end{array} \right )\kom \langle X_{35}\rangle=\left (\begin{array}{c}
x_{3} 
\end{array} \right )\kom \langle X_{45}\rangle=\left (\begin{array}{c}
z_{3} 
\end{array} \right )
\end{align}
where each entry corresponds to a $2\times 2$-matrix. We also fix the $\gamma$-matrices in \eqref{eq:OAModel1VEVs} through
\begin{equation}
\gamma_{5}=\I\sigma_{2}=\left (\begin{array}{cc}
0 & 1 \\ 
-1 & 0
\end{array} \right )\kom\gamma_{1}=\gamma_{2}=\left (\begin{array}{cc}
\I\sigma_{2} & 0\\ 
0 & \I\sigma_{2}
\end{array} \right )\, .
\end{equation}
Analogously to the previous bound state, we checked that the VEVs indeed give rise to the breaking pattern
\begin{equation}
\mathrm{U}(2)^{5}\times \mathrm{U}(4)^{2}\raw \mathrm{U}(2)\xrightarrow{\text{Orientifolding }}\mathrm{USp}(2)\, .
\end{equation}
As before, the oriented model contains $64$ real scalars in adjoints of $\mathrm{U}(2)$.

\subsection*{Higgsing -- the full quiver}

\begin{figure}[t!]
\centering
\includegraphics[scale=0.32]{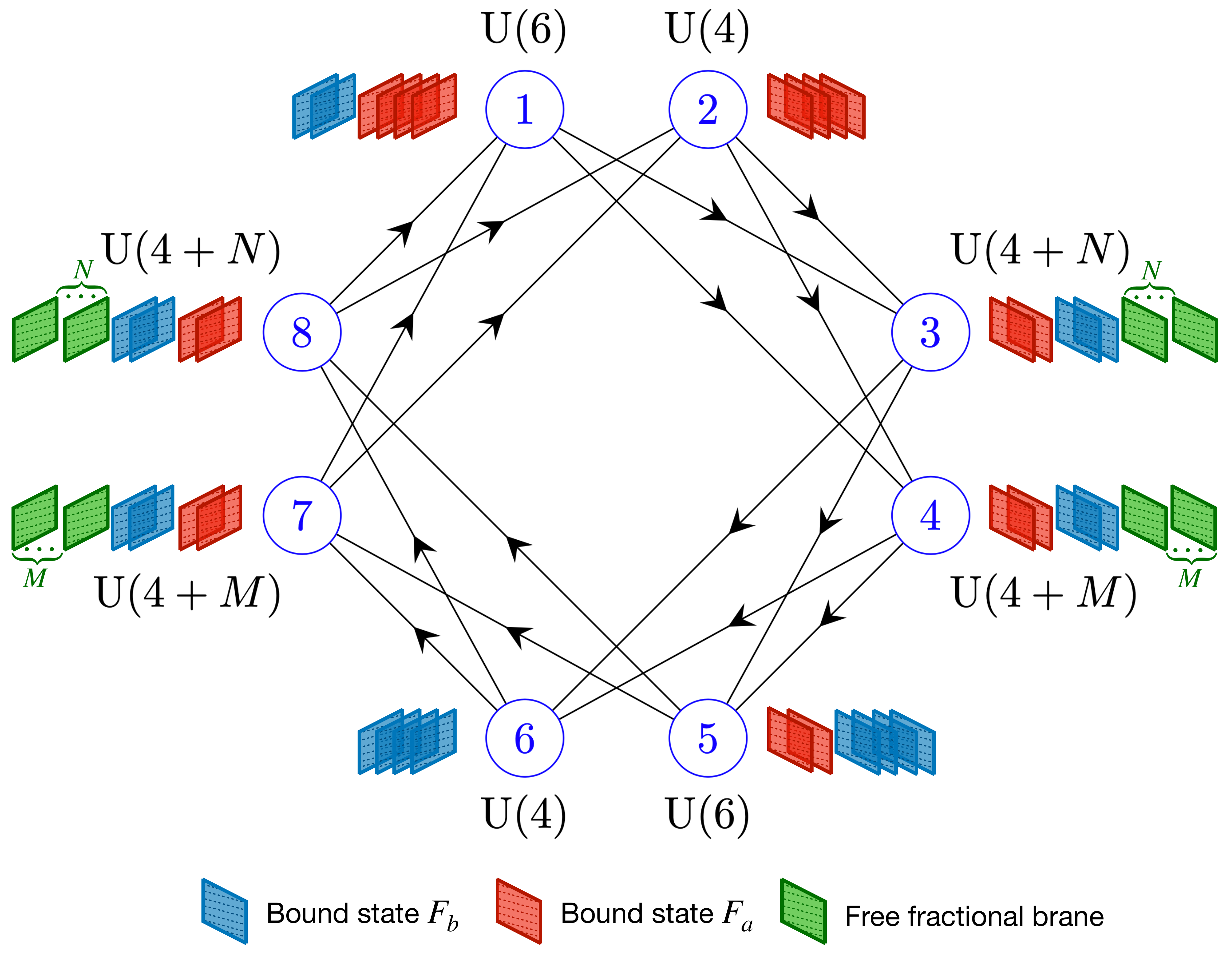}
\caption{Model I quiver diagram $\mathrm{dP}_{5}$ with the bound state $2F_{a}$ indicated in {\color{red}red} and $2F_{b}$ in {\color{blue}blue}. The {\color{green} green} fractional branes do not participate in any bound state.}\label{fig:QuivdP5Mod1Small} 
\end{figure}

Having built suitable bound states, we are now ready to construct the breaking pattern of the full $\mathrm{dP}_{5}$-quiver.
We require two copies of each bound state $F_{a},F_{b}$ to ensure the existence of two new $\mathrm{U}(2)$ factors.
With these guidelines,
we construct a quiver specified by $(2F_{a},NF_{3},MF_{4},2F_{b},MF_{7},NF_{8})$ corresponding to multiplicities $(6,4,4+N,4+M,6,4,4+M,4+N)$ of fractional branes in the exceptional collection $\lbrace F_{1},\ldots ,F_{8}\rbrace$ as illustrated in Fig.~\ref{fig:QuivdP5Mod1Small}. 
The choice $N=3$, $M=1$ has already been discussed in \cite{Wijnholt:2007vn}.
The bound state $2F_{a}$ is depicted in red in Fig.~\ref{fig:QuivdP5Mod1Small}, whereas $2F_{b}$ in blue.

The breaking of the full $\mathrm{dP}_{5}$-quiver is achieved by embedding the VEVs for the bound states $2F_{b}$ and $2F_{a}$ into representations of the full quiver.
To this end, we define the VEVs
\begin{align}\label{eq:HiggsingSmallMod1} 
\langle X_{13}\rangle &= \left( \begin{array}{ccc}
 a_1 & 0 & \cdots \\
 0 & a_2 & \cdots \\
 0 & a_3 & \cdots 
 \end{array}\right) \kom &
 \langle X_{14}\rangle &= \left( \begin{array}{ccc}
 b_1  & 0 & \cdots \\
 0 & b_2 & \cdots \\
 0 & b_3 & \cdots 
 \end{array}\right) \, ,
  \nonumber\\
  \langle X_{23} \rangle &= \left( \begin{array}{ccc}
 0 & c_1 & \cdots \\
 0 & c_2 & \cdots 
 \end{array}\right) \kom &
 \langle X_{24} \rangle &= \left( \begin{array}{ccc}
 0  & d_1 & \cdots \\
 0 & d_2 & \cdots 
 \end{array}\right) \, ,
  \nonumber\\
  \langle X_{35}^{^T}\rangle &= \left( \begin{array}{ccc}
 x_1 & 0 & \cdots \\
 x_2 & 0 & \cdots \\
 0 & x_3 & \cdots 
 \end{array}\right)\kom  &
 \langle X_{45}^{^T}\rangle &= \left( \begin{array}{ccc}
 z_1  & 0 & \cdots \\
 z_2 & 0 & \cdots \\
 0 & z_3 & \cdots 
 \end{array}\right)\, ,\nn\\
 \langle X_{36}^{^T}\rangle& = \left( \begin{array}{ccc}
 y_1 & 0 & \cdots \\
 y_2 & 0 & \cdots 
 \end{array}\right)\kom &
 \langle X_{46}^{^T}\rangle&= \left( \begin{array}{ccc}
 w_1  & 0 & \cdots \\
 w_2 & 0 & \cdots 
 \end{array}\right)
\end{align}
 where all entries are $2\times 2$-matrices.
The $\cdots$ represent $N$ zeros for the matrices on the left side and $M$ zeros for the matrices on the right.
The remaining VEVs for $\langle X_{81}\rangle$ etc. are fixed via the orientifold condition \eqref{eq:OAModel1VEVs} with
\begin{equation}
\gamma_{1}=\gamma_{5}=\I\sigma_{2}\otimes \mathds{1}_{3\times 3}\kom \gamma_{2}=\gamma_{6}=\I\sigma_{2}\otimes \mathds{1}_{2\times 2}\, .
\end{equation}
With our previous arguments, the breaking pattern is equivalently achieved by the choice of \cite{Wijnholt:2007vn} given by
\begin{align}\label{eq:VEVsAS} 
&\langle X_{13}\rangle\bigl |_{a_{3}=0}\kom \langle X_{14}\rangle\bigl |_{b_{2}=0}\kom \langle X_{23}\rangle\bigl |_{c_{2}=0}\kom \langle X_{24}\rangle\bigl |_{d_{1}=0}\, ,\nn\\
& \langle X_{35}^{T}\rangle\bigl |_{x_{2}=0}\kom \langle X_{45}^{T}\rangle\bigl |_{z_{1}=0}\kom \langle X_{36}^{T}\rangle\bigl |_{y_{2}=0} \kom \langle X_{46}^{T}\rangle\bigl |_{w_{1}=0}\, .
\end{align}

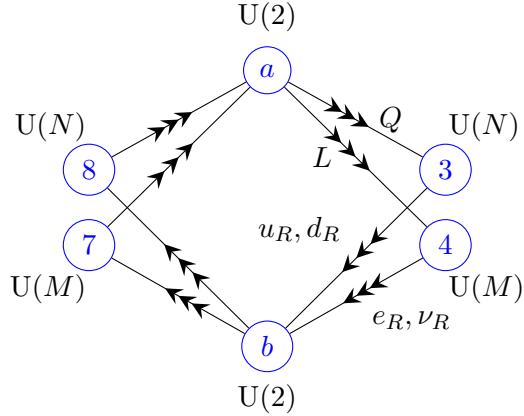
\begin{figure}
\centering
\begin{tikzpicture}[scale=1.]
\draw (0.,7.5) node[above=0pt] {\large\bf Higgsed quiver};
\draw (0,6) node(1)[anchor=south,circle,draw,blue]{$a$}node[above=20pt] {$\mathrm{U}(2)$};
\draw (-2,5) node(8)[anchor=east,circle,draw,blue]{8}node[above left=10pt] {$\mathrm{U}(N)$};
\draw (-2,4) node(7)[anchor=east,circle,draw,blue]{7}node[below left=10pt] {$\mathrm{U}(M)$};
\draw (2,5) node(3)[anchor=west,circle,draw,blue]{3}node[above right=10pt] {$\mathrm{U}(N)$};
\draw (2,4) node(4)[anchor=west,circle,draw,blue]{4}node[below right=10pt] {$\mathrm{U}(M)$};
\draw (0,3) node(5)[anchor=north,circle,draw,blue]{$b$}node[below=20pt] {$\mathrm{U}(2)$};
\draw[directedThree,black] (1) --node[right=5pt]{$Q$} (3);
\draw[directedThree,black] (1) --node[left=5pt]{$L$} (4);
\draw[directedThree,black] (3) --node[above left=3pt]{$u_{R},d_{R}$} (5);
\draw[directedThree,black] (4) --node[below right=3pt]{$e_{R},\nu_{R}$} (5);
\draw[directedThree,black] (5) -- (7);
\draw[directedThree,black] (5) -- (8);
\draw[directedThree,black] (7) -- (1);
\draw[directedThree,black] (8) -- (1);
\end{tikzpicture}
\caption{Higgsed quiver diagram $\mathrm{dP}_{5}$ showing only the chiral matter spectrum. }\label{fig:QuivdP5Mod1Higgsed} 
\end{figure}

We claim that both of these choices of VEVs break the original gauge group $G$ as
\begin{equation}
G=(\mathrm{U}(6)\times \mathrm{U}(4+N) \times \mathrm{U}(4+M) \times \mathrm{U}(4))^{2} \raw H=(\mathrm{U}(2)\times \mathrm{U}(N)\times \mathrm{U}(M))^{2} \, .
\end{equation}
The resulting quiver is shown in Fig.~\ref{fig:QuivdP5Mod1Higgsed}.
Let us try to see this more explicitly applying the observations from Tab.~\ref{tab:BPBifFund}. First notice that each $2\times 2$ matrix breaks a  $\mathrm{U}(2)^2$ to a diagonal $\mathrm{U}(2)$. Then   $\langle X_{13}\rangle$  breaks the original $\mathrm{U}(6)\times \mathrm{U}(4+N)$ of nodes $1$ and $3$ to a $\mathrm{U}(2)^3\times \mathrm{U}(N)$ with the first two $\mathrm{U}(2)$'s corresponding to diagonal $\mathrm{U}(2)$'s. Similarly $\langle X_{14} \rangle $ breaks $\mathrm{U}(6)\times \mathrm{U}(4+M)$ of nodes 1 and 4 to $\mathrm{U}(2)^3\times \mathrm{U}(M)$ with the first and third $\mathrm{U}(2)$ being diagonal (and the first combining with the first of the previous $\mathrm{U}(2)^3$).
Going through all the breaking patterns from these VEVs we can see that there are only two independent $\mathrm{U}(2)$'s surviving associated with the bound states $2F_{b}$ and $2F_{a}$.
In terms of the full quiver, the former one is a diagonal combination of the first $\mathrm{U}(2)$ on nodes 1,3,4,5,6 and the second of nodes 5 and 6. The second $\mathrm{U}(2)$ is a combination of the second and third of the $\mathrm{U}(2)$'s of node 1, the first and second $\mathrm{U}(2)$ of node 2 the second $\mathrm{U}(2)$ of nodes 3, 4 and the third of node 6.

\subsection{Matter spectrum for Model I}

\subsection*{Chiral matter}

Regarding the matter spectrum we have to concentrate on the decomposition of the original states.
For the fields connected to node $3$, the breaking pattern decomposes bi-fundamentals of $\mathrm{U}(6)\times\mathrm{U}(4+N)$ into suitable representations under $\mathrm{U}(2)\times\mathrm{U}(2)\times\mathrm{U}(N)$ such that
\begin{align}
&X_{13}:\quad (\mathbf{6},\overline{\mathbf{4+N}})\quad\longrightarrow\quad (\mathbf{2},\textbf{1},\overline{\mathbf{N}})+2(\mathbf{1},\textbf{2},\overline{\mathbf{N}})+\ldots\, ,\nn\\
&X_{23}:\quad (\mathbf{4},\overline{\mathbf{4+N}})\quad\longrightarrow\quad 2(\mathbf{1},\textbf{2},\overline{\mathbf{N}})+\ldots\, ,\nn\\
&X_{35}:\quad (\mathbf{4+N},\overline{\mathbf{4}})\quad\longrightarrow\quad 2(\mathbf{2},\textbf{1},{\mathbf{N}})+\ldots\, ,\nn\\
&X_{36}:\quad (\mathbf{4+N},\overline{\mathbf{6}})\quad\longrightarrow\quad 2(\mathbf{2},\textbf{1},{\mathbf{N}})+(\mathbf{1},\textbf{2},{\mathbf{N}})+\ldots\, .
\end{align}
Here, the $\ldots$ are additional fields charged under $\mathrm{U}(2)$, but not under $\mathrm{U}(N)$. Altogether, the matter content between $\mathrm{U}(N)$ and any of the $\mathrm{U}(2)$'s corresponds to
\begin{align}
&3\left ((\mathbf{2},\textbf{1},{\mathbf{N}})+(\mathbf{1},\textbf{2},\overline{\mathbf{N}})\right )\quad\longrightarrow\quad \text{3 chiral families}\, ,\nn\\
&(\mathbf{2},\textbf{1},{\mathbf{N}})+(\mathbf{2},\textbf{1},\overline{\mathbf{N}})\quad\longrightarrow\quad \text{vector-like pair}\, ,\nn\\
&(\mathbf{1},\textbf{2},{\mathbf{N}})+(\mathbf{1},\textbf{2},\overline{\mathbf{N}})\quad\longrightarrow\quad \text{vector-like pair}\, .
\end{align}
This means there are three chiral families as indicated on the left of Fig.~\ref{fig:QuivdP5Mod1Higgsed}.
On top of that, we find pairs of vector-like states that can pair up to get a mass through superpotential couplings.

Another way to obtain the chiral matter spectrum is using the symmetrised intersection formula \eqref{eq:AsymEulerChar} for fractional branes \cite{Verlinde:2005jr}.
Indeed, we obtain for the intersection matrix of fractional branes $\lbrace F_{a},F_{3},F_{4},F_{b},F_{7},F_{8}\rbrace$
\begin{equation}\label{eq:IntMatChiralSpecdP5M1} 
\chi_{-}(F_{i},F_{j})=\left (\begin{array}{cccccc}
0 & 3 & 3 & 0 & -3 & -3 \\ 
-3 & 0 & 0 & 3 & 0 & 0 \\ 
-3 & 0 & 0 & 3 & 0 & 0 \\ 
0 & -3 & -3 & 0 & 3 & 3 \\ 
3 & 0 & 0 & -3 & 0 & 0 \\ 
3 & 0 & 0 & -3 & 0 & 0
\end{array} \right )\, .
\end{equation}
The resulting 3 family chiral spectrum agrees with our field theory expectation on the left of Fig.~\ref{fig:QuivdP5Mod1Higgsed}.

To summarise, we find a quiver with three chiral families of
$\left(\mathrm{U}(2)\times \mathrm{U}(N)\times \mathrm{U}(M)\right)^2$ which upon orientifolding is the covering quiver of a three-family left-right symmetric model $\mathrm{USp}(2)_{L}\times\mathrm{USp}(2)_{R}\times \mathrm{U}(N)\times \mathrm{U}(M)$.

\subsection*{Non-chiral matter}

To find the excess non-chiral matter, we count the number of Goldstone modes
\begin{equation}
\dim(G/H)=160+16(N+M)\, .
\end{equation}
The number of complex scalar fields in bi-fundamentals in the original quiver is
\begin{equation}\label{eq:NumChiralOrigM1} 
N_{\text{chiral}}=320+40(N+M)\, ,
\end{equation}
while after Higgsing the three chiral families amount to
\begin{equation}
N_{\text{chiral}}^{\text{Higgsed}}=24(N+M)
\end{equation}
complex scalars.
The Goldstone modes are being eaten by the massive gauge potentials to account for the longitudinal polarisation. In addition, the new massive vector multiplet absorbs one extra real massive scalar so that the number of (potentially light) complex scalars is
\begin{equation}
N_{\text{chiral}}-\dim(G/H)=160+24(N+M)=160+N_{\text{chiral}}^{\text{Higgsed}}\, .
\end{equation}
As expected, the massless chiral spectrum is specified as outlined in the preceding section.
The remaining $160$ complex scalars are either in adjoints of a single $\mathrm{U}(2)$ or in bi-fundamental representations between the two $\mathrm{U}(2)$ factors.
Indeed, we expect to find for each of the two bound states four adjoints so that $160-64=96$ complex scalars remain as bi-fundamentals combining into vector-like pairs.
However, we also need non-chiral\footnote{In fact, it has been argued in \cite{Wijnholt:2007vn} that spectrum is completely chiral. That is, all vector-like states become massive through superpotential couplings.} matter states between $\mathrm{U}(2)_{R}$ and $\mathrm{U}(M=1)$ to break the left-right symmetric model down to the SM gauge group.

\subsection*{Extended model I quiver}

This is achieved by constructing a new bound state of fractional branes generating yet another $\mathrm{U}(2)$.
The Higgses between the corresponding $\mathrm{U}(2)$ and $\mathrm{U}(M)$ are obtained by replacing (among others) $\mathrm{U}(4+N)$ and $\mathrm{U}(4+M)$ by $\mathrm{U}(10+N)$ and $\mathrm{U}(8+M)$.
In this way, we obtain $\mathrm{dim}(G/H)\supset 32M+40M$ which is not symmetric in $N$ and $M$.
We summarised the necessary choice of VEVs in appendix~\ref{sec:LargeModel1}.
The counting of degrees of freedom leads to
\begin{align}
G&=(\mathrm{U}(12)\times \mathrm{U}(10+N) \times \mathrm{U}(8+M) \times \mathrm{U}(8))^{2} \, ,\nn\\
H&=(\mathrm{U}(2)\times \mathrm{U}(N)\times \mathrm{U}(M))^{2} \, ,\nn\\
\dim(G/H)&=736 + 32 M + 40 N\, ,\nn\\
N_{\text{chiral}}&=1440 + 80 (N+M)\, ,\nn\\
N_{\text{chiral}}^{\text{Higgsed}}&=24(N+M)\, ,\nn\\
N_{\text{chiral}}-\dim(G/H)&=696+16(N+M)+N_{\text{chiral}}^{\text{Higgsed}}+8M+8\, .
\end{align}
We expect the $696+16(N+M)$ fields\footnote{We find from the analysis of bound states that $400+32$ complex scalars are in adjoints at the $\mathrm{U}(2)$ factors so that $696-(400+32)=264$ states remain between the two $\mathrm{U}(2)$'s which need to get massive.} to gain masses.
According to \cite{Wijnholt:2007vn}, this is achieved by looking at superpotentials with quartic and octic terms.

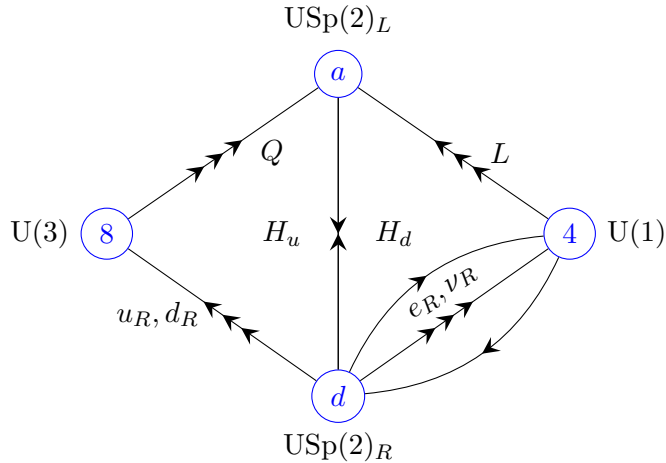
\begin{figure}
\centering
\begin{tikzpicture}[scale=0.9]
\draw (0,6) node(1)[anchor=south,circle,draw,blue]{$a$}node[above=20pt] {$\mathrm{USp}(2)_{L}$};
\draw (-3,4) node(8)[anchor=east,circle,draw,blue]{8}node[left=20pt] {$\mathrm{U}(3)$};
\draw (3,4) node(4)[anchor=west,circle,draw,blue]{4}node[right=20pt] {$\mathrm{U}(1)$};
\draw (0,2) node(5)[anchor=north,circle,draw,blue]{$d$}node[below=20pt] {$\mathrm{USp}(2)_{R}$};
\draw[directed,black] (1) --node[right=10pt] {$H_{d}$} (5);
\draw[directed,black] (5) --node[left=10pt] {$H_{u}$} (1);
\draw[directedThree,black] (4) --node[right=10pt] {$L$} (1);
\draw[directed,black] (4)  to[bend left=30]node[right=10pt]{} (5);
\draw[directed,black] (5)  to[bend left=30]node[left=10pt]{} (4);
\draw[directedThree,black] (5) --node[above =2pt,rotate=35]{$e_{R},\nu_{R}$} (4);
\draw[directedThree,black] (5) --node[left=5pt]{$u_{R},d_{R}$} (8);
\draw[directedThree,black] (8) --node[right=10pt] {$Q$} (1);
\end{tikzpicture}
\caption{Orientifolded and Higgsed quiver diagram $\mathrm{dP}_{5}$ for the larger version of model I.
After identifying $\mathrm{USp}(2)\cong \mathrm{SU}(2)$, this mirrors a supersymmetric version of the left-right symmetric model.}\label{fig:QuivdP5Mod1HiggsedOrie} 
\end{figure}

Ultimately, the unoriented quiver is the one in Fig.~\ref{fig:QuivdP5Mod1HiggsedOrie}.
The massless content consists of the chiral families of the MSSM plus two non-chiral pairs between $\mathrm{U}(2)_{R}$ and $\mathrm{U}(M)$ (additional Higgses to break $\mathrm{U}(2)_{R}\times \mathrm{U}(M=1)\raw \mathrm{U}(1)$), two non-chiral pairs charged under $\mathrm{U}(2)_{R}\times\mathrm{U}(2)_{L}$ (the Higgs fields $H_{u}$, $H_{d}$)
and right-handed neutrinos $\nu_{R}$. Secondly, there are two additional gauge bosons associated to $\mathrm{U}(1)_{\text{B}}$ (left node) and $\mathrm{U}(1)_{\text{L}}$ (right node).
One combination of $\mathrm{U}(1)$'s is anomalous with the corresponding gauge field gaining a Stückelberg mass through the Green-Schwarz mechanism.
The non-anomalous combination $\mathrm{U}(1)_{\text{B-L}}$ together with $\mathrm{SU}(2)_{R}$ is broken to the hypercharge $\mathrm{U}(1)_{\text{Y}}$ via conventional Higgsing which leads to the quiver depicted in Fig.~\ref{fig:SMQuiver}.

\subsection{$F$- and $D$-term Flatness}\label{sec:QuiverDFFlatness}

For simplicity, we discuss the D- and F-flatness conditions for the small version of Model I in this section.
Details on the extended Model I can be found in App.~\ref{sec:LargeModel1}.

\subsection*{$D$-term conditions}

For each non-abelian factor $\mathrm{SU}(N_{i})$ or each node $i$ in the quiver, there are D-flatness conditions for each generator of the form
\begin{equation}\label{eq:QUiverDTNA} 
D_{i}^{a}=X_{ij}T^{a}_{jk}X_{ki}^{\dagger}-Y_{ij}T^{a}_{jk}Y_{ki}^{\dagger}=0
\end{equation}
with ingoing arrows $X_{ij}$ and outgoing $Y_{ij}$ (for details see appendix~\ref{sec:d-terms-suN_and_spN}). The D-term conditions associated with nodes on the top or bottom in Fig.~\ref{fig:QuivdP5Mod1Small} are trivially satisfied since the legs to left and right are identical due to the orientifold symmetry. The only non-trivial conditions are given by
\begin{equation}
|a_{1}|^{2}+|a_{2}|^{2}+|a_{3}|^{2}+|c_{1}|^{2}+|c_{2}|^{2}-|x_{1}|^{2}-|x_{2}|^{2}-|x_{3}|^{2}-|y_{1}|^{2}-|y_{2}|^{2}=0 
\end{equation}
which can be solved by choosing
\begin{equation}\label{eq:CondNonAbD3} 
|a_{1}|^{2}=|x_{1}|^{2}+|x_{2}|^{2}+|y_{1}|^{2}+|y_{2}|^{2}\kom |x_{3}|^{2}=|a_{2}|^{2}+|a_{3}|^{2}+|c_{1}|^{2}+|c_{2}|^{2}\, .
\end{equation}
Furthermore, we obtain
\begin{equation}
|b_{1}|^{2}+|b_{2}|^{2}+|b_{3}|^{2}+|d_{1}|^{2}+|d_{2}|^{2}-|z_{1}|^{2}-|z_{2}|^{2}-|z_{3}|^{2}-|w_{1}|^{2}-|w_{2}|^{2}=0
\end{equation}
which is solved for
\begin{equation}\label{eq:CondNonAbD4} 
|b_{1}|^{2}=|z_{1}|^{2}+|z_{2}|^{2}+|w_{1}|^{2}+|w_{2}|^{2}\kom |z_{3}|^{2}=|b_{2}|^{2}+|b_{3}|^{2}+|d_{1}|^{2}+|d_{2}|^{2}\, .
\end{equation}
All other D-flatness conditions vanish by the symmetry of the quiver. For the choice \eqref{eq:VEVsAS} of VEVs, these conditions are equivalent to the ones given in \cite{Wijnholt:2007vn}.

For the two $\mathrm{U}(2)$ groups and the second $\mathrm{U}(4+M)$ and $\mathrm{U}(4+N)$ the $\mathbb{Z}_2$ symmetry of the quiver implies automatic cancellation.
We have 4 real conditions leaving $36$ unfixed real parameters out of the original 20 complex parameters $a_i, b_i, c_j, d_j, x_i, z_i, y_j, w_j$.
The other $20$ parameters coming from the other half of the quiver are fixed by keeping the $\mathbb{Z}_2$ symmetry of the quiver to be orientifolded.

Next, we consider the D-term conditions of the abelian $\mathrm{U}(1)$'s inside $\mathrm{U}(N_{i})$, that is,
\begin{equation}\label{eq:QUiverDTA} 
D_{i}=Q_{i}^{(ab)}\, |X_{ab}|^{2}=Q_{i}^{(ab)}\, \tr(X_{ab}^{\dagger}X_{ab})=0\, .
\end{equation}
Explicitly, we find
\begin{align}
&D_{1}=D_{2}=D_{5}=D_{6}=0\, ,\nn\\
\label{eq:EquivAbDTs} &D_{8}=-D_{3}\kom D_{7}=-D_{8} \, ,\\
&D_{3}=-2 (|a_{1}|^2 + |a_{2}|^2 + |a_{3}|^2 + |c_{1}|^2 + |c_{2}|^2 - |x_{1}|^2 - |x_{2}|^2 - |x_{3}|^2 - |y_{1}|^2 - |y_{2}|^2)\, ,\nn \\
&D_{4}=-2 (|b_{1}|^2 + |b_{2}|^2 + |b_{3}|^2 + |d_{1}|^2 + |d_{2}|^2 - |w_{1}|^2 - |w_{2}|^2 - |z_{1}|^2 - |z_{2}|^2 - |z_{3}|^2)\nn \, .
\end{align}
There are only two distinguishable non-trivial D-terms $D_{3}$ and $D_{4}$ which are related to $D_{7}$ and $D_{8}$ due to the symmetry of the quiver.
These conditions are trivially satisfied using the non-abelian D-flatness conditions \eqref{eq:CondNonAbD3} and \eqref{eq:CondNonAbD4}.
This also implies that both D-terms for the two anomalous $\mathrm{U}(1)$'s vanish.
Therefore, the two FI parameters $\xi_{7,8}$ are identically zero which immediately sets the model at the singularity.


\subsection*{$F$-term Flatness}

Finally, we claim that our choice of VEVs is sufficient to ensure $F$-term flatness in the vacuum.
The $F$-term conditions are given by
\be
D_{X_{ab}}W=\frac{\partial W}{\partial X_{ab}} + \frac{\partial K}{\partial X_{ab}} W =  0\, .
\label{Fterm}
\ee
They need to be studied together with the dependence of the full superpotential $W$ and K\"ahler potential on the complex structure moduli $U$.
Despite this, we argue that \eqref{Fterm} can always be satisfied through suitable complex structure deformations along the lines of \cite{Wijnholt:2005mp,Buican:2006sn}.

The argument is the standard that dP$_n$ is defined by $\mathbb{P}^2$ with $n$ blow-up points. The automorphism group of $\mathbb{P}^2$,  $\mathrm{PGL}(3,n)$, has  $3^2-1=8$ parameters.
Therefore the $n$ blow-up points for dP$_n$ are determine by $2n-8$ complex parameters and the number of complex structure deformations is encoded in
\begin{equation}
\dim\left (H^{1}(\mathrm{dP}_{n},T\mathrm{dP}_{n})\right )=\begin{cases}
2n-8 & 5\leq n\leq 8\\
0 & n\leq 4\, .
\end{cases}
\end{equation}
For $n=5$, there are $2$ complex structure parameters that appear in the superpotential
\begin{align}\label{eq:QuiverSuperpotential} 
W_Q & = & \alpha_1 X_{13} X_{35} X_{58} X_{81} - \alpha_2  X_{14} X_{46} X_{68} X_{81} + \alpha_3 X_{14} X_{45} X_{57} X_{71} - \alpha_4 X_{13} X_{36} X_{67} X_{71}  + \nonumber \\
& &  \alpha_5 X_{24} X_{46} X_{67} X_{72}  - \alpha_6 X_{23} X_{35} X_{57} X_{72} + \alpha_7  X_{23} X_{36} X_{68} X_{82} - \alpha_8 X_{24} X_{45} X_{58} X_{82} + \nonumber \\
& &  \beta_1 X_{13} X_{35} X_{57} X_{71} - \beta_2 X_{14} X_{46} X_{67} X_{71} + \beta_3 X_{14} X_{45} X_{58} X_{81} - \beta_4 X_{13} X_{36} X_{68} X_{81}  + \nonumber \\
& &  \beta_5 X_{24} X_{46} X_{68} X_{82}  - \beta_6 X_{23} X_{35} X_{58} X_{82} + \beta_7 X_{23} X_{36} X_{67} X_{72} - \beta_8 X_{24} X_{45} X_{57} X_{72} 
\end{align}
with the coefficients $\alpha_i, \beta_i$ functions of complex structure moduli $U$.
It has been suggested in \cite{Wijnholt:2005mp,Buican:2006sn} that tuning the complex structure parameters can allow to fix all $F$-term conditions.
For a globally embedded model the situation is more involved since we have to consider all the fields in the full superpotential.
Hence, we will argue that once we have a global embedding of the quiver model, there is plenty of freedom to satisfy the $F$-term conditions $D_{S,U,X}W=0$ coming from the rich structure provided by the fluxes and also the $18$ free complex parameters of our ansatz \eqref{eq:HiggsingSmallMod1}. 
This will be discussed further in Sect.~\ref{ModStab}.

\section{Calabi-Yau Threefolds with Diagonal $\mathrm{dP}_5$ Divisors}\label{sec:CYThreefoldScan} 

\subsection{Embedding the local model in a compact CY threefold}

We aim to embed the local model described in Section~\ref{sec:SMQuiver} into a global CY orientifold compactification. Of course, the first feature that the compact CY should have is the presence of a $\mathrm{dP}_5$ singularity. This singularity is obtained by taking a limit from a smooth CY with a $\mathrm{dP}_5$ divisor. The singularity is generated when the volume of the $\mathrm{dP}_5$ divisor goes to zero. 

Since we want a global model with moduli that are stabilised at a dS minimum, we need to ask for other properties of the CY.
In particular the desired features of the model are the following:
\begin{enumerate}
\item As we said, we need CY threefolds with $\mathrm{dP}_{5}$ divisors in order to embed the local model of Section~\ref{sec:SMQuiver}. 
In particular we need the $\mathrm{dP}_{5}$ to be `diagonal', such that shrinking it to a point does not force other divisors to shrink (generating a different singularity with respect to the one considered in the local model). 
\item There must be an involution such that the $\mathrm{dP}_{5}$ divisor is transversely invariant and that it intersects the O7-plane (like in the local model, see also App.~\ref{App:InvolLocal}).\footnote{A systematic formalism to constructing CY orientifolds with fluxed branes wrapping shrinkable del Pezzo divisors has been laid out in \cite{Blumenhagen:2008zz}. In addition, it contains a classification of involutions of del Pezzo surfaces which is highly relevant for our considerations.}
\item It is desirable to possibly have $O7$-planes with large $\chi(O7)$ to have large negative $D3$ charge. This would allow to easily satisfy the $D3$-tadpole cancellation condition.
\item In order to have a $T$-brane uplifting to de Sitter \cite{Cicoli:2015ylx, Cicoli:2017shd}, we also need the involution to be such that some D7-branes wrap large (in the LVS sense) divisors. On the other hand, if the dS uplift is realised via anti-$D3$ brane, we may want an involution that generates  $O3$-planes at some appropriate locations \cite{Garcia-Etxebarria:2015lif}.
\item One needs to check the tadpole/anomaly cancellation conditions and that the non-perturbative superpotential contribution is generated. In particular, we require the presence of at least one additional diagonal $\mathrm{dP}_{n}$ to support the  LVS construction~\cite{Balasubramanian:2005zx,AbdusSalam:2020ywo}.
\end{enumerate}

\subsection{Calabi-Yau threefolds with diagonal $\mathrm{dP}_n$}\label{sec:CYDiagdP5} 

\subsection*{Requirements for having a diagonal $\mathrm{dP}_n$ divisor}

As mentioned above, we need to search for Calabi Yau threefolds $X$ which have at least one (diagonal) $\mathrm{dP}_5$ divisor. We will work with CYs that are embedded into toric ambient spaces. Here  $\mathrm{dP}_5$ divisors are usually obtained by the  so-called ``coordinate divisors" $D_i$ which are defined by intersecting the CY equation with the loci given by setting a toric coordinate to zero: $x_i = 0$. This is sufficient for capturing the del Pezzo surfaces in CYs $X$ with $h^{1,1}(X)=h^{1,1}(A)$, where $A$ is the ambient space. We only consider such spaces.\footnote{In fact, in these cases the divisors are given by the intersection of a divisor of $A$ with the equation defining $X$; the rigid divisors (like the del Pezzo divisors) are among the coordinates divisors.}
 
A del Pezzo divisor must satisfy the following topological conditions:
\bea
\label{eq:dP}
& & \int_{X} D_s^3 = k_{sss} > 0\, , \qquad \int_{X} D_s^2 \, D_i \leq 0 \qquad \forall \, i \neq s \,.
\eea
Here $k_{sss} = 9 - n$ for a $\mathrm{dP}_n$ divisor is the degree of $\mathrm{dP}_n$.
We moreover look for divisors $D_s$ that satisfy the following `diagonality' condition \cite{Cicoli:2018tcq}
\bea
\label{eq:diagdP}
& & k_{sss} \, \, k_{s i j } = k_{ss i} \, \, k_{ss j} \, \qquad \qquad \forall \, \, \, i, j \:.
\eea
If this condition is satisfied, then the volume of the four-cycle $D_s$ is a complete-square:
\bea
& & \tau_s = \frac{1}{2}\, k_{s ij} t^i \, t^j = \frac{1}{2 \, k_{sss}}\, k_{ssi} \, k_{s s j} t^i \, t^j = \frac{1}{2 \, k_{sss}}\, \left(k_{ss i} \,t^i \, \right)^2\,,
\eea
where we sum over $i,j$ but not over $s$.
One can then shrink the del Pezzo divisor to a point along one direction of the K\"ahler moduli space, simply by setting to zero the combination of the $t_i$ that appear on the RHS.

\subsection*{A conjecture for diagonal del Pezzo ${dP}_n$ with $1\leq n \leq 5$}

As explained in Appendix~\ref{app:KSScan}, we performed a scan over hypersurface CY threefolds obtained from polytope triangulations listed in \cite{Altman:2014bfa}.\footnote{These were based on the four-dimensional reflexive polytopes listed in the Kreuzer-Skarke (KS) database \cite{Kreuzer:2000xy} with $h^{1,1}\leq 6$.} For CYs in this database, we found that the diagonality condition (\ref{eq:diagdP}) could never be satisfied for the $\mathrm{dP}_5$ divisors.\footnote{For example, most of the times the volume of $\mathrm{dP}_5$ four-cycle takes the form
$\tau_{\mathrm{dP}_5} = \left(\sum_i \, a_i t^i\right) \, \left(\sum_j b_j t^j \right)$ for some $i \neq j$.
Now setting one of the two linear combinations to zero, makes the size of the dP$_5$ go to zero like $t$ instead of $t^2$. This is a signal that the divisor has not shrunk to a point, but rather to a curve. In order to shrink it to a point one needs to set to zero both combinations of the $t_i$'s.
This is what we call a `non-diagonal' del Pezzo. }

The analysis made in Appendix~\ref{app:KSScan} led us to the following conjecture:
\vskip0.2cm
\noindent
``{\it The Calabi Yau threefolds arising from the four-dimensional reflexive polytopes listed in the Kreuzer-Skarke database do not exhibit a `diagonal' del Pezzo divisor $\mathrm{dP}_n$ for $1 \leq n \leq 5$, in the sense of satisfying the eqn. (\ref{eq:diagdP}).}"
\vskip0.2cm
\noindent
Using the topological data of the CY threefolds collected in the AGHJN-database \cite{Altman:2014bfa}, we have checked this conjecture to hold for $1 \leq h^{1,1}(X) \leq 5$. It would be interesting to explore its validity further or find a counter example against our claim.
To begin with, we scanned further $300.000$ geometries with $6\leq h^{(1,1)}\leq 40$ using the software package \texttt{CYTools} \cite{Demirtas:2020dbm} providing evidence that the conjecture might even hold at large values of $h^{(1,1)}$.

Due to this result, we were forced to explore CYs embedded in toric ambient spaces of dimension larger than 4. We have actually been able to construct several CYs (see App.~\ref{Sec:AppCICYdP5}) that are given by two equations in a 5 dimensional toric ambient space. These have diagonal $\mathrm{dP}_5$ divisors.

\subsection*{$\mathrm{dP}_5$ surface as a bi-quadric in ${\mathbb P}^4$}

As just mentioned, it is possible to construct explicit Calabi Yau threefolds which have a diagonal $\mathrm{dP}_5$ divisor. 
A $\mathrm{dP}_5$ surface can be represented by a bi-quadric in ${\mathbb P}^4$ which is given by the following toric data,
\begin{equation} \label{CIdP5}
 \begin{tabular}{|c|c||ccccc|}
\hline
  $HY_1$  & $HY_2$ & $x_1$  & $x_2$  & $x_3$  & $x_4$  & $x_5$   \\
    \hline
 2 & 2 & 1 & 1  & 1 & 1 & 1   \\
 \hline
  \end{tabular}
\end{equation}
 \noindent with the SR ideal being given as $\{x_1 x_2 x_3 x_4 x_5 \}$. Using \texttt{cohomCalg} \cite{Blumenhagen:2010pv, Blumenhagen:2011xn} it is easy to confirm that this has the topology of a $\mathrm{dP}_5$ surface.

We then expect to find $\mathrm{dP}_5$ divisors in CYs that are complete intersections of two equations with a five-dimensional toric space. 
Setting one coordinate to zero and properly gauge fixing all the $\mathbb{C}^\ast$ action except one, we should end up with the surface \eqref{CIdP5} (we will see this explicitly in the concrete model in Section~\ref{sec:GlobalModelConstruction}.

\section{Global Embedding of $\mathrm{dP}_5$ Model}\label{sec:GlobalModelConstruction} 

We have worked out few examples of complete intersection CY's (CICY) that have a $\mathrm{dP}_5$ singularity in some region of their moduli space. In this section we analyse in detail one of them, in order to provide an example of global embedding of our local model. The other CY's  can be found in App.~\ref{Sec:AppCICYdP5}.

\subsection{Geometric data}

Here we consider the following CICY threefold $X$ which has three diagonal $\mathrm{dP}_5$ divisors. As observed above, it is given by two equations intersecting  a five dimensional toric space. 
The toric data for such a CICY threefold are
\begin{table}[H]
  \centering
 \begin{tabular}{|c|c||ccccccccc|}
\hline
  $HY_1$  & $HY_2$ & $x_1$  & $x_2$  & $x_3$  & $x_4$  & $x_5$ & $x_6$ &  $x_7$ &  $x_8$ &  $x_9$ \\
    \hline
 4 & 4 & 1 & 0  & 0 & 0 & 2 & 2 & 1 &  1 & 1  \\
 2 & 2 & 0 & 1  & 0 & 0 & 1 & 1 & 0  & 0 & 1 \\
 2 & 2 & 0 & 0  & 1 & 0 & 1 & 1 & 0  & 1 & 0 \\
 2 & 2 & 0 & 0  & 0 & 1 & 1 & 1 & 1  & 0 & 0 \\
 \hline
  &  & NdP$_{17}$ & dP$_5$ & dP$_5$ & dP$_5$ & SD1  & SD1 & SD2 & SD2 & SD2 \\
    \hline
  \end{tabular}\caption{Toric data of $X$.}\label{ToricDataX}
 \end{table}
 \noindent
with the SR-ideal being given as\footnote{There are  other triangulations giving three diagonal $\mathrm{dP}_5$. We took the one where computations are in a simpler form.}
\bea\label{SR-ideal}
SR =\{x_2 x_3, \, x_2 x_4, \, x_2 x_9, \, x_3 x_4, \, x_4 x_7, \, x_1 x_7 x_9, \, x_3 x_5 x_6 x_8, \, x_1 x_5 x_6 x_7 x_8, \, x_1 x_5 x_6 x_8 x_9 \}. \nn
\eea
This CY threefold has the Hodge numbers $(h^{2,1}, h^{1,1}) = (52, 4)$ and Euler characteristic $\chi=-96$. The first two columns of Table~\ref{ToricDataX} provide the degrees of the polynomial equations defining the CY threefold $X$.
  
An integral basis of $H^{1,1}(X,\mathbb{Z})$ is given by $\{D_b, D_2, D_3, D_4\}$, where $D_b \equiv D_1 + D_2 +  D_3 + D_4$. In this basis, the intersection form is
\bea
I_3 = 4\,D_b^3 \, + 4 \, D_2^3\, + 4 \, D_3^3\, + 4\, D_4^3 \,,
\eea
while the second Chern class is
\be
c_2(X)=10D_b^2+D_2^2+D_3^2+D_4^2\:.
\ee

A detailed divisor analysis using \texttt{cohomCalg} \cite{Blumenhagen:2010pv, Blumenhagen:2011xn} shows that the three divisors $D_2$, $D_3$ and $D_4$ are del Pezzo $\mathrm{dP}_5$ surfaces while the divisor $D_1$ is what we call `rigid but not del Pezzo' NdP$n$. In addition, the divisors $\{D_5,..., \, D_9\}$ are `special deformation' type divisors with the following Hodge diamond:
\bea
{\rm SD1} \equiv
\begin{tabular}{ccccc}
    & & 1 & & \\
   & 0 & & 0 & \\
  8 & & 70 & & 8 \\
   & 0 & & 0 & \\
    & & 1 & & \\
  \end{tabular} , \qquad
		{\rm SD2} \equiv
\begin{tabular}{ccccc}
    & & 1 & & \\
   & 0 & & 0 & \\
  1 & & 24 & & 1 \\
   & 0 & & 0 & \\
    & & 1 & & \\
  \end{tabular} \,.	\nn
  \eea
  
Expanding the K\"ahler form in the basis $\{D_b,D_2,D_3,D_4\}$, $J=t_bD_b+t_2D_2+t_3D_3+t_4D_4$, one obtains the following volumes for the basis divisors
\begin{equation}
\tau_i\equiv  \mbox{vol}(D_i) = \tfrac12 \int_{D_i}J^2 = 2t_i^2\,, \qquad\mbox{with}\qquad i=b,2,3,4 \:.
\end{equation}
The volume of the CY threefold is then
\bea
& & \vo = \tfrac16\int_X J^3 = \frac{1}{3\sqrt{2}} \, \left(\tau_b^{3/2} - \, \tau_2^{3/2} - \, \tau_3^{3/2} - \, \tau_4^{3/2} \right)\,,
\eea
In particular, from this expression we notice that all three $\mathrm{dP}_5$ divisors are diagonal.
Moreover, the K\"ahler cone is given by\footnote{It has been computed by the union of ambient space K\"ahler cones relative to triangulations leading to the same intersection form on the CY threefold.}
\bea
& & \hskip-1.5cm t_2 < 0, \quad t_3 < 0, \quad t_4 < 0, \quad t_2 + t_3 + t_b > 0, \quad t_2 + t_4 + t_b > 0, \quad t_3 + t_4 + t_b > 0. 
\eea
Hence, we can equivalently shrink any of the $\mathrm{dP}_5$ to a point-like singularity by squeezing along a single direction. We will make the choice to shrink $D_2$, by taking $t_2 \to 0$. 

\subsection{Orientifold involution}

We consider the involution 
\begin{equation}\label{OrInvolX}
\sigma\,:\qquad x_5 \mapsto - x_5
\end{equation}
The CY defining equations that respect this involution can be written as
\bea\label{DefEqX}
a^\lambda x_5^2 = P_{4,2,2,2}^\lambda(x_1,x_2,x_3,x_4,x_6,x_7,x_8,x_9) \:, \qquad\qquad 
\qquad \lambda=1,2 \:
\eea  
where the RHS does not depend on $x_5$.

The fixed point set of the involution is given by the codimension-1 locus $\{ x_5=0 \}$. There are no isolated fixed points. We then have a single O7-plane wrapping the divisor $D_5=2D_b-D_2-D_3-D_4$. In particular we have $O7^3=D_5^3=20$ and $\chi(O7)=88$. 

The involution $\sigma$ splits the cohomology groups into eigenspaces, whose dimensions are $h^{p,q}_\pm$, with $h^{p,q}=h^{p,q}_++h^{p,q}_-$. For our CICY $X$, it is easy to see that $h^{1,1}_+=4$, while $h^{1,1}_-=0$. It is less trivial to obtain $h^{1,2}_\pm$. However, we can do it by means
of Lefschetz fixed point theorem,
after having derived the fixed point set of our involution. For a CY threefold, this theorem says that
\begin{equation}\label{eq:LefTheor}
2 + 2 \left( h^{1,1}_+-h^{1,1}_-\right) - 2 \left( h^{1,2}_+-h^{1,2}_- -1\right) = \chi(O_\sigma)
\end{equation}
with $O_\sigma$ the fixed point locus; in our case $O_\sigma=O7$. Combining the relation $h^{1,2}_++h^{1,2}_-=h^{1,2}=52$ with \eqref{eq:LefTheor}, one obtains $h^{1,2}_+=7$ and $h^{1,2}_-=45$.

\subsection{Embedding of the local model}

We now focus on the region in $X$ close to the `diagonal' $\mathrm{dP}_5$ divisor $D_2$.  When we shrink this divisor to zero size, the open patch around it becomes a non-compact CY with a $\mathrm{dP}_5$ singularity. Putting D3-branes on top of the singular point, one obtain a model of D3-branes at singularity. We now show that, if we consider the involution \eqref{OrInvolX}, this model is the same discussed in Section~\ref{sec:SMQuiver} and whose involution is discussed from the geometric point of view in App.~\ref{App:InvolLocal}.

We start by taking an open patch close to $x_2=0$. Because of the SR-ideal \eqref{OrInvolX}, we can gauge fix three of the four $\mathbb{C}^\ast$ action in Table~\ref{ToricDataX}, setting $x_3=1$, $x_4=1$ and $x_9=1$. The local CY is then described by 
\begin{equation}
\begin{array}{|c|c||cccccc|}
\hline
  Eq_1  & Eq_2 & x_1  & x_2  & x_5    & x_6 & x_7 &  x_8 \\
    \hline
 2 & 2 & 1 & -1 & 1   & 1 & 1  &  1   \\
 \hline
\end{array} \qquad \mbox{SR}=\{x_1x_5x_6x_7x_8\}\, .
\end{equation}
We immediately see that $x_2=0$ is given by two quadratic equations in $\mathbb{P}^4$, i.e. it is a $\mathrm{dP}_5$ divisor.

It is moreover easy to blow down the $\mathrm{dP}_5$ divisor. We obtain a three-fold given by two equations in $\mathbb{C}^5$, whose coordinates are $x_1,x_5,x_6,x_7,x_8$. 
The $\mathrm{dP}_5$ singularity is now located at $x_1=x_5=x_6=x_7=x_8=0$. Close to this point, the
CY defining equations are approximated by the following expressions, where we keep only the quadratic monomials discarding the subleading higher order terms:\footnote{One can check that these equations can be completed by adding to each monomial the proper factors of $x_3,x_4,x_9$ to make it of degrees in Table~\ref{ToricDataX}.}
\bea\label{Eq1Eq2LocalCY}
Eq_1 &:& x_1^2  = x_7x_8 + Q_2(x_1,x_5,x_6,x_7,x_8) + ... \nonumber \\
Eq_2 &:& x_1^2 = (x_6+x_5)(x_6-x_5) + R_2(x_1,x_5,x_6,x_7,x_8) + ...
\eea
We can match this local CY with the one in App.~\ref{App:InvolLocal}, where we found out the proper involution in the local model to obtain the quiver theory \eqref{fig:QuivdP5Mod1HiggsedOrie}.
The local CY in App.~\ref{App:InvolLocal} is at a special point in the complex structure moduli space, where the singularity becomes a $\mathbb{Z}_2\times \mathbb{Z}_2$ orbifold of the conifold. The local CY in \eqref{Eq1Eq2LocalCY} reaches that point by specialising the complex structure of $X$ such that 
the polynomials $Q_2$ and $R_2$ are identically zero, i.e.
$$
Q_2(x_1,x_5,x_6,x_7,x_8)\equiv 0 \qquad\mbox{and} \qquad R_2(x_1,x_5,x_6,x_7,x_8)\equiv 0  \:.
$$
When this happens, the defining equations \eqref{Eq1Eq2LocalCY} become
\begin{equation}
\left\{\begin{array}{l}
x_1^2 = x_7x_8 \\
x_1^2 = (x_6+x_5)(x_6-x_5)\\
\end{array}\right.
\end{equation}
that are exactly the equations \eqref{eq:complete-intersection} in App.~\ref{App:InvolLocal} with involution \eqref{eq:meson-action} ($\epsilon=+1$) after identifying the coordinates as $C=x_1$, $X=x_7$, $Y=x_8$, $Z=x_6+x_5$ and $W=x_6-x_5$.

\subsection{Non-perturbative effects}

In order to stabilise the K\"ahler moduli, one needs that the $\mathrm{dP}_5$ divisors at $x_3=0$ and $x_4=0$ host a non-perturbative effect. This divisors are invariant (but not fixed) under the orientifold involution. 

A D3-brane wrapping  an invariant divisor $D$ and having zero flux, i.e. 
$$\mathcal{F}_{E3}\equiv F_{E3} - \iota_D^\ast B=0\:,$$
gives an $O(1)$ instanton that could generate a non-perturbative term in the superpotential. Here $\iota_D^\ast$ is the pullback map from two-forms on $X$ to two-forms on the surface $D$. 

Since the $\mathrm{dP}_5$ surface is non-spin, the gauge flux $F_{E3}$ must be half-integral, since it must satisfy the Freed-Witten quantisation condition \cite{Freed:1999vc}
\begin{equation}
F_{E3} + \frac{c_1(D)}{2} \,\,\,\in \,\,\, H^2(D,\mathbb{Z}) \:.
\end{equation}
In particular, in the present case $c_1(S)=-\iota_{S}^\ast S$ with $S=D_3, D_4$. In order to have zero flux $\mathcal{F}_{E3}$ one needs a B-field such that \cite{Collinucci:2010gz}
\begin{equation}
 \iota^\ast_{S} B = \frac{\iota^\ast_{S} S}{2}
\end{equation}
up to an integral two-form.

On the other hand, if one takes zero B-field, $B=0$, then an $O(1)$ E3-instanton is not allowed on the non-spin surface. However, a rank two instanton can be present  \cite{Berglund:2012gr}. Such a D3-brane supports a vector bundle $\mathcal{E}$ of rank two. This configuration is invariant under the orientifold involution $\sigma$ when
\begin{equation}\label{E3rk2invariance}
 \sigma^\ast \left(\mathcal{E}^\vee\right) \otimes K_S = \mathcal{E}
\end{equation}
where $K_S$ is the canonical line bundle of $S$. 
A solution to this equation is given by the dual of the holomorphic tangent bundle of $S$ (whose first Chern class is again $-c_1(S)=\iota^\ast_S S$).

\subsection{D-brane setup}

The O7-plane wrapping the $D_5$ divisor generates a non-zero D7-tadpole that needs to be cancelled. It is then necessary for the global consistency of the model to introduce D7-branes whose total D7-charge is equal to $-8D_5$. This configuration must be invariant under the orientifold involution (and then the D5-charge is automatically cancelled).

These D7-branes pass through the $\mathrm{dP}_5$ singularity, after taking
$t_2\rightarrow 0$ (in fact, the intersection $D_2\cap D_5$ is a
non-trivial curve). They seem like flavour branes for the fractional
D3-branes and one may wonder whether this generates extra chiral states
for the theory at the singularity.  However a chiral spectrum between D3- and D7-branes would
produce an anomalous spectrum (since the local model was anomaly
free). However, since a globally consistent model with all tadpoles
cancelled must be anomaly free, we expect that the D7-branes do not
introduce extra chiral states to the D3-model.

In the following we consider two different D7-brane configurations: The first one is consistent with an $O(1)$ E3-instanton on top of the $\mathrm{dP}_5$ divisors at $x_3=0$ and $x_4=0$ and will consequently have non-zero B-field. The second one is the configurations with the maximal D3-charge but still allowing a T-brane (necessary for dS uplift);  it is consistent with the proper structure of zero modes for a rank 2 E3-instanton on the $\mathrm{dP}_5$'s at $x_3=0$ and $x_{4}=0$ and one can take zero B-field.

\subsubsection*{SO(8) D7-brane configuration}

We will consider a stack of four D7-branes (plus their four orientifold images) wrapping the locus $x_5=0$. This produces an $\mathrm{SO}(8)$ gauge group living on this locus. Remember that $D_5=2D_b-D_2-D_3-D_4$.

In order to have an $O(1)$ instanton on $D_3$  and $D_4$ we choose the B-field as
\begin{equation}\label{BfieldModel}
B = -\frac{D_3}{2} -\frac{D_4}{2} + \frac{D_b}{2} \:.
\end{equation} 
The last term is not necessary to make the E3-instanton orientifold invariant, as $\iota^\ast_{D_{3,4}}D_b=0$. However, it will be necessary to generate the wanted T-brane.

The following flux $\mathcal{F}$ on each one of the four D7-branes  (and $-\sigma^\ast \mathcal{F} $ on the four images) is consistent with flux quantisation:
\begin{equation}\label{FD7Ex2}
\mathcal{F} = F - B = \left(n_b-\frac12\right)D_b + \left(n_2-\frac12\right)D_2+n_3D_3+n_4D_4  \,\,\,\mbox{with}\,\,\, n_b,n_2,n_3,n_4\in\mathbb{Z}\:,
\end{equation}
where the symbol $\iota^\ast_{D_5}$ is implicit.

In order for the E3-instanton to generate a non-perturbative term in
the superpotential, it should have zero chiral modes\footnote{It
should also have no \emph{non}-chiral zero modes, but this also
generically holds in our case: non-chiral zero modes living on the
curve $\ccC$ where a D7 brane and a D3 intersect arise from elements
of $H^i(\ccC,E|_\ccC\otimes F^\vee|_\ccC\otimes N_{\ccC|X})$ where
$E|_\ccC$ and $F|_\ccC$ are the bundles on the intersecting branes
restricted to $\ccC$, and $N_{\ccC|X}$ is the normal bundle of
$\ccC$ in the ambient space \cite{Katz:2002gh}. In our case $\ccC$ is
topologically a $T^2$ (for both $\kappa=1$ and $\kappa=2$), and the bundle
$E|_\ccC\otimes F^\vee|_\ccC\otimes N_{\ccC|X}$ is degree zero
(because there are no chiral zero modes). A generic degree zero
bundle on $T^2$ has no sections, and therefore there are no
non-chiral zero modes for generic choices of flux.} at the
intersection with the D7-branes, i.e.
\begin{eqnarray}
  0 &=& \int_{D7\cap E3} \mathcal{F}-\mathcal{F}_{E3} =  D_5 \cdot D_\kappa \cdot \mathcal{F}= -D_\kappa^3n_\kappa=-4n_\kappa  \qquad\mbox{with }\kappa=3,4\:.
\end{eqnarray}

Hence the flux on the D7-branes reduces to
\begin{equation}
\mathcal{F} = F - B = \left(n_b-\frac12\right)D_b + \left(n_2-\frac12\right)D_2 \qquad\mbox{with}\qquad n_b,n_2\in\mathbb{Z}\:.
\end{equation}
This breaks the $\mathrm{SO}(8)$ gauge group to $\mathrm{U}(4)$ (the diagonal $\mathrm{U}(1)$ is actually massive due to a St\"uckelberg mechanism) and it generates the following FI-term:
\begin{eqnarray}\label{FItermSO8}
\xi_{D7} &=& \frac{1}{4\pi\mathcal{V}} \int_{D7}\mathcal{F}\wedge J = \frac{1}{4\pi\mathcal{V}} D_5\cdot \mathcal{F}\cdot (t_bD_b+t_2D_2+t_3D_3+t_4D_4) \\
 &=&  \frac{1}{4\pi\mathcal{V}} \left\{ 4(2n_b-1)t_b - 2(2n_2-1)t_2 \right\}  \xrightarrow[]{t_2 \to 0} 
  \frac{1}{\pi\mathcal{V}} (2n_b-1)t_b   \simeq  \frac{1}{\pi\mathcal{V}^{2/3}} \left(\frac32 \right)^{1/3}(2n_b-1)   \nn
\end{eqnarray}

This FI-term is non-zero. This implies that a non-zero VEV must be switched on for the adjoint complex scalar $\Phi$ living on the D7-brane stack. In particular we will consider a T-brane background \cite{Gomez:2000zm,Donagi:2003hh,Donagi:2008ca,Cecotti:2010bp}. For more detail on what we need in this context, see Section 3.4 of  \cite{Cicoli:2017shd}. 

Under the breaking of $\mathrm{SO}(8)$ to $\mathrm{U}(4)$ (due to non-zero $\mathcal{F}$), the adjoint  representation of $\mathrm{SO}(8)$ is broken as:
\be
{\bf 28} \rightarrow {\bf 16}_0 \oplus {\bf 6}_{+2} \oplus {\bf 6}_{-2} \,,
\label{phicharge}
\ee
where ${\bf R}_q$ is in the representation ${\bf R}$ for $SU(4)$ and has charge $q$ with respect to the diagonal $\mathrm{U}(1)$. Here ${\bf 16}_0={\bf 15}_0 \oplus {\bf 1}_0 $ is the reducible adjoint representation of $\mathrm{U}(4)$. According to \eqref{phicharge}, the scalar field $\Phi$ can be written as:\footnote{We use a different basis with respect to the usual matrix notation for the adjoint of $\mathrm{SO}(8)$ where the matrices are simply antisymmetric.}
\be
\label{Ex2PhiSO8}
 \Phi = \left(\begin{array}{cc}
     \phi_{{\bf 16}_0} & \phi_{{\bf 6}_{+2}} \\   \phi_{{\bf 6}_{-2}} &  -\phi_{{\bf 16}_0}^T \\
 \end{array}\right) \:.
\ee
The first four rows (and columns) refer to the four D7-branes, while rows (and columns) from the fifth to the eighth refer to their images: the upper right block corresponds to strings going from the four D7-branes to their images, while the lower left block corresponds to strings with opposite orientation (in fact, they have opposite charges with respect to the diagonal $\mathrm{U}(1)$). Giving a VEV to both $\phi_{{\bf 6}_{+2}}$ and $\phi_{{\bf 6}_{-2}}$ recombines some of the four D7-branes with some of the image D7-branes. On the other hand, $\phi_{{\bf 16}_0}$, that is in the adjoint of $\mathrm{U}(4)$,
describes deformations and the recombinations of the $\mathrm{U}(4)$ stacks (with  the analogous process in the image stack).

The D7-branes, after switching on a non-zero VEV for $\Phi$, is described by the Tachyon matrix
\begin{equation}
T+\Phi \:,
\end{equation}
where $T$ is the tachyon describing the $\mathrm{SO}(8)$ stack. The knowledge of the tachyon matrix (with its domain and codomain) allows  to derive the D-brane charges of the stack, as we will show shortly.

In presence of an orientifold projection with involution $\xi\mapsto -\xi$ for some coordinate $\xi$, the full tachyon (describing the invariant D7-brane configuration that cancels the O7-plane tadpole) must satisfy the condition \cite{Collinucci:2008pf}:
\be
 T = \xi S + A\,,
\ee
where $S$ and $A$  take the following form:\footnote{Here we use a different basis with respect to \cite{Collinucci:2008pf}, where $S$ ($A$) is a symmetric (antisymmetric) matrix.}
\be
S = \left(\begin{array}{cc}
M_S&S_1\\S_2&M_S^T\\
\end{array} \right) \qquad\mbox{ and }\qquad 
A = \left(\begin{array}{cc}
M_A&A_1\\A_2&-M_A^T\\
\end{array} \right) \:,
\ee
where $M_{S,A}$ are generic $N\times N$ matrices, $S_{1,2}$ are symmetric $N\times N$ matrices and $A_{1,2}$ are antisymmetric $N\times N$ matrices.\footnote{The first $N$ lines (and columns) refer to a set of $N$ branes, while the last $N$ lines (and columns) refer to their $N$ images.}

Let us come back to our setup, where we have an orientifold plane at $x_5=0$ and four D7-branes (plus their four images) on the same locus. Before giving VEV to $\Phi$, the tachyon of this configuration is  given by (in our case $x_5=\xi$):
\be
\label{TSO8}
T = \left(\begin{array}{cc}
x_5 {\bf 1}_4& 0 \\ 0 & x_5 {\bf 1}_4 \\
\end{array} \right) \:.
\ee
We need to specify also the domain and codomain of this map. These are related to  the flux on the branes \cite{Collinucci:2008pf,Collinucci:2010gz}. In the chosen setup, where all the four D7-branes have the same flux, we have 
\be
\label{TSO8domcodom}
T :  \qquad \begin{array}{c}
\mathcal{O} (-\frac{D_5}{2}-F_{}+2B)^{\oplus 4}\\ \oplus \\ \mathcal{O} (-\frac{D_5}{2}+F_{})^{\oplus 4} \\
\end{array}  \qquad \rightarrow \qquad \begin{array}{c}
\mathcal{O} (\frac{D_5}{2}-F_{}+2B)^{\oplus 4}\\ \oplus \\  \mathcal{O} (\frac{D_5}{2}+F_{})^{\oplus 4} \\
\end{array} 
\ee
where $F$ and the $B$-field are defined in (\ref{FD7Ex2}) and (\ref{BfieldModel}).\footnote{The orientifold symmetry imposes constraints also on domain and codomain, that involve also the $B$-field.}

We now want to switch on a T-brane background, i.e. a VEV for $\Phi$ where either only $ \phi_{{\bf 6}_{+2}} $ or only $ \phi_{{\bf 6}_{-2}} $ gets a non-zero VEV. 
In \cite{Marchesano:2017kke,Marchesano:2020idg} the authors studied what are the conditions that allow a stable T-brane configuration. These are compatible with what studied in \cite{Collinucci:2014qfa} with a different language.

Let us say we want to give VEV only to $ \phi_{{\bf 6}_{+2}} $. This field is a section  of
$\mathcal{O}(D_5-2F+2B)=\mathcal{O}(D_5-2\mathcal{F})$ (with values in the representation ${\bf 6}_{+2}$). We can switch on a holomorphic VEV only when this line bundle is effective, that is for $J$ in the K\"ahler cone \cite{Marchesano:2017kke,Marchesano:2020idg}
\begin{equation}
0 \leq \int_{D_5} J\wedge (D_5-2\mathcal{F}) = 4\left( (6-4n_b)t_b +    t_3 +    t_4 \right)
\end{equation}
after taking the limit $t_2\rightarrow 0$. The RHS is always positive when $n_b\leq 1$. 

In order for this VEV to make the D-term vanish, one needs a proper sign for the FI term.\footnote{From the 4D point of view, the two off-diagonal blocks are related to modes with different charges with respect to the $\mathrm{U}(1)$; for a given sign of the FI term, only one charge can get VEV if the other has zero VEV. } The proper sign requirement is again given by \cite{Marchesano:2017kke,Marchesano:2020idg, Collinucci:2014qfa} and is $\xi>0$ for non-zero VEV to $ \phi_{{\bf 6}_{+2}} $ (of course, the opposite sign holds for $ \phi_{{\bf 6}_{-2}} $). Looking at \eqref{FItermSO8}, we see that this is realised for $n_b\geq 1$. 

We immediately see that these conditions fix the flux along $D_b$ to be $n_b=1$. As one can check, in order to find a non-zero VEV satisfying these constraints it was necessary to have a half-integral B-field along the $D_b$ direction. Applying the same reasoning to $ \phi_{{\bf 6}_{-2}} $ one obtains $n_b=0$.

The deformation $\Phi$ does not change the D-brane charges that are given by \eqref{TSO8domcodom}. If $T: E_{\rm domain} \rightarrow E_{\rm codomain}$, the D-brane charge of the D7-stack is
\be
\label{eq:charge-vector}
\Gamma_{D7}= e^{-B} \left(  \mbox{ch}(E_{\rm codomain}) - \mbox{ch}(E_{\rm domain}) \right) \left( 1+ \frac{c_2(X)}{24} \right)\,.
\ee 
In our case 
\begin{equation}
\begin{array}{lcl}
\mbox{ch}(E_{\rm domain}) &=& 4\left( e^{-\frac{D_5}{2}-F+2B}+e^{-\frac{D_5}{2}+F}\right), \\
\mbox{ch}(E_{\rm codomain}) &=& 4\left( e^{\frac{D_5}{2}-F+2B}+e^{\frac{D_5}{2}+F}\right).
\end{array}\nonumber
\end{equation}

The D-brane charge of the O7-plane at $x_5=0$ is:
\be
\Gamma_{O7} = -8 D_5 + D_5\frac{D_5^2+c_2(X)}{6} \:.
\ee
Summing the D7 and the O7 contributions,  $\Gamma = \Gamma_{D7}+\Gamma_{O7} $, we actually see that all charges cancel except the D3-charge, that is computed to be (for both $n_b=1$ and $n_b=0$)
\be
Q_{D3} = - \int_X \Gamma|_{\rm 6-form} = - (48  + 16 n_2 -16  n_2^2)\:.
\ee
This number should be added to the positive D3-charge of the D3-branes at the dP$_5$ singularity
which is given by $8$ for the large version of model I.
This leaves space for switching on 3-form fluxes necessary for stabilising dilaton and complex structure moduli.

\subsubsection*{$\mathrm{Sp}(1)$ D7-brane configuration}

We now consider zero B-field, i.e. B=0. We cancel the D7-charge of the O7-plane by a stack of two branes wrapping an invariant locus $P(x)=0$ (with $P(x)$ a polynomial of degrees $(8,4,4,4)$ in all coordinates except $x_5$) in the class $D_{D7}=4D_5$. Moreover we consider a flux $F$ on one brane and a flux $-F$ on the second brane, that is the orientifold image of the first one. The configuration is then orientifold invariant. The corresponding tachyon matrix is 
\begin{equation}
T = \left(\begin{array}{cc}
P(x) & 0 \\0 & -P(x) \\
\end{array}\right)  \:,
\end{equation}
with
\be
\label{TSp1domcodom}
T :  \qquad \begin{array}{c}
\mathcal{O} (-2D_5-F_{})\\ \oplus \\ \mathcal{O} (-2D_5+F_{}) \\
\end{array}  \qquad \rightarrow \qquad \begin{array}{c}
\mathcal{O} (2D_5-F_{})\\ \oplus \\  \mathcal{O} (2D_5+F_{}) \\
\end{array} 
\ee

Since the divisor wrapped by the branes is even (hence spin), a properly quantised flux $F$ is
\begin{equation}\label{FD7Ex2rk2}
F = m_b D_b + m_2 D_2 + m_3 D_3+ m_4D_4  \qquad\mbox{with}\qquad m_b,m_2,m_3,m_4\in\mathbb{Z}\:.
\end{equation}

In order for the rank-2 E3-instantons to generate a non-perturbative term in the superpotential, they should have zero chiral modes at the intersection with the D7-branes. The rank-2 bundle $\mathcal{E}_\kappa$ supported on the E3-instantons wrapping $D_\kappa$ ($\kappa=3,4$) has the charge vector given by 
\begin{equation}
\Gamma_{E3_\kappa} = D_\kappa \,  \mbox{ch}(\mathcal{E}_\kappa) \sqrt{\frac{\mbox{Td}(TD_\kappa)}{\mbox{Td}(ND_\kappa)}} \qquad\qquad \kappa=3,4
\end{equation}
where ch$(\mathcal{E})$ is the Chern character of $\mathcal{E}$, Td$(V)$ is the Todd class of the bundle $V$ and $TD_\kappa$ and $ND_\kappa$ are respectively the tangent and the normal bundle of the surface $D_\kappa$.
Due to the condition \eqref{E3rk2invariance}, one has $c_1(\mathcal{E}_\kappa)=D_\kappa$ and consequently
\begin{equation}
\Gamma_{E3_\kappa} = D_\kappa \left( 2 + \omega  \right)
\end{equation}
where $\omega$ is a four-form that depends on the choice of the bundle $\mathcal{E}$.\footnote{If $\mathcal{E}$ is the dual of the holomorphic tangent bundle of $D_4$, we have $\omega = \tfrac{11}{12}$ch$_2(\mathcal{E})$.}
From this charge vector we see that the rank-2 instantons have not chiral spectrum at the intersection with the D7-branes when the pullback of $F$ on $D_\kappa$ is equal to zero, i.e. when $m_3=m_4=0$.
Hence the flux on the D7-branes reduces to
\begin{equation}
F = m_b D_b + m_2 D_2  \qquad\mbox{with}\qquad m_b,m_2\in\mathbb{Z}\:.
\end{equation}

The flux generated FI-term is
\begin{eqnarray}\label{FItermSp1}
\xi_{D7} &=& \frac{1}{4\pi\mathcal{V}} \int_{D7}\mathcal{F}\wedge J = \frac{1}{4\pi\mathcal{V}} 4D_5\cdot F\cdot (t_bD_b+t_2D_2+t_3D_3+t_4D_4) \\
 &=&  \frac{1}{\pi\mathcal{V}}  8n_bt_b - 4m_2t_2  \xrightarrow[]{t_2 \to 0} 
  \frac{8}{\pi\mathcal{V}} m_bt_b   \simeq  \frac{8}{\pi\mathcal{V}^{2/3}} \left(\frac32 \right)^{1/3}m_b  \:. \nn
\end{eqnarray}

Let us see what is the T-brane VEV $\Phi$ that we can switch on. Now $\Phi$ is a $2\times 2$ matrix acting on the same spaces as the tachyon, see \eqref{TSp1domcodom}.
The upper-right element must be of the form $x_5Q$, where $Q$ is a holomorphic section, that happens when
\begin{eqnarray}
0 &\leq& \int_{D_{D7}} J\wedge (D_{D7}-2F-D_5) = 16\left( 4(3-m_b)t_b +  3 t_3 + 3 t_4 \right)
\end{eqnarray}
in the limit $t_2\rightarrow 0$. The RHS is always positive when $m_b < 3$ (in the K\"ahler cone $t_{3,4}<0$). 
Switching on an upper-right element is compatible with a positive FI-term, i.e.  $m_b>0$, see \eqref{FItermSp1}. Hence, $m_b$ can take the values $1,2$.

The D-brane charge is given by \eqref{eq:charge-vector}, where now domain and codomain are as in  \eqref{TSp1domcodom}. Hence, 
$\mbox{ch}(E_{\rm domain}) = e^{-2D_5-F}+e^{-2D_5+F}$, while $\mbox{ch}(E_{\rm codomain}) = e^{2D_5-F}+e^{2D_5+F}$. 

Summing the D7 and the O7 contributions and integrating over the CY $X$, we obtain
\begin{align}
Q_{D3}& = - \int_X ( \Gamma_{D7}+\Gamma_{O7} )_{\rm 6-form} \nn\\[0.5em]
&= - (144+32m_b^2-16m_2^2)= \left\{\begin{array}{lcl}
 - ( 176 - 16m_2^2 ) &{\rm for}& m_b=1 \\
 - ( 272 - 16m_2^2 ) &{\rm for}& m_b=2 \\
\end{array}\right.
\:.
\end{align}
We see that with this configuration we obtain a much larger (negative) D3-charge than considering the $\mathrm{SO}(8)$ stack. This is good because it allows a bigger choice in the fluxes needed to stabilise the complex structure moduli.
The other difference is the multiplicity of the instanton contributing to the superpotential and that affects the exponential relation between the volume of $X$ and the volume of the divisor $D_4$ in the LVS minimum.

\section{Moduli Stabilisation}
\label{ModStab}

As a final step towards a fully fledged string compactification, we stabilise all closed string moduli in a de Sitter minimum via the LARGE Volume Scenario (LVS) \cite{Balasubramanian:2005zx, Conlon:2005ki, Cicoli:2008va}.
The relevant moduli fields are $h^{1,2}_-$ complex structure moduli $U_\alpha$, the axio-dilaton $S =g_s^{-1} + {\rm i} C_0$ and $h^{1,1}_+$ K\"ahler moduli $T_i = \tau_i + {\rm i} \rho_i$. Here, the $\tau_{i}$ measure $4$-cycle volumes of $X$ and the associated axions are given by $\rho_i=\int_{\cD_i} C_4$. 
As seen in Sect.~\ref{sec:GlobalModelConstruction}, for our model we have $h^{1,1}_-=0$, $h^{1,2}_+=7$, $h^{1,2}_-=45$ and $h^{1,1}_+=h^{1,1}(X)=4$.
Here, we work with the volume expression in the singular limit $t_{2}\raw 0$:
\begin{equation}
\cV=d_{1}\tau_{b}^{3/2}-d_{3}\tau_{3}^{3/2}-d_{4}\tau_{4}^{3/2}\kom d_{1}=d_{3}=d_{4}=\dfrac{1}{3\sqrt{2}} \, .
\end{equation}

For a globally embedded model we have to consider all the fields in the full superpotential
\be
W=W_{\text{flux}} (U,S)+ W_Q(X,U) + W_{\text{np}}(T,U,S,\varphi)\, .
\ee
Here, $W_{\text{flux}}$ is the typical Gukov-Vafa-Witten flux superpotential \cite{Gukov:1999ya} and $W_{Q}$ the quiver superpotential defined in \eqref{eq:QuiverSuperpotential}.
Further, we introduce a non-perturbative superpotential $W_{\text{np}}$ depending on K\"ahler moduli $T$ and extra D7 matter fields $\varphi$.
In a similar spirit, we define the complete K\"ahler potential as
\begin{equation}
K=K_{S}+K_{\text{cs}}-2\, {\rm ln} \,\left({\cal V} +\, \frac{\xi}{2} \left [\dfrac{S+\bar{S}}{2}\right ]^{3/2} \right)+K_Q(U,\bar{U},X,\bar{X},\vo)+K_{D7}(S,\bar{S},\varphi,\bar{\varphi})
\end{equation}
where
\begin{equation}
K_{S}(S,\bar{S})=- \ln\left(S+{\bar S}\right)\kom K_{\text{cs}}(U,\bar{U})= -\ln\left(-i\int_X\Omega  \wedge{\bar\Omega}\right)\kom \xi = - \frac{\chi(X) \, \zeta(3)}{2 (2 \pi)^3}\, .
\end{equation}
We take into account the $(\alpha^{\prime})^{3}$-corrections derived in \cite{Becker:2002nn} which are required for the LVS \cite{Balasubramanian:2005zx}.
In addition, the K\"ahler potentials $K_{Q}$ and $K_{D7}$ for the matter fields $X_{ab}$ and $\varphi_{i}$ are \cite{Conlon:2006tj,Aparicio:2008wh}
\be
K_Q=\dfrac{A(U,\bar{U})}{\mathcal{V}^{2/3}}\mathrm{Tr}(X_{ab}{X_{ab}}^\dagger)\kom K_{D7}=\dfrac{1}{S+\bar{S}}\, \sum_{i}\tr(|\varphi_{i}|^{2}) \, .
\ee
In particular, $A(U,\bar{U})$ is an unknown function of the complex structure moduli.

For moduli stabilisation purposes, we are interested in the $\cN=1$ scalar potential
\begin{equation}\label{eq:FtermSP}
V=V_{F}+V_{D}\qquad\text{with}\qquad V_{F}=\mathrm{e}^{K}\left (K^{A\bar{B}}\, D_{A}W\, \overline{D_{B}W}-3|W|^{2}\right )\, .
\end{equation}
In the subsequent analysis,
the 4D scalar potential can be treated as an expansion in $\vo^{-1}\ll 1$ starting with $\mc{O}(\vo^{-2})$ at leading order.
In the full minimisation of the scalar potential the F-term conditions $D_SW=0$, $D_U W=0$ and $D_XW=0$ come at the leading $\mathcal{O}(\mathcal{V}^{-2})$ order together with $D$-terms.
The minimisation with respect to the $T$ fields comes at next order $\mathcal{O}(\mathcal{V}^{-3})$ where SUSY is broken spontaneously through non-vanishing F-terms.
Due to the extended no-scale structure, additional perturbative corrections to $K$ such as from Kaluza-Klein string loops $\mc{O}(g_s^2 \alpha'^2)$ or from winding loops $\mc{O}(g_s^2 \alpha'^4)$  appear effectively at higher order in $\gs$ and $1/\vo$ \cite{vonGersdorff:2005bf,Cicoli:2007xp,Berg:2005ja, Berg:2007wt}. Further corrections from higher derivative $F^{4}$ terms at $\mc{O}(\alpha'^3)$ are again suppressed by additional factor of $1/\vo$ \cite{Ciupke:2015msa}.
We refer to \cite{Cicoli:2021rub} for a systematic analysis of perturbative corrections to the low-energy scalar potential of F-theory/IIB compactifications.

\subsection{Background fluxes and D-terms}
\label{secD}

To leading order in $\vo^{-1}\ll 1$, we find a scalar potential induced by three-form background fluxes $G_3$ and D-terms.
At this order, it suffices to consider the tree level K\"ahler potential $K$ and flux induced superpotential $W_{\text{flux}} $ \cite{Gukov:1999ya} (setting $M_p=1$)
\be
K_{\text{tree}} = - \ln\left(S+{\bar S}\right) -\ln\left(-i\int_X\Omega  \wedge{\bar\Omega}\right)-2\ln\vo \qquad\qquad W_{\text{flux}} =\int_X G_3 \wedge \Omega
\label{Ktree}
\ee
inducing a supergravity F-term scalar potential of no-scale type 
\be
V_F^{\rm flux}=\mathrm{e}^{K_{\text{tree}}}\left(|D_S W_{\text{flux}} |^2+\sum_{\alpha=1}^{h^{1,2}_-} |D_{U_\alpha} W_{\text{flux}} |^2+\sum_{a,b}\, |D_{X_{ab}}(W_{Q}+W_{\text{flux}})|^{2}\right)\,.
\ee
The axio-dilaton and all complex structure moduli are fixed at a Minkowski minimum by solving $D_S W_{\text{flux}}  = D_{U_\alpha} W_{\text{flux}}  = 0$ which is ensured by positive semi-definiteness of $V_F^{\rm flux}$.\footnote{At large complex structure,
general expressions for flux vacua have recently been analysed in \cite{Marchesano:2021gyv}.}
In fact, fluxes also enter the quiver superpotential $W_{Q}$ \eqref{eq:QuiverSuperpotential} accompanied by additional $18$ free complex parameters stemming from bi-fundamental VEVs \eqref{eq:HiggsingSmallMod1}. 
Therefore we conclude that there are sufficiently many degrees of freedom to satisfy the quiver $F$-term conditions $D_{X_{ab}}W=0$.
At this order of approximation, the minimum leaves the K\"ahler moduli directions flat and generically breaks supersymmetry because $D_{T_{i}}W_{\text{flux}} =K_{T_{i}}W_{\text{flux}} \neq 0$ whenever $G_{3}$ has a non-trivial $(0,3)$ component \cite{Giddings:2001yu}.

Further $\mc{O}(\vo^{-2})$ contributions arise from D-terms which split into a bulk and local (quiver) potential
\be
V_D = V_D^{\rm bulk} + V_D^{\rm quiver}\,.
\ee
The former is associated with the anomalous $\mathrm{U}(1)$'s living on the D7-stack wrapped around the O7-plane, whereas the latter stems from the D3-brane at the dP$_5$ singularity.
The bulk D-term potential in the convention of \cite{Cicoli:2015ylx} is given by 
\be
V_D^{\rm bulk} =\frac{1}{2{\rm Re}(f_{D7})} \left(\sum_i q_{\varphi_i} \frac{|\varphi_i|^2}{{\rm Re}(S)}  - \xi_{D7}\right)^2 \,,
\ee
where the FI-parameters $\xi_{D7}$ have been defined in \eqref{FItermSO8} for the $\mathrm{SO}(8)$
and in \eqref{FItermSp1} for the $\mathrm{Sp}(1)$ configuration.
Furthermore, we defined the $\mathrm{U}(1)$ charges $q_{\varphi_i}$ of the $\varphi_i$ as well as the hidden sector gauge kinetic function
\begin{equation}
f_{D7}=\dfrac{2T_{b}-T_{2}-T_{3}-T_{4}}{2\pi}\, .
\end{equation}
We approximate
\begin{equation}
{\rm Re}(f_{D7})\simeq\dfrac{2\tau_{b}}{2\pi} \simeq \dfrac{1}{\pi}\, \dfrac{1}{d_{1}^{\frac{2}{3}}}\, \vo^{2/3}
\end{equation}
and consider without loss of generality a single canonically normalised charged matter field $\varphi$ so that
\be
V_D^{\rm bulk} = \frac{c_1}{\vo^{2/3}} \left(q_\varphi |\varphi|^2 - \frac{c_2}{\vo^{2/3}} \right)^2\,.
\label{VDbulk}
\ee
where the coefficients $c_{1}$ and $c_{2}$ are given by
\be
c_1 = \frac{\pi d_{1}^{\frac{2}{3}}}{2} \qquad\text{and}\qquad c_2 =\frac{1}{\sqrt{2}\pi d_{1}^{1/3}} \begin{cases}
(2n_b-1)&\mathrm{SO}(8)\\
8m_{b}&\mathrm{Sp}(1)\,.
\end{cases}
\ee 
As for $V_F^{\rm flux}$, the bulk D-term potential is positive semi-definite with a minimum at $V_D^{\rm bulk}=0$ stabilising $\varphi$ as
\begin{equation}\label{Dfix} 
|\varphi|^2 = \frac{c_2}{q_\varphi \vo^{2/3}} \, .
\end{equation}

The quiver D-term potential for the anomalous $\mathrm{U}(1)$ and canonically normalised matter fields $X_{ab}$ reads
\begin{equation}
V_D^{\rm quiver}=\dfrac{1}{2\mathrm{Re}(f_{D3})} \left (Q^{(ab)} \tr(X^{\dagger}_{ab}X_{ab})-\xi_{D3}\right )^{2}\kom f_{D3}=\dfrac{S}{2\pi}\kom \xi_{D3}\simeq \dfrac{\tau_{2}}{\cV}\, .
\end{equation}
As discussed in Sect.~\ref{sec:QuiverDFFlatness}, the cancellation of the non-abelian quiver D-terms already implies $Q_{i}^{(ab)} \tr(X^{\dagger}_{ab}X_{ab})=0$ for all $\mathrm{U}(1)$ charges $Q_{i}$.
Thus $\xi_{D3}=0$ for the anomalous $\mathrm{U}(1)$ at the minimum which puts the $\mathrm{dP}_{5}$ volume to zero, $\tau_{2}=0$.
Hence, the local model is set at the singularity.

\subsection{Non-perturbative and $\alpha'$ effects}
\label{secF}

Stabilisation of the remaining flat directions necessitates effects breaking the no-scale structure induced by perturbative contributions to the tree-level K\"ahler potential as well as non-perturbative corrections to the superpotential.
In what follows, we assume that the $S$ and $U$-moduli are stabilised at their tree-level minimum which is only minorly affected by quantum corrections.

In the remainder of this section,
we work with the full $(\alpha^{\prime})^{3}$-corrected K\"ahler potential $K=K_{S}+K_{\text{cs}}+K_{\alpha^{\prime}}$ where \cite{Becker:2002nn}
\be
K_{\alpha^{\prime}}= -2\, {\rm ln} \,\left({\cal V} +\, \frac{\zeta}{2} \right)\qquad\text{with}\qquad \zeta = - \frac{\chi(X) \, \zeta(3)}{2 (2 \pi)^3\, g_s^{3/2}}\,.
\label{Ka}
\ee
In the model of Sec.~\ref{sec:GlobalModelConstruction}, the E3-instantons wrapping the other two dP$_5$ cycles introduce exponential terms in the superpotential:
\be
W = W_0 + A_3\, e^{ - a_{3}T_3}+ A_4\, e^{ - a_{4}T_4}\:, 
\label{Wtot}
\ee
where $W_0 = \biggl  \langle \int_X G_3\wedge \Omega \biggl \rangle$ denotes the VEV of the flux superpotential and
$a_{3}=a_{4}=2\pi $ for the $\mathrm{SO}(8)$ D7-brane configuration, whereas $a_{3}=a_{4}=4\pi $ for two rank-2 instantons compatible with the $\mathrm{Sp}(1)$ D7-configuration.
Plugging (\ref{Ka}) and (\ref{Wtot}) into \eqref{eq:FtermSP}
gives rise to the F-term scalar potential,
\be
V_F = V_{\alpha'} + V_{\rm np1} + V_{\rm np2}\,,
\label{oddVgen1}
\ee
where (writing $W_0 = |W_0|\,e^{{\rm i}\theta_0}$ and $A_i = |A_i|\,e^{{\rm i}\theta_i}$):
\bea
\label{Vtot1}
V_{\alpha'} &=& \, \frac{12 \, \zeta \,  |W_0|^2}{\left(2\vo+ \zeta \right)^2 \left(4\vo- \zeta \right)} \\
V_{\rm np1}  &=&\sum_{i=3}^{4} \, \frac{8\, |W_0|\, |A_i| \, e^{-a_{i}\tau_i} \, \cos\left( a_{i} \rho_i +\theta_0-\theta_i\right)}{\left(2\vo+ \zeta \right) \left(4\vo - \zeta \right)} \left(4a_{i}\tau_i + \frac{3\, \zeta}{\left(2\vo+\zeta \right)} \right) \\
V_{\rm np2} &=& \sum_{i=3}^{4}\, \biggl\{  \frac{16 a_{i}^{2}\sqrt{\tau_i} \, |A_i|^2\, e^{-2a_{i}\tau_i}}{3d_{i} \left(2\vo + \zeta \right)} + \frac{4\, |A_i|^2 \, e^{- 2a_{i}\, \tau_i}}{\left(2\vo+ \zeta \right) \left(4\vo- \zeta \right)} \left( 8a_{i} \tau_i \left(a_{i}\tau_i+1\right) + \frac{3 \zeta}{\left(2\vo+\zeta \right)} \right )\biggl \} \nn\\
&\quad+&\dfrac{8|A_{3}|\, |A_{4}|\, e^{-a_{3}\tau_{3}-a_{4}\tau_{4}} \,  \cos\left( a_{3} \rho_3 -a_{4}\rho_{4}+\theta_0-\theta_3-\theta_{4}\right)}{\left(2\vo+ \zeta \right) \left(4\vo- \zeta \right)}\nn\\
&\quad\times&\left( 4(a_{3}\tau_{3}+a_{4}\tau_{4}+2a_{3}a_{4}\tau_{3}\tau_{4}) + \frac{3 \zeta}{\left(2\vo+\zeta \right)}\right ) \,. 
\label{Vtot2}
\eea
In the large volume limit $\vo\gg \zeta$, the above potential can be approximated as a typical LVS scalar potential of the form
\be
V_\LVS =\sum_{i=3}^{4}\left( \frac{8 a_{i}^{2}\sqrt{\tau_{i}} \, |A_i|^2\, e^{-2a_{i}\tau_i}}{3d_{i} \vo}+\frac{4a_{i}|W_0|\, |A_i| \,\tau_i\, e^{-a_{i}\tau_i} \, \cos\left(a_{i} \rho_i +\theta_0-\theta_i\right)}{\vo^2} \right) 
+ \frac{3 \, \zeta \,  |W_0|^2}{4\vo^3}\,.
\label{VLVS}
\ee
The axion is fixed at
\be
\dfrac{a_{i}}{2\pi}\rho_i = k+\frac12+\frac{(\theta_i-\theta_0)}{2\pi} \qquad\text{with}\quad k\in\mathbb{Z}\,.
\label{axMin}
\ee

In addition, the LVS potential (\ref{VLVS}) receives contributions due to soft scalar masses of the open string modes $\varphi$ which read
\be
V_{\rm soft} = m_\varphi^2 |\varphi|^2 \,.
\ee
Without loss of generality, we restrict to a single canonically normalised visible sector matter field $X$.
Generally, the soft scalar masses $m_{0}$ can be written as
\be
m_0^2 = m_{3/2}^2 - F^I F^{\bar{I}}\partial_I\partial_{\bar{J}}\ln\tilde{K}
\label{m0}
\ee
in terms of the gravitino mass $m_{3/2}$, the moduli F-terms and the K\"ahler metric for matter fields $\tilde{K}$.
For $\varphi$, we simply have $\tilde{K}_\varphi = 1/{\rm Re}(S)$ and, since $S$ is stabilised supersymmetrically at leading order, $F^{S}=0$ which ensures that the hidden sector matter field $\varphi$ has a mass of the order of the gravitino mass
\be
m_\varphi^2 = m_{3/2}^2 = e^K |W|^2 = \frac{e^{K_{\rm cs}}\, |W_0|^2}{2\, {\rm Re}(S)\, \vo^2}\,.
\label{m32}
\ee
All in all, the contribution from soft scalar masses becomes
\be
V_{\rm soft} = \frac{c_2\,m_{3/2}^2}{q_\varphi \vo^{2/3}} \,,
\label{Vsoft}
\ee
where we plugged in the D-term stabilisation condition (\ref{Dfix}) for $\varphi$.

Collecting all the formulas, the total F-term scalar potential becomes
\be
V_{\rm tot} =  \frac{e^{K_{\rm cs}}}{2\,{\rm Re}(S)} \left( V_\LVS + \frac{\cC_{\rm up}\, |W_0|^2}{\vo^{8/3}}\right)
\qquad\text{with}\qquad \cC_{\rm up} =  \frac{c_2}{q_\varphi}  >0\, .
\label{Vtot}
\ee
Notice that $\cC_{\rm up}>0$ can be ensured for $n_{b},m_{b}\geq 1$ for the respective D7-brane flux configuration (assuming $q_{\varphi}>0$).
In the limit $\epsilon_{i} = \frac{1}{4a_{i}\tau_i}\ll 1$, the global minimum of (\ref{Vtot}) is given by
\bea
\label{Tstab1}
\vo &=& \frac{3d_{i} \, \sqrt{\tau_i}\,(1-4\epsilon_{i})}{4a_{i}  (1 - \epsilon_{i})}\,\frac{|W_0|}{|A_{i}|}  \,e^{a_{i} \tau_i}
\simeq \frac{3d_{i} \, \sqrt{\tau_i}}{4a_{i}}\,\frac{|W_0|}{|A_{i}|}  \,e^{a_{i}\tau_{i}} \,, \\
\label{eq:RelBlowUps} e^{a_{3}\tau_{3}-a_{4}\tau_{4}}&=&\dfrac{a_{3}|A_{3}|d_{4}}{a_{4}|A_{4}|d_{3}}\dfrac{1-\epsilon_{3}}{1-4\epsilon_{3}}\dfrac{1-4\epsilon_{4}}{1-\epsilon_{4}} \dfrac{\sqrt{\tau_{4}}}{\sqrt{\tau_{3}}}\simeq \dfrac{a_{3}|A_{3}|d_{4}}{a_{4}|A_{4}|d_{3}} \dfrac{\sqrt{\tau_{4}}}{\sqrt{\tau_{3}}} \, ,\\
\dfrac{\zeta}{2}  &=& \sum_{i=3}^{4} \frac{d_{i}(1-4\epsilon_{i})}{(1-\epsilon_{i})^2}\,\tau_{i}^{3/2}-\frac{16\,\cC_{\rm up}}{27}\, \vo^{1/3}
\simeq d_{3}\tau_{3}^{3/2}+d_{4}\tau_{4}^{3/2}-\frac{16\,\cC_{\rm up}}{27}\,\vo^{1/3}\,.
\label{Tstab2}
\eea
At this minimum, we determine the vacuum energy as
\be
\langle V_{\rm tot} \rangle \simeq \, \frac{e^{K_{\rm cs}} |W_0|^2}{18\,{\rm Re}(S)\, \vo^3} 
\left[\cC_{\rm up} \vo^{1/3} - \sum_{i=3}^{4} \dfrac{27d_{i}}{4a_{i}}\dfrac{(1-4\epsilon_{i})}{(1-\epsilon_{i})^{2}}\,  \sqrt{\tau_{i}}\right]\,.
\label{CC}
\ee
A Minkowski or dS vacuum is achieved by tuning the gauge and background fluxes so that
\begin{equation}\label{CCtuningExact} 
\cC_{\rm up} \vo^{1/3}\geq  \sum_{i=3}^{4} \dfrac{27d_{i}}{4a_{i}}\dfrac{(1-4\epsilon_{i})}{(1-\epsilon_{i})^{2}}\,  \sqrt{\tau_{i}}
\end{equation}
Plugging this result with equality sign back in (\ref{Tstab2}) we find
\be
\frac{\zeta}{2 } =\sum_{i=3}^{4} \dfrac{d_{i}(1-4\epsilon_{i})(1-16\epsilon_{i})}{(1-\epsilon_{i})^{2}}\,  \tau_{i}^{3/2} \,.
\label{Tstab2newExact}
\ee
At leading order in $\epsilon_{i}\ll 1$, we obtain
\be
\cC_{\rm up} \vo^{1/3} \gtrsim\dfrac{27}{4}\left (  \frac{d_{3}\sqrt{\tau_{3}}}{a_{3}}+\frac{d_{4}\sqrt{\tau_{4}}}{a_{4}} \right )\kom \frac{\zeta}{2 } \simeq d_{3}\tau_{3}^{3/2}+ d_{4}\tau_{4}^{3/2}\,.
\label{CCtuning}
\ee
We conclude that the volumes of the two blow up $\mathrm{dP}_{5}$'s at the minimum only depend on the $\alpha^{\prime}$-parameter $\zeta$ and, hence, on the Euler characteristic $\chi(X)$ and the string coupling $g_{\text{s}}$.

We can solve \eqref{CCtuningExact} for $\cV$ to find the volume at the Minkowski minimum
\begin{equation}\label{VolMinkowskiExact} 
\langle\vo\rangle_{\text{Mink.}} =\dfrac{1}{\cF_{\text{up}}^{3}}\left ( \sum_{i=3}^{4} \dfrac{27d_{i}}{4a_{i}}\dfrac{(1-4\epsilon_{i})}{(1-\epsilon_{i})^{2}}\,  \sqrt{\tau_{i}}\right )^{3}\, .
\end{equation}
The required value for $|W_{0}|/|A_{i}|$ can be determined from the combination with \eqref{Tstab1}.

Since $\epsilon_{i}\ll 1$ is a good approximation at sufficiently small $g_{s}$
(cf. Fig.~\ref{fig:MinSolGS}), we may combine \eqref{VolMinkowskiExact} with \eqref{Tstab1} and \eqref{CCtuning} to find
\begin{equation}
\langle\vo\rangle_{\text{Mink.}} \simeq\dfrac{1}{\cF_{\text{up}}^{3}}\left ( \sum_{i=2}^{3} \dfrac{27d_{i}}{4a_{i}}\,  \sqrt{\tau_{i}}\right )^{3}\simeq\frac{3d_{3}\sqrt{\tau_{3}} }{4a_{3}}\,\frac{|W_0|}{|A_{3}|}  \,e^{a_{3} \tau_{3}}\, .
\end{equation}
This can be solved for explicitly using
\begin{equation}
A_{3}=A_{4}\kom a_{3}=a_{4}\kom d_{3}=d_{4}
\end{equation}
which enforces $\tau_{3}=\tau_{4}=(\zeta/(4d_{3}))^{2/3}$ and thus
\begin{equation}
\langle\vo\rangle_{\text{Mink.}} \simeq\left ( \dfrac{27d_{3}}{4a_{3}\cC_{\rm up}}\right )^{3}\,  \frac{2\zeta}{d_{3}}\simeq\frac{3d_{3} }{4a_{3}}\, \left (\frac{\zeta}{4 d_{3}}\right )^{1/3}\,\frac{|W_0|}{|A_{3}|}  \,e^{a_{3} \,[\zeta/(4 d_{3})]^{2/3}}\, .
\end{equation}
Solving for $|W_{0}|/|A_{3}|$ leads to
\begin{equation}
\frac{|W_0|}{|A_{3}|} \simeq\dfrac{4a_{3}}{3d_{3}}\left ( \dfrac{27d_{3}}{2a_{3}\cC_{\rm up}}\right )^{3}\,  \left (\frac{\zeta}{4 d_{3}}\right )^{2/3}e^{-a_{3} \,[\zeta/(4 d_{3})]^{2/3}}\, .
\end{equation}
Generally, we expect $|W_{0}|/|A_{3}|$ to be exponentially suppressed at the Minkowski minimum.
In fact, given that $e^{-a_{3} \,[\zeta/(4 d_{3})]^{2/3}}\sim e^{-\frac{2.47}{g_{s}}}$, this behaviour resembles the expression for $|W_{0}|$ from flux choices proposed in \cite{Demirtas:2019sip}.

\subsection*{A comment on the $h^{1,1}=N>4$ case}

Let us briefly comment on scenarios with larger number of K\"ahler moduli.
We assume we extend the model to $h^{1,1}=N$ with $N-2$ additional $\mathrm{dP}_{5}$ divisors without changing any of the other parameters.
Assuming as above
\begin{equation}
A_{i}=A_{j}\kom a_{i}=a_{j}\kom d_{i}=d_{j}\kom \forall\; i,j=3,\ldots,N
\end{equation}
we would find that
\begin{equation}
\dfrac{\zeta}{2}\simeq \sum_{i=3}^{N}\, d_{i}\tau_{i}^{3/2}=(N-2)d_{N}\tau_{N}^{3/2}\qquad\Rightarrow\qquad \tau_{N}\simeq\left (\dfrac{\zeta}{2(N-2)d_{N}}\right )^{2/3}\, .
\end{equation}
Then, the VEV for the volume at the Minkowski minimum is given by
\begin{equation}
\langle\vo\rangle_{\text{Mink.}} \simeq\left (\dfrac{27d_{N}}{4a_{N}\cF_{\text{up}}}\right )^{3}\left ( \sum_{i=3}^{N}   \sqrt{\tau_{i}}\right )^{3}=\left (\dfrac{27d_{N}}{4a_{N}\cF_{\text{up}}}\right )^{3} (N-2)^{3}\, \dfrac{\zeta}{2(N-2)d_{N}}\, .
\end{equation}
So at sufficiently small $g_{s}$ where $\epsilon_{N}\ll 1$, the volume at the Minkowski minimum is increased by a factor of $(N-2)^{2}$.
Similarly, \eqref{Tstab1} implies
\begin{equation}
\vo\simeq\frac{3d_{N} }{4a_{N}}\, \left (\frac{\zeta}{(N-2) d_{N}}\right )^{1/3}\,\frac{|W_0|}{|A_{N}|}  \,e^{a_{N} \,[\zeta/((N-2) d_{3})]^{2/3}}
\end{equation}
Finally, we obtain
\begin{equation}\label{eq:W0AGenN} 
\frac{|W_0|}{|A_{N}|} \simeq\dfrac{4a_{N}}{3d_{N}}\left ( \dfrac{27d_{N}}{4a_{N}} \dfrac{(N-2)}{\cC_{\rm up}}\right )^{3}\,  \left (\frac{\zeta}{(N-2) d_{N}}\right )^{2/3}e^{-a_{N} \,[\zeta/((N-2) d_{N})]^{2/3}}\, .
\end{equation}
Under the assumption that $\cF_{\text{up}}$ remains constant and nothing else changes dramatically (tadpole, $\zeta$ etc.), there might not be as much tuning required for a large number of moduli.

Notice that in going from $h^{1,1}=3\raw N$, we effectively replace $\zeta\raw \zeta/(N-2)$ and $\cF_{\text{up}}\raw \cF_{\text{up}}/(N-2)$. The former helps with the tuning of $W_{0}$ (it appears in the exponential), whereas the latter helps increasing the volume at the Minkowski minimum. So overall this means a step in the right direction, albeit relying on a few strict assumptions.
Another possibility to reduce the tuning on $W_0$ would be to consider constructions with gaugino condensation which would reduce the coefficient $a_N$ in the exponent in \eqref{eq:W0AGenN} to $a_N/P$ with $P \in\mathbb{N}$.

\subsection{Choices of underlying parameters}
\label{dSnumerics}

We close our discussion of moduli stabilisation with presenting explicit choices of parameters stabilising all K\"ahler moduli in a dS or Minkowski minimum.
The Euler characteristic $\chi(X)=-96$ potentially receives a $\mathcal{N}=1$ correction from O7/D7 contributions  \cite{Minasian:2015bxa} that leads to an effective Euler characteristic\footnote{One should however keep in mind that this has only been computed for a configuration with one O7-plane and one fully recombined invariant D7-brane.
Here, we take the perspective that such a correction must also persist in our situation.}
\be
\chi_{\rm eff} = \chi(X) +2 \int_X D_{O7}^3 =-96+ 40=-56\,,
\label{eq:chi_eff}
\ee

\begin {table}[t!]
\begin{center}
 \begin{tabular}{|c || c | c | c || c | c | c |} 
 \hline
 & & &  & & & \\[-0.6em]
$g_s$ & $|W_0|/|A_s|$ & $\langle\tau_s\rangle$ & $\langle\vo\rangle$ & $|W_0|/|A_s|$ & $\langle\tau_s\rangle$ &$\langle\vo\rangle$ \\ 
 & & & & & &\\[-0.6em]
 \hline\hline
  & & & & & &\\[-0.9em]
 0.10 & $3.57\cdot 10^{-6}$  &3.23  &115.6  & $2.88\cdot10^{-9}$  &4.42 &  188.6\\ [0.2em]
 \hline
   & & & & & &\\[-0.9em]
 0.05 & $2.22\cdot 10^{-13}$ & 5.98 & 301.1 & $1.05\cdot 10^{-19}$ &  8.35&  503.2 \\[0.2em]
 \hline
   & & & & & &\\[-0.9em]
 0.03 & $3.72\cdot 10^{-23}$ & 9.63 & 626.4 &$8.38\cdot 10^{-34}$  & 13.59 & 1057.4  \\[0.2em]
 \hline
   & & & & & &\\[-0.9em]
 0.02 &$1.79\cdot 10^{-35}$  & 14.21 & 1131.3 & $1.61\cdot 10^{-51}$ & 20.15 &  1919.3 \\[0.2em]
 \hline
   & & & & & &\\[-0.9em]
 0.01 & $1.21\cdot 10^{-72}$ & 27.94 & 3145.2  & $6.85\cdot 10^{-105}$ &  39.82& 5363.7\\[0.2em]
 \hline
\end{tabular}
\caption{Numerical analysis for $\mathrm{SO}(8)$-configuration. \emph{Left}: Minkowski minima for the effective $\chi_{\text{eff}}$. \emph{Right}: Minkowski minima for the full $\chi$.  }
\label{TabdSSO8}
\end{center}
\end {table}

\begin{figure}[t!]
\centering
\includegraphics[scale=0.25]{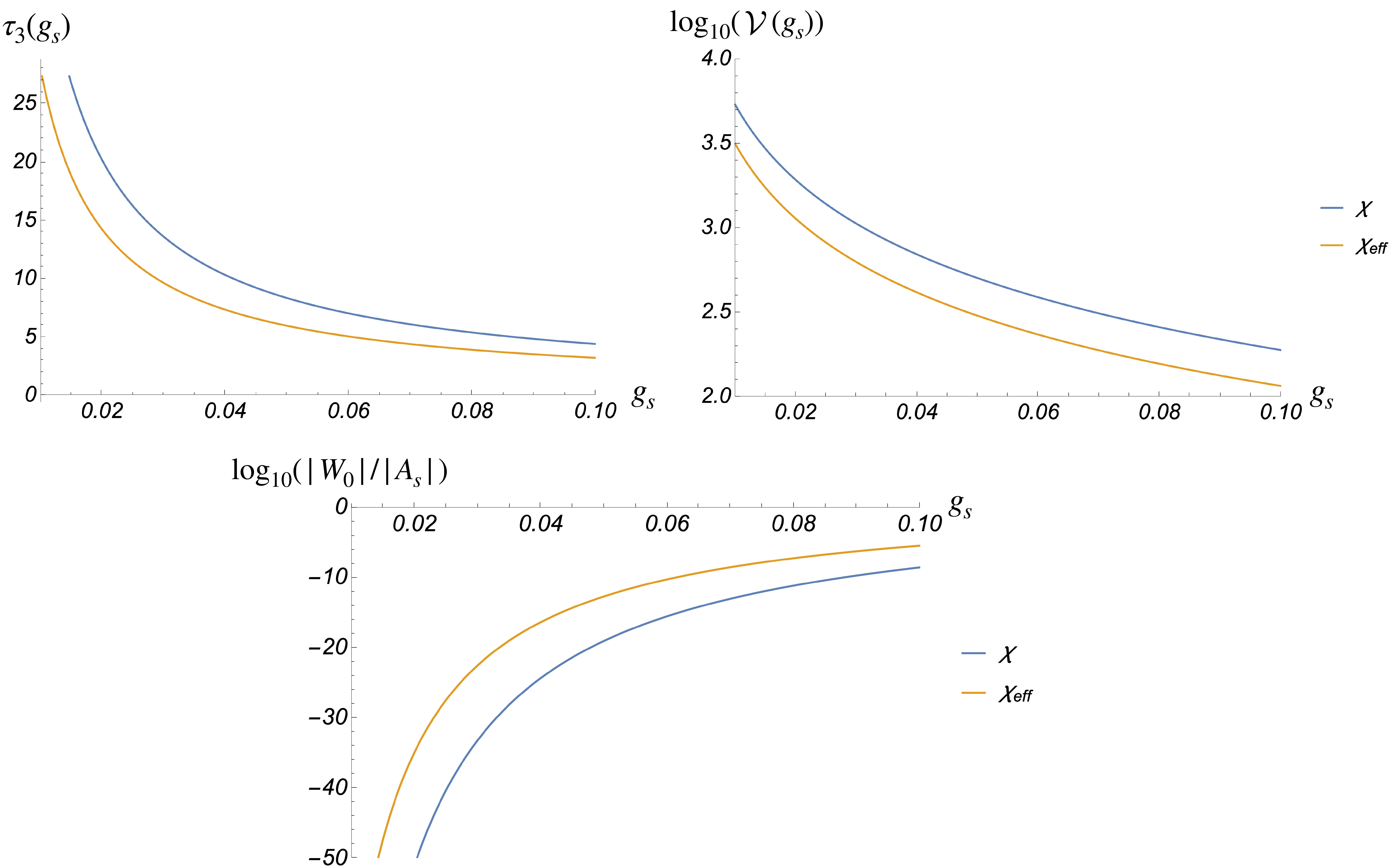}
\caption{Values for $\tau_{s}$, $\cV$ and $|W_{0}|/|A_{s}|$ for Minkowski minima for the full and effective Euler characteristic.}\label{fig:MinSolGS}
\end{figure}

We set $|A_{s}|=|A_{3}|=|A_{4}|$ which fixes $\tau_{s}=\tau_{3}=\tau_{4}$ upon using \eqref{eq:RelBlowUps}.
Overall, we can thus tune only three parameters $g_s$, $|W_0|$, $|A_{s}|$.
The condition for a Minkowski minimum \eqref{CCtuning} fixes one combination of them which leaves us with two free parameters.
For the $\mathrm{SO}(8)$ configuration, we set $n_{b}=1$ to find
\begin{equation}
\cC_{\rm up} =\frac{1}{4\pi} \left(\dfrac{1}{d_{1}} \right)^{1/3}\approx 0.1822
\end{equation}
using that according to (\ref{phicharge}) the $\mathrm{U}(1)$ charge of $\varphi$ is $q_\varphi=2$.\footnote{For the $\mathrm{Sp}(1)$ configuration, we have $\cC_{\rm up} = 1.4575$ which together with the two rank-2 instantons leads to even smaller VEVs for the volume at the Minkowski minimum.}
We summarise our results for Minkowski minima for five values of $g_{s}$ in Tab.~\ref{TabdSSO8}.
The general behaviour of $\tau_{s}$, $\cV$ and $|W_{0}|/|A_{s}|$ as a function of $g_{s}$ is depicted in Fig.~\ref{fig:MinSolGS}.

The numerical results for Minkowski minima require tuning in the ratio $|W_{0}|/|A_{s}|$.
While generically $|A_{s}|\sim \cO(1)$ is expected, there are no explicit expressions available for $|A_{s}|$.
In contrast, $|W_{0}|$ can be computed exactly when stabilising complex structure moduli through fluxes.
Over the past few years, progress has been made in finding flux vacua with
$|W_{0}|\ll 1$ such as in \cite{Demirtas:2019sip} by studying perturbatively flat vacua so that $|W_{0}|\sim \mathrm{e}^{-1/g_{s}}$.
Similarly, computer based methods from stochastic search optimisation like Genetic Algorithm have proven useful in finding solutions to $F$-term conditions with small flux superpotential \cite{Cole:2019enn}.

\section{Phenomenological and Cosmological Implications}
\label{sec:PhenoImp} 

The global dP$_5$ model presented in the previous sections represents the first type IIB example of a realistic global model which can successfully combine: ($i$) an explicit compact CY threefold, orientifold involution and D-brane set with tadpole cancellation; ($ii$) closed string moduli stabilisation in a dS minimum thanks to hidden sector T-branes \cite{Cicoli:2015ylx}; ($iii$) a mechanism of dynamical supersymmetry breaking by the non-zero F-terms of the K\"ahler moduli; and ($iv$) exactly the MSSM gauge group and chiral matter spectrum at low energies. Moreover it features two blow-up modes which are stabilised by non-perturbative effects. Hence the CY volume form has the right structure to realise K\"ahler moduli inflation \cite{Conlon:2005jm}, similarly to the analysis performed in \cite{Cicoli:2017shd}. 

Therefore our model, after more detailed  studies, could represent the first example which is both theoretically robust and fully phenomenologically viable from both the particle physics and the cosmological point of view. Notice that previous type IIB global models with dS moduli stabilisation, based on D7-branes \cite{Cicoli:2011qg}, D3-branes at singularities away from the orientifold \cite{Cicoli:2012vw,Cicoli:2013cha}, systems with flavour D7-branes and D3-branes at singularities far from O7-planes \cite{Cicoli:2013mpa}, D3-branes at orientifolded singularities \cite{Cicoli:2017shd}, could reproduce points ($i$), ($ii$) and ($iii$) but not ($iv$) which is the major step forward of our construction. 

\subsection*{Moduli mass spectrum}

The mass spectrum of the closed string moduli is summarised in Tab. \ref{MassSpectrum}. The first contributions to the 4D scalar potential arise at  $\mc{O}(1/\vo^2)$ through fluxes and D-terms. The moduli stabilised at this order of approximation are the axio-dilaton $S$, the complex structure moduli $U_\alpha$, $\alpha=1,...,h^{1,2}_-$, and the K\"ahler modulus $T_2=\tau_2+{\rm i}\rho_2$. The first two acquire a mass of order the gravitino mass $m_{3/2}$, while the mass of the $\mathrm{dP}_{5}$ volume modulus $\tau_2$ is of order the string scale $M_s$. Its axionic partner $\rho_2$ is instead eaten up by the anomalous $\mathrm{U}(1)$ at the singularity which develops a mass also around the string scale.

\begin{table}
\centering
\begin{tabular}{|c|c|c|}
\hline 
 &  &  \\ [-0.7em]
Field & Name & Mass \\ [0.5em]
\hline 
\hline 
 &  &  \\ [-0.7em]
$\mathrm{dP}_{5}$ modulus &  $\tau_{2},\rho_2$& $\sim M_{s}$ \\ [0.5em]
\hline 
 &  &  \\ [-0.7em]
cx str moduli & $U_{\alpha}$ & $\sim m_{3/2}$ \\ [0.5em]
\hline 
 &  &  \\ [-0.7em]
dilaton & $S$ & $\sim m_{3/2}$ \\ [0.5em]
\hline 
 &  &  \\ [-0.7em]
blow-up cycles & $\tau_{3},\tau_{4}$ & $\sim m_{3/2}$ \\ [0.5em]
\hline 
 &  &  \\ [-0.7em]
blow-up axions & $\rho_{3},\rho_{4}$ & $\sim m_{3/2}$ \\ [0.5em]
\hline 
 &  &  \\ [-0.7em]
volume modulus & $\tau_{b}$ & $\sim m_{3/2}/\sqrt{\vo}$ \\ [0.5em]
\hline 
 &  &  \\ [-0.7em]
volume axion & $\rho_{b}$& $\sim M_p\,e^{-\vo^{2/3}}$ \\ [0.5em]
\hline 
\end{tabular} 
\caption{Mass spectrum of the closed string moduli where the axion $\rho_2$ becomes the longitudinal component of the massive anomalous $\mathrm{U}(1)$ at the dP$_5$ singularity.}
\label{MassSpectrum}
\end{table}

The remaining closed string moduli $T_{b}$, $T_{3}$ and $T_{4}$ are stabilised at order $\mc{O}(1/\vo^3)$ and below by $\alpha^{\prime}$- and non-perturbative effects.
The fields $\tau_{3}$, $\tau_{4}$, $\rho_{3}$ and $\rho_{4}$, associated with the blow-up modes, also receive a mass of order $m_{3/2}$. Even though this is of the same order as the mass of $S$ and $U_{\alpha}$, the decoupling at leading order as a result of the factorised structure of the K\"ahler potential \eqref{Ktree} guarantees that dilaton and complex structure can be safely integrated out. Finally, the overall volume modulus $\tau_{b}$ is fixed by perturbative $\alpha^{\prime}$ corrections acquiring a mass of order $m_{3/2}/\sqrt{\vo}$. Therefore, the remaining axion $\rho_{b}$ is massless at this level of approximation, but receives an exponentially small mass of order $M_p\,e^{-\vo^{2/3}}$ once non-perturbative effects $\sim A_b\, e^{ - 2\pi T_b}$ are included in the superpotential \eqref{Wtot}. This axionic field tends therefore to be ultra-light.

\subsection*{Supersymmetry breaking}

The dS vacuum breaks supersymmetry dynamically due to non-vanishing F-terms
\begin{equation}
\dfrac{F^{T_b}}{\tau_b} \sim \dfrac{F^{T_3}}{\tau_{3}} \sim \dfrac{F^{T_4}}{\tau_{4}} \sim m_{3/2}\quad ,\quad F^S\sim F^{U_{\alpha}} \sim \dfrac{m_{3/2}}{\vo}\, ,
\end{equation}
while $F^{T_{2}}=0$ (up to volume suppressed subleading corrections) for the $\mathrm{dP}_{5}$ modulus. Given that $\tau_b\gg\tau_3\simeq \tau_4$, the largest F-term is $F^{T_b}$, signalling that the Goldstino eaten up by the gravitino is the $T_b$-modulino. Gravitational interactions mediate supersymmetry breaking to the visible sector at the $\mathrm{dP}_5$ singularity. At first sight, given that the local K\"ahler modulus $T_2$ has a vanishing F-term, the resulting soft terms are expected to be suppressed with respect to the gravitino mass, as typical of sequestered scenarios with D3-branes at singularities \cite{Blumenhagen:2009gk, Aparicio:2014wxa}. However, threshold corrections to the gauge kinetic function should induce a redefinition of the local $\mathrm{dP}_5$ modulus of the form \cite{Conlon:2009kt, Conlon:2009qa}:
\begin{equation}
\tau_2 \quad \to \quad \tau_2^{\rm new} = \tau_2 - \alpha\ln\vo\,,
\label{Redef}
\end{equation}
where $\alpha$ can be expressed in terms of the 1-loop $\beta$-function coefficient of the local gauge theory. Notice that this redefinition should occur for orientifolded singularities, but not for orbifolded ones. The presence of flavour D7-branes is also expected to break sequestering inducing logarithmic corrections similar to (\ref{Redef}). The authors of \cite{Conlon:2010ji} noticed that the redefinition (\ref{Redef}) induces non-zero F-terms for the local $\mathrm{dP}_5$ modulus
\begin{equation}
F^{T_2}=0 \quad \to \quad F^{T_2} \sim \alpha m_{3/2}\,.
\end{equation}
This effect breaks sequestering and all soft masses turn out to be of order the gravitino mass, $M_{\text{soft}}\sim m_{3/2} \sim |W_0| M_p/\vo$. The choice of the underlying parameters corresponding to the first line of Tab. \ref{TabdSSO8} would give rise to intermediate scale soft terms, $M_{\text{soft}}\sim m_{3/2}\sim 10^{10}$ GeV, which can be compatible with the observed value of the Higgs mass if $\tan\beta\sim 1$. Notice, in addition, that intermediate scale supersymmetry could also be motivated by the fact that the Higgs quartic coupling in the SM vanishes exactly around $10^{10}$ GeV which is the energy scale where new physics should arise to guarantee the stability of the Higgs potential. On the other hand, the second line would yield TeV-scale superpartners, $M_{\text{soft}}\sim m_{3/2}\sim 1$ TeV which would provide a standard solution to the hierarchy problem due to low energy supersymmetry. All the other parameter choices do not seem to be phenomenologically viable since they would give rise to soft terms below LHC scales.\footnote{Unless the prefactor of the non-perturbative effects $|A_s|$ can be tuned to very large values, which we consider however a very contrived situation.} Notice that more examples with phenomenologically viable soft terms and gravitino mass could be obtained by focusing on CY threefolds with more blow-up modes, since this would reduce the tuning in $|W_0|$ and increase $m_{3/2}$, as explained in Sec. \ref{secF}.

\subsection*{Cosmology}

Realising inflation in scenarios with a TeV-scale gravitino mass is rather hard due to the well-known difficulty to combine inflation with low scale supersymmetry \cite{Kallosh:2004yh}. Moreover, in this case the volume mode would suffer from the cosmological moduli problem, unless its mass is raised above $50$ TeV which would however increase also the gravitino mass around $m_{3/2}\sim 1000$ TeV. 

We therefore focus on the case with $m_{3/2}\sim 10^{10}$ GeV corresponding to the first line of Tab. \ref{TabdSSO8}. This value of the gravitino mass is exactly in the right ballpark to reproduce the observed amplitude of the density perturbations in inflationary models where inflation is driven by one blow-up mode, say $\tau_3$, which is slow-rolling towards the minimum of its potential, while the overall volume is kept approximately constant by the other blow-up mode, $\tau_4$, which is kept at its minimum, i.e. $\vo\sim |W_0| e^{a_4\tau_4}$. The post-inflationary evolution of this kind of inflation models has been already analysed in several papers and can lead to non-standard thermal histories. Preheating effects have been studied in \cite{Barnaby:2009wr, Krippendorf:2018tei} while perturbative reheating has been analysed in \cite{Cicoli:2010ha, Cicoli:2010yj}. A crucial modulus whose dynamics controls the post-inflationary evolution is $\tau_b$. During inflation this field gets slightly displaced from its minimum due to the inflationary energy density \cite{Cicoli:2016olq}. When the Hubble scale becomes of order of its mass, the volume mode starts oscillating and gives rise very quickly to an early epoch of matter domination. When $\tau_b$ decays it dilutes everything that has been produced before. This dilution mechanism can be very useful to have viable super-heavy dark matter scenarios \cite{Allahverdi:2020uax} and Affleck-Dine baryogenesis \cite{Allahverdi:2016yws} that would otherwise lead to an overproduction of either WIMP dark matter or matter-antimatter asymmetry. Moreover, the decay of the volume mode tends to produce axionic dark radiation \cite{Cicoli:2012aq, Higaki:2012ar, Cicoli:2015bpq} which can be within observational constraints and can represent an interesting experimental signature of these constructions.

\subsection*{Comments on dS vacua}

The dS vacua obtained in Sec. \ref{secF} rely on a full stabilisation of all K\"ahler moduli in detail. We did not perform an explicit fixing of the axio-dilaton and the complex structure moduli even if we checked that the D3 tadpole cancellation condition leaves enough freedom to turn on appropriate 3-form background fluxes which should lift these directions at semi-classical level. We therefore argue that the stabilisation of these moduli should be under control. In \cite{Cicoli:2013cha} we exploited symmetries of the complex structure moduli space to reduce the effective number of these moduli to just a few. Similar techniques can be combined with the ones of \cite{Demirtas:2019sip, Demirtas:2020ffz, Blumenhagen:2020ire} to perform a full stabilisation of the dilaton and the $U$-moduli with an exponentially small flux superpotential $|W_0|$. Notice that the need to tune $|W_0|$ is model-dependent since it depends on microscopic quantities like CY intersection numbers and gauge flux quanta. In fact, in \cite{Cicoli:2017shd} we obtained similar dS vacua without the need to tune the flux superpotential. Moreover, as already stressed above, cases with more blow-up modes would reduce the tuning in $|W_0|$. Moreover we have shown that open string moduli can be fixed via D- and F-flatness conditions, even if a complete fixing of all these modes has still to be achieved. However we expect this to be possible via a combination of gauge and bulk fluxes, together with supersymmetry breaking effects. 

Let us also mention that we managed to obtain dS minima at values of the volume of order $\vo\sim 10^2-10^3$ which are not extremely large, but still large enough to keep a numerical, even if not parametric, control over quantum corrections. This result confirms the expectation that dS minima cannot appear at arbitrarily large volume in agreement with current swampland considerations \cite{Ooguri:2018wrx}. This is because the barrier becomes increasingly small when the volume is increased, and the dS minima eventually disappear.  Larger values of the volume are expected for more generic compactifications with the volume expected to increase with the square of $h^{1,1}$ as argued in Sec. \ref{secF}.

It is worth stressing that uplifting with T-branes is a feature that arises very naturally in type IIB string compactifications since it relies on generic features of these constructions: ($i$) the presence of hidden sector D7-branes which is typically forced by D7 tadpole cancellation; ($ii$) non-zero worldvolume fluxes which are in general required by Freed-Witten anomaly cancellation \cite{Freed:1999vc}; ($iii$) the need to turn on 3-form bulk fluxes for dilaton and complex structure moduli stabilisation. As explained in \cite{Cicoli:2015ylx}, points ($i$) and ($ii$) naturally induce a T-brane background, or equivalently charged matter fields fixed in terms of moduli-dependent FI terms, while point ($iii$) gives rise to the positive uplifting term, corresponding to non-zero F-terms of these matter fields.

We finally stress that our global constructions provide fully explicit setups of type IIB flux compactifications where the visible sector is exactly the MSSM and the moduli can be stabilised in a dS vacuum. While achieving full control is always challenging, there is a coherent picture of viable string vacua emerging from these investigations. We strongly believe that further exploring these phenomenologically preferred string scenarios opens up new avenues to scrutinising physical implications of string compactifications.

\section{Conclusions}
\label{sec:Conclusions} 

Quiver gauge theories from fractional branes constitute quintessential realisations of local models with viable particle phenomenology in string compactifications.
Despite that,
their embedding into compact CY backgrounds remains largely unexplored.
Here, the challenge is building a fully trustable and consistent global model including brane setups satisfying tadpole cancellation conditions combined with moduli stabilisation and dS uplifting.
In recent years, some of the authors of this paper made progress in this direction by providing global embeddings of oriented \cite{Cicoli:2012vw,Cicoli:2013mpa,Cicoli:2013zha,Cicoli:2013cha} and unoriented \cite{Cicoli:2017shd} quiver gauge theories.
In the case of the former, two identical del Pezzo divisors are exchanged under the orientifold involution.
In the latter scenario, there is only a single del Pezzo divisor transversely invariant under the orientifold action.
D-term stabilisation forces the shrinking of the divisor volume to zero size yielding a CY singularity.

The visible sector is realised on the worldvolume of D3-branes sitting at the tip of this singularity and consists of realistic extensions of the Standard Model such as trinification, Pati-Salam or $\mathrm{SU}(5)$ models.
Additional rigid divisors in the CY threefold host non-perturbative effects which are imperative to stabilise closed string moduli together with $\alpha^{\prime}$-corrections at exponentially large volume \cite{Balasubramanian:2005zx}.
The cancellation of Freed-Witten anomalies may induce a non-trivial flux background on the hidden sector D7-branes.
In this way, a T-brane background is naturally generated leading to Minkowski or slightly de Sitter vacua \cite{Cicoli:2015ylx}.
In the models with orientifolded fractional branes, soft masses turn out to be of order the gravitino mass
\cite{Conlon:2010ji} which can be either at intermediate or at LHC scales. In the former case, these models can also provide a viable description of cosmic inflation as in \cite{Cicoli:2017shd}.

This paper concerned the first construction of the Minimal Quiver Standard Model (MQSM) \cite{Anastasopoulos:2006da,Berenstein:2006pk} from D3-branes at del Pezzo singularities in a fully fledged type IIB CY threefold flux compactification.
Specifically, we employed ideas first proposed in \cite{Wijnholt:2007vn} to construct supersymmetric versions of the MQSM from a single orientifolded D3-brane at a $\mathrm{dP}_{5}$ singularity.
We primarily focussed on a particular setup which led to the Minimal Supersymmetric Left-Right Symmetric Model with an additional vector like pair of Higgs doublets as an intermediate quiver gauge theory.
This extension of the Standard Model addresses several open questions such as the origin of parity violation or the strong CP problem making it phenomenologically highly attractive, see for example \cite{Babu:2008ep} and references therein.
Critically, the local model stems from a single D3-brane without the need to introduce flavour D7-branes,
although the global embedding contained D7-branes passing through the singularity.
While they did not introduce any additional chiral states in the quiver gauge theory,
this could represent another source of desequestering of the visible sector realisation.

Subsequently, we searched for compact CY geometries suited for accommodating the local model. 
Among the many constraints on such a global embedding,
the diagonality constraints imposed on
the collapsing $\mathrm{dP}_{5}$ divisor turned out to be more restrictive than initially anticipated.
This condition on the triple intersection numbers was necessary to ensure that the shrinking of the $\mathrm{dP}_{5}$ volume does not force any other divisors to shrink in which case the visible sector would look rather different.

In a first attempt,
we studied the KS database \cite{Kreuzer:2000xy} of $4$-dimensional reflexive polytopes where CY threefolds are represented as fine, regular, star triangulations.
We randomly scanned over a range of Hodge numbers $h^{1,1}\leq 40$ utilising the software package \texttt{CYTools} \cite{Demirtas:2020dbm}.
None of the $\approx 350.000$ distinct geometries exhibited diagonal $\mathrm{dP}_{n}$ divisors with $1\leq n\leq 5$.
In fact, these results were exhaustive for $h^{1,1}\leq 5$ possibly hinting at a deeper underlying reason.
This observation led us to formulate a conjecture about the absence of such divisor structures in the KS database in general.
Clearly, a much more thorough analysis is required to quantify the validity of this conjecture.
We stress however that we found $\approx 3000$ distinct geometries involving diagonal $\mathrm{dP}_{n}$ divisors with $n\geq 6$ whose quiver gauge theories can be related to the one of $\mathrm{dP}_{5}$ via Higgsing, see e.g. \cite{Wijnholt:2002qz}.
In this sense, the KS database contains CY threefolds implicitly related to the types of constructions studied throughout this paper.
We hope to come back to these models in the near future.

To proceed, we utilised the basic fact that $\mathrm{dP}_{5}$ surfaces can be constructed as bi-quadrics in $\mathbb{P}^{4}$.
Thus, the strategy to constructing CY threefolds with (diagonal) $\mathrm{dP}_{5}$ divisors became looking for complete intersections of two equations with five-dimensional toric spaces.
The bi-quadric is recovered upon setting one coordinate to zero and properly gauge fixing all but one of the $\mathbb{C}^\ast$ actions.
In this way, we obtained several CY threefolds for $2\leq h^{1,1}\leq 4$ with diagonal $\mathrm{dP}_{5}$ divisors together with additional del Pezzo divisors supporting Euclidean D3-instantons.

In the remainder of the paper, we focussed on one particular geometry and presented a concrete embedding of the local model.
We studied a particular involution that required only a single O7-plane on a large $4$-cycle.
Here, we showed using the techniques of \cite{Franco:2007ii} that
in a local neighbourhood around the singularity the global orientifold involution can be related to the well-understood line orientifold of complex cones over $\mathrm{dP}_{5}$ at a special locus in complex structure moduli space \cite{Franco:2007ii,Garcia-Etxebarria:2015lif,Collinucci:2016hgh}.

The D7-tadpole induced by the O7-plane was cancelled by adding four D7-branes plus their images on top of the O7-plane.
The resulting hidden sector $\mathrm{SO}(8)$ gauge group was broken to $\mathrm{U}(4)$ by worldvolume fluxes which were required by Freed-Witten anomaly cancellation \cite{Freed:1999vc}.
The induced FI-term forced the adjoint scalar on the D7 worldvolume to gain a VEV generating
a T-brane background thereby ensuring bulk D-term cancellation.
Ultimately, this led to a positive contribution to the $4$d scalar potential from scalar soft masses and therewith to a well-controlled de Sitter uplift \cite{Cicoli:2015ylx}.
The additional $\mathrm{dP}_{5}$ divisors were wrapped by rank-$1$ ED3-instantons which were critical for closed string moduli stabilisation.
We used the standard LVS \cite{Balasubramanian:2005zx} to find SUSY-breaking minima by balancing non-perturbative effects against perturbative $\alpha^{\prime}$ corrections.
As usual, supersymmetry was broken spontaneously in the hidden sector by non-vanishing F-terms for bulk K\"ahler moduli mediating SUSY breaking to the visible sector via gravitational couplings.
Our model has also all the right features to realise K\"ahler moduli inflation \cite{Conlon:2005jm} and to give rise to a viable post-inflationary evolution with interesting observational implications, as described in Sect.~\ref{sec:PhenoImp}.

In summary, this paper is a first step towards having fully-fledged string constructions with geometric moduli stabilised and the MSSM or its Left-Right extension. We have followed the three steps mentioned in the introduction and in some sense it is a culmination of accumulated progress over the years on both the model building and moduli stabilisation. We have illustrated that combining the two into global string constructions with concrete CY compactifications is highly non-trivial but achievable. We also illustrated the richness of these constructions that somehow complements and generalises the F-theory constructions to the cases where it is not the fibre but the base that  is singular. This is a very large and promising class of models that has not been much explored so far.

In the future, we aim at considering a systematic approach to more generic models with gaugino condensation and a higher number of Kähler moduli that would naturally provide further realistic properties and could open new avenues towards inflation. This may also help the numerics to allow larger volumes with not so small values of $W_0$, as argued in Sec. \ref{secF}.
Furthermore, our realisation of the left-right symmetric model could provide a golden opportunity to study all the interesting properties of the left-right symmetric models within concrete string models, providing a UV completion for these interesting phenomenological models (see \cite{Maleknejad:2016qjz,Maleknejad:2020yys,Maleknejad:2021nqi} for recent discussions). Having explicit low-energy string models, there are plenty of flavour questions that may need to be studied before claiming to have a fully realistic model.

On the local side, it remains an unresolved issue to construct explicit orientifold actions on exceptional collections of fractional branes.
While the construction of exceptional collections is systemically possible using the techniques of \cite{Hanany:2006nm},
the resulting collections are generically incompatible with the standard ``large volume'' orientifold action.
This has been partially explored in \cite{Garcia-Etxebarria:2013tba} for the case of $\mathbb{C}^{3}/\mathbb{Z}_{3}$.
However, the situation is far from clear for del Pezzo singularities.

Similarly, on the global embedding side, there continue to be open challenges like the inclusion of U(1) instantons.
In addition, recent software developments such as \texttt{cohomCalg} \cite{Blumenhagen:2010pv, Blumenhagen:2011xn} and \texttt{CYTools} \cite{Demirtas:2020dbm} should allow for a
systematic classification of viable global models with D-branes at singularities.
While most model building strategies thus far have been based on a case by case study,
this would constitute a huge leap towards treating large classes of string compactifications simultaneously.

\vspace{0.5cm}

\noindent \textbf{\Large Acknowledgements}

\vspace{0.25cm}

\noindent 

We thank Christoph Mayrhofer for initial collaboration on this project and valuable input.
We gratefully acknowledge discussions with Alex Cole, Arthur Hebecker, Sven Krippendorf, Jakob Moritz and Gary Shiu.
We would like to thank the SISSA/ICTP HPC Cluster for allowing the access, and in particular Benvenuto Bazzo, Ivan Girotto and Johannes Grassberger at ICTP for their crucial technical help in scanning geometries. The work of FQ has been partially supported by STFC consolidated grants ST/P000681/1, ST/T000694/1.
AS acknowledges support by the German Academic Scholarship Foundation, by DAMTP through an STFC studentship as well as by the Cambridge Trust through a Helen Stone Scholarship.
AS also thanks ICTP, Trieste, for hospitality during the initial stages of this work.
I.G.E. is supported in part by STFC through grant ST/T000708/1.
M.C. and R.V. acknowledge support by INFN Iniziativa Specifica ST\&FI. 


\appendix

\section{Analysing the CY Threefolds from the KS Database}\label{app:KSScan}

\subsection{Diagonal dP divisors in CY hypersurfaces in toric ambient spaces }

In this section, we discuss the challenges for the global embedding of our local $\mathrm{dP}_5$ model using the CY threefolds arising from the four-dimensional reflexive polytopes listed in the KS database \cite{Kreuzer:2000xy}. As mentioned in the first requirement of the list, we need to search for CY threefolds ($X_3$) which could have at least one (diagonal) $\mathrm{dP}_5$ divisor. For that purpose, we utilized the topological data of CY threefolds from the polytope triangulations presented in \cite{Altman:2014bfa}, which we refer as AGHJN-database. In the search of $\mathrm{dP}_5$ divisor, we focus only on looking at the topology of the so-called ``coordinate divisors" $D_i$ which are defined through setting $x_i = 0$. However this is sufficient for capturing the del Pezzo surfaces as the non-coordinate divisors, which could arise from considering the combinations of various coordinate divisors, would not be rigid. With this underlying strategy, our plan for looking at the suitable del Pezzo divisors is twofold:
First we scan for the divisors $D_s$ which satisfy the topological conditions \eqref{eq:dP}.
Using this information, we subsequently impose the `diagonality' condition \eqref{eq:diagdP} on each of the del Pezzo divisor $D_s$.
For all the $\mathrm{dP}_5$ toric divisors which we have obtained for the CY threefolds in the KS database using the triangulation of the AGHJN-database \cite{Altman:2014bfa}, we find that the diagonality condition (\ref{eq:diagdP}) could never be satisfied for the $\mathrm{dP}_5$ divisors. In fact, most of the times we find that the volume of $\mathrm{dP}_5$ four-cycle takes the following form,
\bea
\label{eq:voldP5}
& & \tau_{\mathrm{dP}_5} = \left(\sum_i \, a_i t^i\right) \, \left(\sum_j b_j t^j \right) \qquad {\rm for \, \, some} \, \, i \neq j \, .
\eea
To illustrate the volume form for a $\mathrm{dP}_5$ divisor to take of the form as given in Eq.~(\ref{eq:voldP5}), one can consider the explicit CY threefolds with a $\mathrm{dP}_5$ divisor presented in \cite{AbdusSalam:2020ywo, Cicoli:2017axo}.
The main problem with the volume of the form (\ref{eq:voldP5}) is that one cannot shrink such $\mathrm{dP}_5$ divisors to a point-like singularity via squeezing along a single direction. This is what we call a `non-diagonal' del Pezzo. In this case, squeezing along a single direction results in a line-like singularity, and one has to squeeze the $\mathrm{dP}_5$ divisor from two directions to get a point-like singularity.

\subsection{Scanning results for (diagonal) del Pezzo divisors}

All the various scans which we will present in this article will correspond to the so-called `favourable' triangulations (Triang$^*$) and `favourable' geometries (Geom$^*$) \cite{Altman:2014bfa}. In fact, in non-favourable CY threefolds, the number of toric divisors in the basis is less than $h^{1,1}(CY)$, and subsequently there is always at least one coordinate divisor which is non-smooth, and often turns out to be a disjoint union of two del Pezzo surfaces. We  exclude such spaces from our scan; several of them can be described as complete intersections in higher dimensional toric spaces. 
Moreover, we look for the presence of del Pezzo surfaces of any degree (not just $\mathrm{dP}_5$).\footnote{This will be useful in future work for embedding generic local $\mathrm{dP}_n$ models.}
The results of our search are collected in the following tables\footnote{Let us mention that the sum of the individual counting of spaces having a particular type of $\mathrm{dP}_n$ is reflected to be quite large, e.g. as can be seen from Tab.~\ref{tab_dPns-Gstar}, this sum is even larger than the total number of CY threefolds for $h^{1,1}=4$. This is because of the fact that there can be multiple (types of) del Pezzo divisors within the same CY threefold.}.
\noindent
\begin{table}[H]
	\centering
	\hskip0.11cm  \begin{tabular}{|c||c|c||c|c|c|c|c|c|c|c|c|}
		\hline
		$h^{1,1}$ & Poly$^*$ & Geom$^*$ & $\mathrm{dP}_0$  & $\mathrm{dP}_1$ or  & $\mathrm{dP}_2$  & $\mathrm{dP}_3$  & $\mathrm{dP}_4$ & $\mathrm{dP}_5$  & $\mathrm{dP}_6$  & $\mathrm{dP}_7$  & $\mathrm{dP}_8$   \\
	        &  &  &   & ${\mathbb F}_0$  &   &   &  &   &   &  &   \\
		\hline
		1 & 5 & 5 & 0 & 0  & 0 & 0 & 0  & 0 & 0 & 0 & 0 \\
		2 & 36 & 39 & 9 & 4 & 0 & 0 & 0  & 0 & 2 & 4 & 5 \\
		3 & 243 & 305 & 55 & 88 & 4 & 4 & 2 & 9 & 20 & 62 & 64 \\
		4 & 1185 & 2000 & 304 & 767  & 146 & 135 &  52 & 175  & 213 & 566 &  506 \\
		5 & 4897 & 13494 & 2107 &  6518 & 1960 & 2094 & 880  & 2005  & 2011 & 4358 & 3837 \\
		\hline
	\end{tabular}
	\caption{Number of CY geometries with a particular type of del Pezzo divisor.}
	\label{tab_dPns-Gstar}
\end{table}
 \noindent
\begin{table}[H]
	\centering
	\hskip0.11cm \begin{tabular}{|c||c|c||c|c|c|c|c|c||c|}
		\hline
		$h^{1,1}$ & Poly$^*$ &  Geom$^*$ & $\mathrm{ddP}_0$  & $\mathrm{ddP}_1$ or  & $\mathrm{ddP}_n$ & $\mathrm{ddP}_6$  & $\mathrm{ddP}_7$  & $\mathrm{ddP}_8$  & $n_{\rm LVS}$ \\
		&  & ($n_{\rm CY}$) &  & $d{\mathbb F}_0$  & $2\leq n \leq 5$ &  &  &  & ($\mathrm{ddP}_n\geq 1$) \\
		\hline
		1 & 5 & 5 & 0 & 0  & 0  & 0 & 0 & 0 & 0 \\
		2 & 36 & 39 & 9 & 2 & 0 & 2  & 4  & 5 & 22 \\
		3 & 243 & 305 & 55 & 16 & 0  & 16 & 37 & 34 & 132\\
		4 & 1185 & 2000 & 304 & 140 & 0 & 97 & 210 & 126 &  750 \\
		5 & 4897 & 13494 & 2107 & 901  & 0  & 486 & 731 & 374 & 4104 \\
		\hline
	\end{tabular}
	\caption{Number of CY geometries with a `diagonal' del Pezzo divisor suitable for LVS.}
	\label{tab_ddPns-Gstar}
\end{table}
\noindent
To present the number of CY threefolds which could support the standard LVS \cite{Balasubramanian:2005zx}, we have created separate column ``LVS" in Table \ref{tab_ddPns-Gstar} which correspond to  the CY threefolds having at least one `diagonal' del Pezzo divisor. 

Note that both the scanning approaches regarding the divisor topologies, either by looking at the Intersection Tensor satisfying the Eq. (\ref{eq:dP}) or by considering the Hodge Diamond using the \texttt{cohomCalg} \cite{Blumenhagen:2010pv, Blumenhagen:2011xn}, do not distinguish between the divisors ${\mathbb F}_0 = {\mathbb P}^1 \times {\mathbb P}^1$ and $\mathrm{dP}_1$ as both the surfaces are described by a set of Hodge numbers $\{h^{0,0} =1, h^{0,1} =0, h^{0,2} =0, h^{1,1} =2\}$. In order to make this distinction, we have looked at the circumstances when a surface can be a $\mathrm{dP}_1$ and when it cannot. This can be checked by considering the following triple-intersections for a given divisor $D_s$ with the above mentioned Hodge numbers, which could either be a ${\mathbb F}_0 $ or  a $\mathrm{dP}_1$ surface, 
\bea
& & \hskip-1cm \int_{CY} {D_s}^2 \, D_i  = m, \qquad  \int_{CY} {D_s} \, D_i^2  = n \qquad \qquad s \neq i.
\eea
Now
\begin{itemize}
\item{if at least one of the intersection numbers $m, \, n  \in \{2 k + 1: k \in {\mathbb Z}\}$ for some divisor $D_i$, then the corresponding divisor $D_s$ is a $\mathrm{dP}_1$. This simple condition is sufficient however not necessary, and it  turns out to be quite strong to capture the $\mathrm{dP}_1$ surfaces. }
\item{for the cases where all the $m \in \{2 k: k \in {\mathbb Z}\}$, the situation can get a bit subtle to make a conclusion as it might be possible that all divisors in the CY restrict to the same
(even) homology class in the surface. However one can determine the cases when the divisor cannot be a $\mathrm{dP}_1$. This can be checked by failing to find a solution of the following conditions for a given even number $m = 2k$:
\bea
& & 2k = p^2 - q^2 \qquad  {\rm for} \, \, {\rm some} \, \, p , q  \in {\mathbb Z}, \quad k \in {\mathbb Z}^\ast 
\eea
Or
\bea
& & \exists \, \, k_1, k_2 \in 2{\mathbb Z}^\ast : \quad \frac{k_2}{k_1} = \left(\frac{p}{q} \right)^2 \qquad  {\rm for} \, \, {\rm some} \, \, p , q  \in {\mathbb Q}.
\eea
In our case, it turns out that there are no solutions to each of these conditions and hence ensuring that the corresponding divisor cannot be a $\mathrm{dP}_1$.}
\item{assuming that the divisors with $\{h^{0,0} =1, h^{0,1} =0, h^{0,2} =0, h^{1,1} =2\}$ appearing in our scan are either $\mathrm{dP}_1$ or ${\mathbb F}_0 = {\mathbb P}^1 \times {\mathbb P}^1$, we consider the number of ${\mathbb F}_0$ divisors as those which are guaranteed to be not a $\mathrm{dP}_1$ surface.}
\end{itemize}
 \noindent
\begin{table}[H]
	\centering
	\hskip0.11cm  \begin{tabular}{|c||c|c||c|c|c|c|c|c|c|c|c|c|}
		\hline
		$h^{1,1}$ & Poly$^*$ & Geom$^*$ & $\mathrm{dP}_0$  & ${\mathbb F}_0$ & $\mathrm{dP}_1$ & $\mathrm{dP}_2$  & $\mathrm{dP}_3$  & $\mathrm{dP}_4$ & $\mathrm{dP}_5$  & $\mathrm{dP}_6$  & $\mathrm{dP}_7$  & $\mathrm{dP}_8$   \\
		\hline
		1 & 5 & 5 & 0 & 0 & 0  & 0 & 0 & 0  & 0 & 0 & 0 & 0 \\
		2 & 36 & 39 & 9 & 2 & 2 & 0 & 0 & 0  & 0 & 2 & 4 & 5 \\
		3 & 243 & 305 & 55 & 20 & 68 & 4 & 4 & 2 & 9 & 20 & 62 & 64 \\
		4 & 1185 & 2000 & 304 & 166 & 601 & 146 & 135 &  52 & 175  & 213 & 566 &  506 \\
		5 & 4897 & 13494 & 2107 & 1203 & 5315 & 1960 & 2094 & 880  & 2005  & 2011 & 4358 & 3837 \\
		\hline
	\end{tabular}
	\caption{Number of CY geometries with a particular type of del Pezzo divisor.}
	\label{tab_dPns-GstarM}
\end{table}
 \noindent
\begin{table}[H]
	\centering
	\hskip0.11cm \begin{tabular}{|c||c|c||c|c|c|c|c|c||c|}
		\hline
		$h^{1,1}$ & Poly$^*$ &  Geom$^*$ & $\mathrm{ddP}_0$  & $d{\mathbb F}_0$  & $\mathrm{ddP}_n$ & $\mathrm{ddP}_6$  & $\mathrm{ddP}_7$  & $\mathrm{ddP}_8$  & $n_{\rm LVS}$ \\
		&  & ($n_{\rm CY}$) &  &  & $1\leq n \leq 5$ &  &  &  & ($\mathrm{ddP}_n\geq 1$) \\
		\hline
		1 & 5 & 5 & 0 & 0  & 0  & 0 & 0 & 0 & 0 \\
		2 & 36 & 39 & 9 & 2 & 0 & 2  & 4  & 5 & 22 \\
		3 & 243 & 305 & 55 & 16 & 0  & 16 & 37 & 34 & 132\\
		4 & 1185 & 2000 & 304 & 140 & 0 & 97 & 210 & 126 &  750 \\
		5 & 4897 & 13494 & 2107 & 901  & 0  & 486 & 731 & 374 & 4104 \\
		\hline
	\end{tabular}
	\caption{Number of CY geometries with a `diagonal' del Pezzo divisor suitable for LVS.}
	\label{tab_ddPns-GstarM}
\end{table}
\noindent
Let us also note here that we have extended our notation to denote a diagonal ${\mathbb F}_0 = {\mathbb P}^1 \times {\mathbb P}^1$ as $d{\mathbb F}_0$ which satisfy the condition given in Eq. (\ref{eq:diagdP}).

In course of making this study we have analysed all the divisor topologies for more than 60,000 CY threefolds arising from the triangulations of the four-dimensional polytopes collected in the KS database. In this process, we have encountered around 2000 $\mathrm{dP}_5$ divisors while considering the distinct CY geometries, and around 9000 $\mathrm{dP}_5$ divisors while considering the CY triangulations, and interestingly none of these $\mathrm{dP}_5$ divisors satisfy the diagonality criteria mentioned in Eq. (\ref{eq:diagdP}). The generic observations made in this analysis made us formulate the conjecture stated in Sect.~\ref{sec:CYDiagdP5}.

\section{CICY Threefolds with Diagonal $\mathrm{dP}_5$ Divisors}\label{Sec:AppCICYdP5}

\subsection{$\mathrm{dP}_5$ embedded in a CICY with $h^{1,1} = 2$}

The toric data for a complete intersection CY threefold which realizes the diagonal $\mathrm{dP}_5$ divisors is given as under,
\begin{table}[H]
  \centering
 \begin{tabular}{|c|c||ccccccc|}
\hline
  $HY_1$  & $HY_2$ & $x_1$  & $x_2$  & $x_3$  & $x_4$  & $x_5$ & $x_6$ &  $x_7$ \\
    \hline
 3 & 3 & 1 & 1  & 1 & 1 & 1 & 1 & 0  \\
 1 & 1 & 0 & 0  & 0 & 0 & 0 & 1 & 1   \\
 \hline
  &  & SD1 &  SD1 & SD1 & SD1 & SD1  & SD2 & dP$_5$ \\
    \hline
  \end{tabular}
 \end{table}
 \noindent
with the SR ideal being given as $\{x_1 x_2 x_3 x_4 x_5, \, x_6 x_7 \}$. This CY threefold has the Hodge numbers $(h^{2,1}, h^{1,1}) = (66, 2)$ and subsequently the Euler number $\chi=-128$. A detailed divisor analysis using \texttt{cohomCalg} \cite{Blumenhagen:2010pv, Blumenhagen:2011xn} shows that the divisor $D_7$ is a indeed a del Pezzo $\mathrm{dP}_5$ while the divisors $\{D_1,..., \, D_6\}$ constitute two `special deformation' divisors with Hodge diamond:
\bea
{\rm SD1} \equiv
\begin{tabular}{ccccc}
    & & 1 & & \\
   & 0 & & 0 & \\
  4 & & 45 & & 4 \\
   & 0 & & 0 & \\
    & & 1 & & \\
  \end{tabular} \qquad \qquad\text{and}\qquad\qquad
		{\rm SD2} \equiv
\begin{tabular}{ccccc}
    & & 1 & & \\
   & 0 & & 0 & \\
  5 & & 51 & & 5 \\
   & 0 & & 0 & \\
    & & 1 & & \\
  \end{tabular} \nn
	\eea
Here following the nomenclature proposed in \cite{Gao:2013pra} we refer the divisors with $h^{2,0}(D) \geq 2$ as ``special deformation" (SD) divisors. These are special in the sense that the simplest divisor which can be deformed has $h^{2,0}(D) = 1$ e.g. the well known $K3$ surfaces. The intersection form in the basis of smooth divisors $\{D_6, D_7\}$ can be written as:
\be
I_3=  9\, D_6^3+  4 \, D_7^3\,.
\label{I3B0}
\ee
Writing the K\"ahler form in the above basis of divisors as $J=t_6\, D_6 + t_7\, D_7$  and using the intersection
polynomial (\ref{I3B0}), the CY overall volume takes the form:
\be
\vo = \frac{3\, t_6^3}{2} + \frac{2\, t_7^3}{3} \, = \frac{\sqrt{2}}{9} \tau_6^{3/2} - \frac{1}{3\sqrt{2}} \tau_7^{3/2} \, ,
\ee
where the divisor volumes are given as $\tau_6 = \frac{9\, t_6^2}{2}$ and $\tau_7 = 2\, t_7^2$.
Unlike the $\mathrm{dP}_5$ divisors embedded into CY hypersurfaces, this divisor is indeed diagonal. In other words, now we can shrink it to a point-like singularity along a single direction, namely by taking $t_7 \to 0$. Unlike the previous conclusion drawn for the whole set of examples in the KS list, this example serves as the proof for the existence of a CICY threefold with a `diagonal' $\mathrm{dP}_5$ divisor. 

The K\"ahler cone conditions are given as below,
\bea
& & 
t_6 + t_7 > 0, \qquad \quad t_7 < 0, 
\eea
which show that shrinking the volume of the $dP_5$ divisor corresponds to approaching the boundary of the K\"ahler cone.

\subsection{CICY examples with $h^{1,1} = 3$}

\subsubsection*{Example A}

Now we present a CICY threefold which has a diagonal $\mathrm{dP}_5$ divisor along with another diagonal $\mathrm{dP}_n$ divisor, in order to support our local $\mathrm{dP}_5$ model within a LVS framework. The toric data for such a CICY threefold is given as under,
\begin{table}[H]
  \centering
 \begin{tabular}{|c|c||cccccccc|}
\hline
  $HY_1$  & $HY_2$ & $x_1$  & $x_2$  & $x_3$  & $x_4$  & $x_5$ & $x_6$ &  $x_7$ &  $x_8$ \\
    \hline
 3 & 3 & 1 & 1  & 1 & 1 & 1 & 1 & 0 &  0  \\
 2 & 0 & 0 & 0  & 0 & 0 & 0 & 1 & 1  & 0  \\
 1 & 1 & 0 & 0  & 0 & 0 & 1 & 0 & 0  & 1 \\
 \hline
  &  & SD1 &  SD1 & SD1 & SD1 & SD2  & SD3 & $\mathrm{dP}_6$ & dP$_5$ \\
    \hline
  \end{tabular}
 \end{table}
 \noindent
with the SR ideal being given as $\{x_1 x_2 x_3 x_4, \,  x_5 \, x_8, \, x_6 x_7 \}$. This CY threefold has the Hodge numbers $(h^{2,1}, h^{1,1}) = (55, 3)$ and subsequently the Euler number $\chi=-104$. A detailed divisor analysis using \texttt{cohomCalg} \cite{Blumenhagen:2010pv, Blumenhagen:2011xn} shows that the divisor $D_8$ is a indeed a del Pezzo $\mathrm{dP}_5$ while the divisor $D_7$ is a $\mathrm{dP}_6$. Further, the divisors $\{D_1,..., \, D_6\}$ constitute three `special deformation' type divisors with Hodge diamond:
\bea
{\rm SD1} \equiv
\begin{tabular}{ccccc}
    & & 1 & & \\
   & 0 & & 0 & \\
  3 & & 38 & & 3 \\
   & 0 & & 0 & \\
    & & 1 & & \\
  \end{tabular} , \quad
		{\rm SD2} \equiv
\begin{tabular}{ccccc}
    & & 1 & & \\
   & 0 & & 0 & \\
  4 & & 44 & & 4 \\
   & 0 & & 0 & \\
    & & 1 & & \\
  \end{tabular}, \quad {\rm SD3} \equiv
\begin{tabular}{ccccc}
    & & 1 & & \\
   & 0 & & 0 & \\
  4 & & 45 & & 4 \\
   & 0 & & 0 & \\
    & & 1 & & \\
  \end{tabular} \,.	\nn
  \eea
The intersection form in the basis of smooth divisors $\{D_6, D_7, D_8\}$ can be written as:
\be
I_3=  5\, D_6^3+  3 \, D_7^3 + 4 \, D_6^2 D_8 - 4 \, D_6 D_8^2 + 4 D_8^3 \,.
\label{I3B00}
\ee
Writing the K\"ahler form in the above basis of divisors as $J=t_6\, D_6 + t_7\, D_7 + t_8\, D_8$  and using the intersection
polynomial (\ref{I3B00}), the CY overall volume takes the form:
\bea
& \vo &= \frac{5}{6}\, t_6^3+  \frac{1}{2} \, t_7^3 + 2 \, t_6^2 t_8 - 2 \, t_6 t_8^2 + \frac{2}{3} t_8^3 \, = \frac{\sqrt{2}}{9} (\tau_6 + \tau_8)^{3/2} - \frac{1}{3} \sqrt{\frac{2}{3}} \, \tau_7^{3/2} - \frac{1}{3\sqrt{2}} \, \tau_8^{3/2} \,, \nonumber
\eea
where in the second step we have used the following expressions for the divisor volumes $\tau_6 = \frac{5\, t_6^2}{2} + 4 t_6 t_8 - 2 t_8^2$, $\tau_7 = \frac{3\, t_7^2}{2}$ and $\tau_8 = 2\, (t_6 - t_8)^2$.
This volume form suggests that working in the different basis of smooth divisors $\{D_x, D_7, D_8\}$, the intersection polynomial reduces into the following form,
\bea
I_3 = 9\,D_x^3 \, + 3 D_7^3 + 4 D_8^3 \,,
\eea
which subsequently gives the following strong swiss-cheese volume form,
\bea
& & \vo = \frac{\sqrt{2}}{9} \tau_x^{3/2} - \frac{1}{3} \sqrt{\frac{2}{3}} \, \tau_7^{3/2} - \frac{1}{3\sqrt{2}} \, \tau_8^{3/2} \,,
\eea
where $\tau_x = \frac{9 t_x^2}{2}, \tau_7 = \frac{3 t_7^2}{2}$ and $\tau_8= 2\, t_8^2$. Thus we ensure that both the $\mathrm{dP}_5$ as well as $\mathrm{dP}_6$ divisors are diagonal. In other words, now we can shrink it to a point-like singularity along a single direction, namely by $t_8 \to 0$.

\subsubsection*{Example B}

Now we refine our construction a bit more by presenting a CICY threefold which has two diagonal $\mathrm{dP}_5$ divisors for $h^{1,1}(X) = 3$. The toric data is given by,
\begin{table}[H]
  \centering
 \begin{tabular}{|c|c||cccccccc|}
\hline
  $HY_1$  & $HY_2$ & $x_1$  & $x_2$  & $x_3$  & $x_4$  & $x_5$ & $x_6$ &  $x_7$ &  $x_8$ \\
    \hline
 4 & 4 & 1 & 0  & 0  & 2 & 2 & 1 &  1 & 1  \\
 2 & 2 & 0 & 1  & 0  & 1 & 1 & 0  & 0 & 1 \\
 2 & 2 & 0 & 0  & 1  & 1 & 1 & 0  & 1 & 0 \\
 \hline
  &  & SD1 & dP$_5$ & dP$_5$ & SD2  & SD2 & SD1 & SD3 & SD3 \\
    \hline
  \end{tabular}
 \end{table}
 \noindent
with the SR ideal being given as 
\bea
SR =\{x_2 x_3, \, x_2 x_8, \, x_1 x_6 x_8, \, x_3 x_4 x_5 x_7, \, x_1 x_4 x_5 x_6 x_7 \}. \nn
\eea
This CY threefold has the Hodge numbers $(h^{2,1}, h^{1,1}) = (59, 3)$ and subsequently the Euler number $\chi=-112$. In fact these GLSM charges results in three triangulations such that the corresponding CICY has two diagonal $\mathrm{dP}_5$ divisors. Other two triangulations corresponds to the following SR ideas:
\bea
& & SR_1 = \{x_2 x_3, \, x_2 x_8, \, x_3 x_7, \, x_1 x_4 x_5 x_6 x_7, \, x_1 x_4 x_5 x_6 x_8 \} \nonumber\\
& & SR_2 = \{x_2 x_3, \, x_3 x_7, \, x_1 x_6 x_7, \, x_2 x_4 x_5 x_8, \, x_1 x_4 x_5 x_6 x_8\}. \nn
\eea 
\noindent
A detailed divisor analysis using \texttt{cohomCalg} \cite{Blumenhagen:2010pv, Blumenhagen:2011xn} shows that the two divisors $D_2$ and $D_3$ are del Pezzo $\mathrm{dP}_5$ surfaces while the remaining divisors constitute three types of what we call `special deformation' divisors with the following Hodge diamonds:
\bea
{\rm SD1} \equiv
\begin{tabular}{ccccc}
    & & 1 & & \\
   & 0 & & 0 & \\
  1 & & 24 & & 1 \\
   & 0 & & 0 & \\
    & & 1 & & \\
  \end{tabular} , \quad
		{\rm SD2} \equiv
\begin{tabular}{ccccc}
    & & 1 & & \\
   & 0 & & 0 & \\
  9 & & 76 & & 9 \\
   & 0 & & 0 & \\
    & & 1 & & \\
  \end{tabular}, \quad
		{\rm SD3} \equiv
\begin{tabular}{ccccc}
    & & 1 & & \\
   & 0 & & 0 & \\
  2 & & 30 & & 2 \\
   & 0 & & 0 & \\
    & & 1 & & \\
  \end{tabular}.	\nn
  \eea
In the basis of smooth divisors $\{D_b, D_2, D_3\}$, where is defined as $D_b = D_1 + D_2 + D_3$, the intersection form can be written as:
\bea
I_3 = 4\,D_b^3 \, + 4 \, D_2^3\, + 4\, D_3^3 \,,
\eea
which subsequently gives the following strong swiss-cheese volume form,
\bea
& & \vo = \frac{1}{3\sqrt{2}} \, \left(\tau_b^{3/2} - \, \tau_2^{3/2} - \, \tau_3^{3/2} \right)\,,
\eea
where the four-cycle volumes are given in terms of the two-cycle volumes as: $\tau_b = 2\, t_b^2, \, \tau_2 = 2\, t_2^2$ and $\tau_3 = 2\, t_3^2$. Thus we ensure that both of the $\mathrm{dP}_5$ divisors are diagonal. So we can equivalently shrink any of the $\mathrm{dP}_5$ to a point-like singularity by squeezing along a single direction, for example say via taking $t_3 \to 0$.

\subsubsection*{Example C}

Now we present a CICY threefold example which has one diagonal $\mathrm{dP}_5$ divisor along with a diagonal  $\mathrm{dP}_8$ divisor for $h^{1,1}(X) = 3$.
Even though this model does not work for our construction,
it might become useful in the future.
The toric data for such a CICY threefold is given as under,
\begin{table}[H]
  \centering
 \begin{tabular}{|c|c||cccccccc|}
\hline
  $HY_1$  & $HY_2$ & $x_1$  & $x_2$  & $x_3$  & $x_4$  & $x_5$ & $x_6$ &  $x_7$ &  $x_8$ \\
    \hline
 4 & 6 & 1 & 0  & 0  & 3 & 2 & 2  &  1 & 1  \\
 2 & 4 & 0 & 1  & 0  & 2 & 1 & 1  & 0 & 1 \\
 2 & 0 & 0 & 0  & 1  & 0 & 0 & 0  & 1 & 0 \\
 \hline
  &  & NdP$_{12}$ & dP$_5$ & dP$_8$ & SD1  & SD2 & SD2 & SD3 & SD4 \\
 \hline
 \end{tabular}
 \end{table}
 \noindent
with the SR ideal being given as 
\bea
SR =\{x_3 x_7,\, x_2 x_3 x_8, \, x_2 x_4 x_8, \, x_1 x_5 x_6 x_7, \, x_1 x_4 x_5 x_6 \}. \nn
\eea
This CY threefold has the Hodge numbers $(h^{2,1}, h^{1,1}) = (43, 3)$ and subsequently the Euler number $\chi=-80$. A detailed divisor analysis using \texttt{cohomCalg} \cite{Blumenhagen:2010pv, Blumenhagen:2011xn} shows that the two divisors $D_2$ and $D_3$ are del Pezzo $\mathrm{dP}_5$ and $\mathrm{dP}_8$ surfaces respectively while the remaining divisors constitute four types of what we call `special deformation' (SD) divisors with the following Hodge diamonds:
\bea
{\rm SD1} \equiv
\begin{tabular}{ccccc}
    & & 1 & & \\
   & 0 & & 0 & \\
  8 & & 67 & & 8 \\
   & 0 & & 0 & \\
    & & 1 & & \\
  \end{tabular} , \qquad \qquad
		{\rm SD2} \equiv
\begin{tabular}{ccccc}
    & & 1 & & \\
   & 0 & & 0 & \\
  3 & & 36 & & 3 \\
   & 0 & & 0 & \\
    & & 1 & & \\
  \end{tabular}, 	\nn
  \eea
\bea
{\rm SD3} \equiv
\begin{tabular}{ccccc}
    & & 1 & & \\
   & 0 & & 0 & \\
  1 & & 22 & & 1 \\
   & 0 & & 0 & \\
    & & 1 & & \\
  \end{tabular}, \qquad \qquad
		{\rm SD4} \equiv
\begin{tabular}{ccccc}
    & & 1 & & \\
   & 0 & & 0 & \\
  1 & & 19 & & 1 \\
   & 0 & & 0 & \\
    & & 1 & & \\
  \end{tabular}.	\nn
  \eea
In the basis of smooth divisors $\{D_b, D_2, D_3\}$, where is defined as $D_b = D_1 + D_2 + D_3$, the intersection form can be written as:
\bea
I_3 = 2\,D_b^3 \, + 4 \, D_2^3\, + \, D_3^3 \,,
\eea
which subsequently gives the following strong swiss-cheese volume form,
\bea
& & \vo = \frac{1}{3} \, \tau_b^{3/2} - \, \frac{1}{3\sqrt{2}} \, \tau_2^{3/2} - \, \frac{\sqrt{2}}{3} \, \tau_3^{3/2} \,,
\eea
where the four-cycle volumes are given in terms of the two-cycle volumes as: $\tau_b = t_b^2, \, \tau_2 = 2\, t_2^2$ and $\tau_3 = \frac{t_3^2}{2}$. Thus we ensure that both of the del Pezzo divisors, namely $\mathrm{dP}_5$ as well as $\mathrm{dP}_8$, are diagonal. So we can shrink the $\mathrm{dP}_5$ to a point-like singularity by squeezing along a single direction via taking $t_2 \to 0$, leading to the following effective volume form: 
\bea
& & \vo = \frac{1}{3} \, \tau_b^{3/2} - \, \frac{\sqrt{2}}{3} \, \tau_3^{3/2} \,,
\eea
The second Chern-class is given as:
\bea
& & c_2(X) = 16 \, D_b^2 + D_2^2 + 10 \, D_3^2.
\eea

\subsection{A CICY example with $h^{1,1} = 4$}

Now we take our construction one step ahead by supporting one of the LVS models of inflation, namely the K\"ahler moduli inflation, in addition to have a global LVS embedding of our local chiral $\mathrm{dP}_5$ model. For that here present a CICY threefold which has a diagonal $\mathrm{dP}_5$ divisor and two other diagonal $\mathrm{dP}_n$ divisor. The toric data for such a CICY threefold is given as under,
\begin{table}[H]
  \centering
 \begin{tabular}{|c|c||ccccccccc|}
\hline
  $HY_1$  & $HY_2$ & $x_1$  & $x_2$  & $x_3$  & $x_4$  & $x_5$ & $x_6$ &  $x_7$ &  $x_8$ &  $x_9$ \\
    \hline
 3 & 3 & 1 & 1  & 1 & 1 & 1 & 1 & 0 &  0 & 0  \\
 2 & 0 & 0 & 0  & 0 & 0 & 0 & 1 & 1  & 0 & 0 \\
 0 & 2 & 0 & 0  & 0 & 0 & 1 & 0 & 0  & 1 & 0 \\
 1 & 1 & 0 & 0  & 0 & 1 & 0 & 0 & 0  & 0 & 1 \\
 \hline
  &  & SD1 &  SD1 & SD1 & SD2 & SD2  & SD2 & $\mathrm{dP}_6$ & $\mathrm{dP}_6$ & dP$_5$ \\
    \hline
  \end{tabular}
 \end{table}
 \noindent
with the SR ideal being given as 
\bea
SR =\{x_4 x_9, \, x_6 x_7, \, x_7 x_8, \, x_7 x_9, \, x_5 x_8, \, x_8 x_9, \, x_1 x_2 x_3 x_4 x_5, \, x_1 x_2 x_3 x_4 x_6, \, x_1 x_2 x_3 x_5 x_6 \}. \nn
\eea
This CY threefold has the Hodge numbers $(h^{2,1}, h^{1,1}) = (44, 4)$ and subsequently the Euler number $\chi=-80$. A detailed divisor analysis using \texttt{cohomCalg} \cite{Blumenhagen:2010pv, Blumenhagen:2011xn} shows that the divisor $D_8$ is a indeed a del Pezzo $\mathrm{dP}_5$ while the divisor $D_6$ and $D_7$ divisors are both a $\mathrm{dP}_6$ surface. Further, the divisors $\{D_1,..., \, D_5\}$ constitute two `special deformation' type divisors with Hodge diamond:
\bea
{\rm SD1} \equiv
\begin{tabular}{ccccc}
    & & 1 & & \\
   & 0 & & 0 & \\
  2 & & 31 & & 2 \\
   & 0 & & 0 & \\
    & & 1 & & \\
  \end{tabular} , \qquad
		{\rm SD2} \equiv
\begin{tabular}{ccccc}
    & & 1 & & \\
   & 0 & & 0 & \\
  3 & & 38 & & 3 \\
   & 0 & & 0 & \\
    & & 1 & & \\
  \end{tabular} \,.	\nn
  \eea
The intersection form in the basis of smooth divisors $\{D_6, D_7, D_8, D_9\}$ can be written as:
\be
I_3=  2\, D_6^3+  3 \, D_7^3 + 9 \, D_6^2 D_8 - 9 \, D_6 D_8^2 + 3 D_8^3 + 4 D_6^2 D_9 - 4 D_6 D_9^2 + 4\, D_9^3 \,.
\label{I3B3}
\ee
Writing the K\"ahler form in the above basis of divisors as $J=t_6\, D_6 + t_7\, D_7 + t_8\, D_8$  and using the intersection
polynomial (\ref{I3B3}), the CY overall volume takes the form:
\bea
& \vo & = \frac{t_6^3}{3}\, +  \frac{t_7^3}{2} + \frac{3 \, t_6^2\, t_8}{2} - \frac{3 \, t_6 \, t_8^2}{2} + \frac{t_8^3 }{2} + 2\, t_6^2\, t_9 - 2\, t_6 \, t_8^2 +\, \frac{2\, t_9^3}{3} \\
& & = \frac{\sqrt{2}}{9} (\tau_6 + \tau_8+ \tau_9)^{3/2} - \frac{1}{3} \sqrt{\frac{2}{3}} \, \tau_7^{3/2} - \frac{1}{3} \sqrt{\frac{2}{3}} \, \tau_8^{3/2} - \frac{1}{3\sqrt{2}} \, \tau_9^{3/2} \,, \nonumber
\eea
where in the second step we have used the following expressions for the divisor volumes 
$\tau_6 = t_6^2 + 3\, t_6\, t_8 - \frac{3\, t_8^2}{2} + 4 t_6 t_9 - 2 t_9^2$, $\tau_7 = \frac{3\, t_7^2}{2}$, $\tau_8 = \frac{3\, (t_6-t_8)^2}{2}$ and $\tau_8 = 2\, (t_6 - t_9)^2$.
This volume form suggests that working in the different basis of smooth divisors $\{D_x, D_7, D_8, D_9\}$, the intersection polynomial reduces into the following form,
\bea
I_3 = 9\,D_x^3 \, + 3 \, D_7^3\, + 3\, D_8^3 + 4 \, D_9^3 \,,
\eea
which subsequently gives the following strong swiss-cheese volume form,
\bea
& & \vo = \frac{\sqrt{2}}{9} \tau_x^{3/2} - \frac{1}{3} \sqrt{\frac{2}{3}} \, \tau_7^{3/2} - \frac{1}{3} \sqrt{\frac{2}{3}} \, \tau_8^{3/2} - \frac{1}{3\sqrt{2}} \, \tau_9^{3/2} \,,
\eea
where $\tau_x = \frac{9 t_x^2}{2}, \tau_7 = \frac{3 t_7^2}{2}, \tau_8 = \frac{3 t_8^2}{2}$ and $\tau_9= 2\, t_9^2$. Thus we ensure that both the $\mathrm{dP}_6$, as well as $\mathrm{dP}_5$ divisors are diagonal. So we can shrink the $\mathrm{dP}_5$ to a point-like singularity along a single direction, namely by taking $t_9 \to 0$.

\section{Additional Material on $\mathrm{dP}_{5}$ Quivers}

\subsection{D-terms for $\mathrm{SU}(N)$ and $\mathrm{USp}(2N)$ gauge groups}\label{sec:d-terms-suN_and_spN}

For the non-abelian part of the gauge group there is no FI-term. Hence the D-term reduces to
\begin{equation}
D_a D^a=\kappa^{ab} D_a D_b
\end{equation} 
with $D_a$ given by
\begin{equation}\label{eq:d-term-1}
D_a=\sum_i\left\langle \varphi_i\right|T_a\left|\varphi_i\right\rangle= \Tr{T_a\sum_i\left|\varphi_i\right\rangle\left\langle \varphi_i\right|}=\Tr{T_a H}\,.
\end{equation}
Here we introduced the hermitian matrix $H$ as an abbreviation for $\sum_i\left|\varphi_i\right\rangle\left\langle \varphi_i\right|$ and the $T_a$'s are the generators of the non-abelian gauge group in the representation of the $\left|\varphi_i\right\rangle$'s.

We denote by $t_\alpha$ the generators of hermitian $N\times N$-matrices, whereupon we chose a basis such that $t_0=\mathds 1$, $\Tr{t_\alpha}=0$ for $\alpha>0$ and $\Tr{t_\alpha t_\beta}=0$ for $\alpha\ne\beta$. We use this basis to expand $H$ as $H=h^\beta t_\beta$ with $h^\beta\in \mathbb R$. Furthermore, the generators of $\mathrm{SU}(N)$ are given by $t_\alpha$ with $\alpha>0$.\footnote{Note that this is true for the fundamental representation. In the case that there are further states transforming under this gauge group, but in a different representation, the trace in \eqref{eq:d-term-1} is obviously a sum of traces over the respective representations. This means in particular, if we have states in the fundamental representation,  its conjugate representation and the adjoint representation then \eqref{eq:d-term-1} will look like $\Tr{t_a H_1}-\Tr{t_a H_2}+\sum_i \Tr{t_a [\phi_i,\phi_i^\dagger]}$, where $H_1$ and $H_2$ encode the states in the fundamental and anti-fundamental representation, respectively, and $\phi_i$ are the fields in the adjoint representation. } Hence we obtain for $D_\alpha$: 
\begin{equation}
D_\alpha=\Tr{t_\alpha h^\beta t_\beta}=h^\beta \Tr{t_\alpha t_\beta} \propto h_\alpha \qquad \textmd{with $\alpha>0$.} 
\end{equation}
Since $\kappa^{ab}$ is definite for simple Lie groups, we need either $H\propto\mathds 1$ or $H=0$ for a vanishing D-term.

For $\mathrm{USp}(2N)$ gauge groups the situation is a bit less restrictive, as we explain in the following. The $2N\times 2N$-matrices $t_a$  generating $\mathrm{USp}(2N)$ must be hermitian and in addition of the form
\begin{equation}
t_a=\left(\begin{array}{cc}
A_a & B_a^\dagger \\ B_a & -A_a^T
\end{array}\right)
\end{equation}  
with ($A=A^\dagger$ and) $B=B^T$ in order to be symplectic too. To simplify notation, we also subdivide the matrix $H$ into $N\times N$-matrices:
\begin{equation}
H=\left(\begin{array}{cc}
O & P \\ P^\dagger & Q^T
\end{array}\right)
\end{equation} 
with $O=O^\dagger$ and $Q=Q^\dagger$. Plugging this into \eqref{eq:d-term-1} we find
\begin{equation}
D_a=\Tr{A_a O+B_a^\dagger P^\dagger}+\Tr{B_aP-A_a^TQ^T}=\Tr{A_a(O-Q)}+2\,\Re(\Tr{B_aP})\,.
\end{equation}
To further evaluate this expression, we need a `good basis' for the generators of $\mathrm{USp}(2N)$. From 
\begin{equation}
\Tr{t_a t_b}=2\,\Tr{A_a A_b}+2\,\Re\left(\Tr{B_a^\dagger B_b}\right)
\end{equation}
we see that $A_\alpha=t_\alpha$, $B_\alpha=0$ for $\alpha=0,\ldots,N^2-1$ and $A_{n+N^2}=0$, $B_{n+N^2}=\tilde{t}_{n}$ for $n=0,\ldots,N^2+N-1$ is a basis of the kind we want.\footnote{The $\tilde{t}_n$'s are   
the $N(N+1)$ generators of the purely real or purely imaginary symmetric matrices, i.e.\ $\left[\tilde{t}_n\right]_{\mu\nu}=\delta_{i_n\mu}\delta_{j_n\nu}+\delta_{j_n\mu}\delta_{i_n\nu}$ with  $i_n$, $j_n$ on of the $N(N+1)/2$ index tuples $i\le j\le N$ and similar for the imaginary ones.} Because for this basis $\Tr{t_at_b}$ is again diagonal, i.e.\ $\Tr{t_at_b}=0$ for $a\ne b$. Since $O$ and $Q$ are hermitian and $P$ is unconstrained, we need for $D_a=0$ that
\begin{equation}
O=Q\qquad\textmd{and} \qquad P=-P^T\,.
\end{equation}
As argued in Sect.~\ref{sec:HiggsingModel1}, these constraints are trivially fulfilled for the constructed orientifold models of the $\mathrm{dP}_{5}$-quiver.

\subsection{Larger Model I}\label{sec:LargeModel1}

\subsection*{Higgsing}

After Higgsing, the quiver in Fig.~\ref{fig:QuivdP5Mod1Small} does not have the required non-chiral matter spectrum.
This is a sign that we have to construct new bound states.
For this reason, we replace $F_{b}$ by $F_{d}$ defined by \cite{Wijnholt:2007vn}
\begin{align}
\text{ch}(F_{d})&=\sum_{i=1}^{8}\, n^{(d)}_{i}\text{ch}(F_{i})=\text{ch}(F_{a})+\text{ch}(F_{3})+2\text{ch}(F_{b})+\text{ch}(F_{8})\kom \mathrm{deg}(F_{d})=-6
\end{align}
where
\begin{equation}
\mathbf{n}^{(d)}=(4, 2, 4, 3, 5, 4, 3, 4)\, .
\end{equation}
Looking at the charge vector in terms of the bound states $F_{a}$ and $F_{b}$,
the VEVs should be constructible from \eqref{eq:VEVs2FB} for $2F_{b}$ and \eqref{eq:Model1BoundStateFaVEVs} for $2F_{a}$.
Indeed, it turns out to be sufficient to choose
\begin{align}\label{eq:VEVs2FD} 
&\langle X_{13}\rangle=\left (\begin{array}{cccc}
a_{1} &0&0 &0\\
0 &a_{2}&0 &0 \\
0&0&a_{3}&0\\
0&0&0&a_{4}
\end{array} \right )\kom\langle X_{14}\rangle=\left (\begin{array}{ccc}
b_{1} &0&0\\
0 &b_{2}&0  \\
0&0&0\\
0&0&b_{3}
\end{array} \right )\, ,\nn\\
&\langle X_{23}\rangle=\left (\begin{array}{cccc}
0 &0&c_{1} &0\\
0 & 0 &0 &0
\end{array} \right )\kom \langle X_{24}\rangle=\left (\begin{array}{cccc}
0 &0&0 \\
0 &0&d_{1}  \\
\end{array} \right )\, ,\nn\\
 &\langle X_{35}^{T}\rangle=\left (\begin{array}{cccc}
x_{1} &0&0 &0\\
0 &x_{2}&0 &0 \\
0&0&x_{3} &0\\
0&0&0&0\\
0&0&0&0
\end{array} \right )\kom \langle X_{36}^{T}\rangle=\left (\begin{array}{cccc}
y_{1} &0&0 &0\\
0 &y_{2}&0 &0 \\
0&0&0&0\\
0&0&0&y_{3}
\end{array} \right )\, ,\nn\\
&\langle X_{45}^{T}\rangle=\left (\begin{array}{ccc}
0 &0 &0\\
0 &0 &0\\
z_{1}&0 &z_{4}\\
0 &z_{2} &0\\
0 &0 &z_{3}\\
\end{array} \right )\kom \langle X_{46}^{T}\rangle=\left (\begin{array}{ccc}
0&0&0\\
0&0&0\\
w_{1}&0&0\\
0&w_{2}&0
\end{array} \right )
\end{align}
where all entries correspond to $2\times 2$ matrices. We computed the mass matrix for the gauge potentials numerically and confirmed the breaking pattern
\begin{equation}
\mathrm{U}(8)^{3}\times\mathrm{U}(4)^{2}\times\mathrm{U}(6)^{2}\times\mathrm{U}(10)\raw \mathrm{U}(2)\, .
\end{equation}

\begin{figure}[t!]
\centering
\includegraphics[scale=0.3]{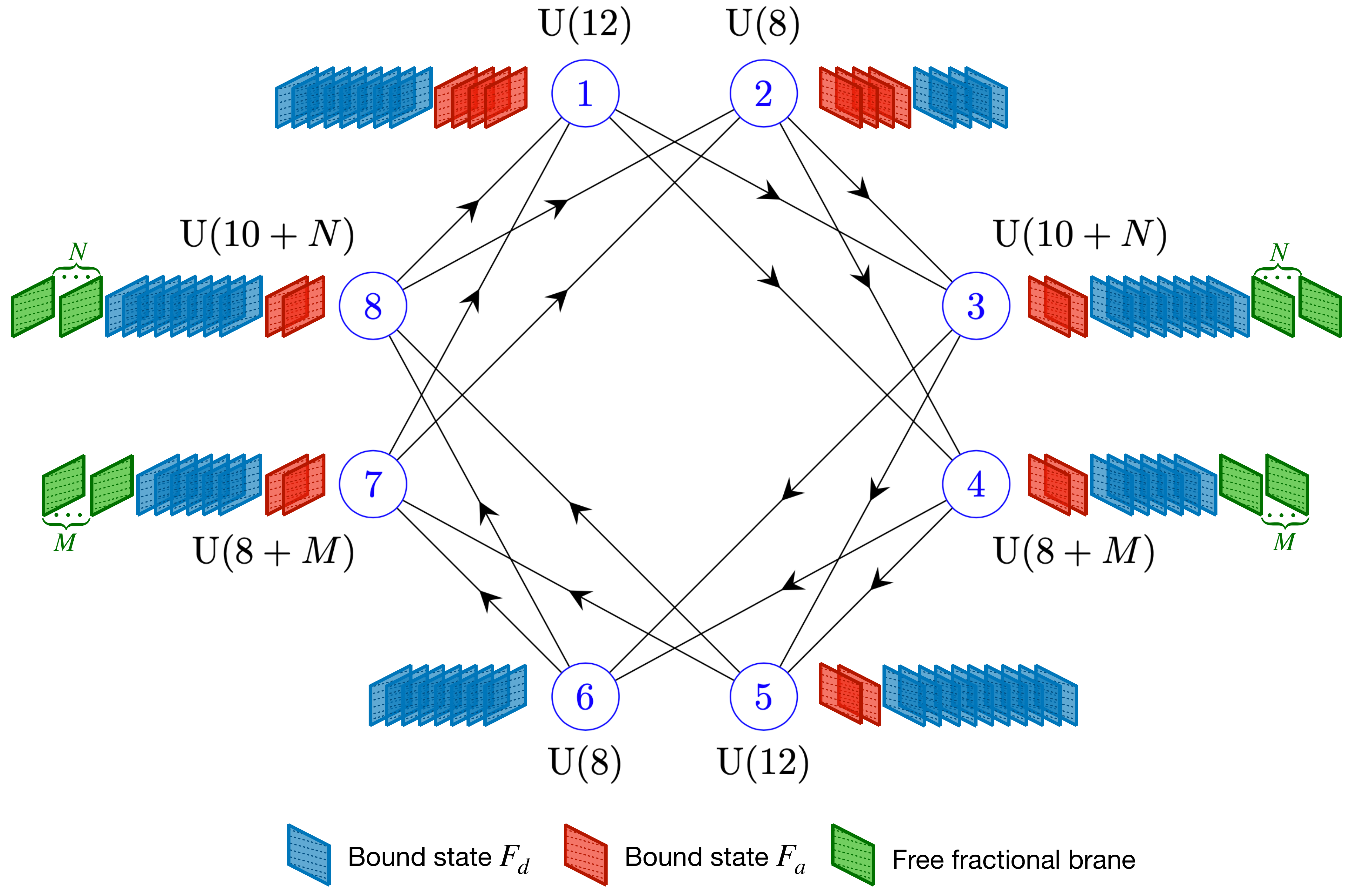}
\caption{Large model I quiver diagram for a D3-brane at a $\mathrm{dP}_{5}$-singularity with the bound state $2F_{a}$ indicated in {\color{red}red} and $2F_{d}$ in {\color{blue}blue}. The {\color{green}green} fractional branes do not contributing to any bound state.}\label{fig:LargeQuivdP5Mod1} 
\end{figure}

We fix the Higgsed quiver to be of the from $(2F_{a},NF_{3},MF_{4},2F_{d},MF_{7},NF_{8})$.
As before, the VEVs to Higgs the full quiver are obtained by embedding the choices \eqref{eq:VEVs2FD} for $2F_{d}$ and \eqref{eq:Model1BoundStateFaVEVs} for $2F_{a}$ into the larger gauge group representations.
This leads to
\begin{align}\label{eq:LMIChoice2} 
&\langle X_{13}\rangle=\left (\begin{array}{cccccc}
a_{1} &0&0 &0&0&\ldots\\
0 &a_{2}&0 &0&0&\ldots \\
0&0&a_{3}&0&0&\ldots\\
0&0&0&a_{4}&0&\ldots\\
0&0&0&0&a_{5}&\ldots\\
0&0&0&0&0&\ldots
\end{array} \right )\kom\langle X_{14}\rangle=\left (\begin{array}{ccccc}
b_{1} &0&0&0&\ldots\\
0 &b_{2}&0&0&\ldots  \\
0&0&0&0&\ldots\\
0&0&b_{3}&0&\ldots\\
0&0&0&b_{4}&\ldots\\
0&0&0&0&\ldots
\end{array} \right )\, ,\nn\\
&\langle X_{23}\rangle=\left (\begin{array}{cccccc}
0 &0&c_{1} &0&0&\ldots\\
0 & 0 &0 &0&0&\ldots\\
0 & 0 &0 &0&c_{2}&\ldots\\
0 & 0 &0 &0&c_{3}&\ldots
\end{array} \right )\kom \langle X_{24}\rangle=\left (\begin{array}{cccccc}
0 &0&0&0&\ldots \\
0 &0&d_{1}&0&\ldots  \\
0 &0&0&d_{2}&\ldots \\
0 &0&0&d_{3}&\ldots \\
\end{array} \right )\, ,\nn\\ 
&\langle X_{35}^{T}\rangle=\left (\begin{array}{cccccc}
x_{1} &0&0 &0&0&\ldots\\
0 &x_{2}&0 &0&0&\ldots \\
0&0&x_{3} &0&0&\ldots\\
0&0&0&0&0&\ldots\\
0&0&0&0&0&\ldots\\
0&0&0&x_{4}&x_{5}&\ldots
\end{array} \right )\kom\langle X_{45}^{T}\rangle=\left (\begin{array}{ccccc}
0 &0 &0&0&\ldots\\
0 &0 &0&0&\ldots\\
z_{1}&0 &z_{4}&0&\ldots\\
0 &z_{2} &0&0&\ldots\\
0 &0 &z_{3}&0&\ldots\\
0 &0 &0&z_{5}&\ldots\\
\end{array} \right )\, ,\nn\\
& \langle X_{36}^{T}\rangle=\left (\begin{array}{cccccc}
y_{1} &0&0 &0&0&\ldots\\
0 &y_{2}&0 &0&0&\ldots \\
0&0&0&0&0&\ldots\\
0&0&y_{3}&y_{4}&0&\ldots
\end{array} \right )\kom \langle X_{46}^{T}\rangle=\left (\begin{array}{ccccc}
0&0&0&0&\ldots\\
0&0&0&0&\ldots\\
w_{1}&0&0&0&\ldots\\
0&w_{2}&w_{3}&0&\ldots
\end{array} \right )
\end{align}
where as usual all entries are $2\times 2$ matrices and the $\ldots$ on the left (right) correspond to $N$ ($M$) zeros.
As shown in Sect.~\ref{sec:HiggsingModel1}, this leads to the correct spectrum with a quiver generated by multiplicities $(12,8,(10+N),(6+M),12,8,(6+M),(10+N))$ of fractional branes $\lbrace F_{1},\ldots ,F_{8}\rbrace$ which is shown in Fig.~\ref{fig:LargeQuivdP5Mod1}. The corresponding bound states $F_{a}$ and $F_{d}$ are depicted in red and blue respectively. 
The quiver after Higgsing together with the field content is shown in Fig.~\ref{fig:QuivdP5Mod1HiggsedOrie}.

\subsection*{$D$-Flatness conditions}

Let us check the non-abelian $D$-term conditions.
Plugging the VEVs \eqref{eq:LMIChoice2} into the expression for the non-abelian D-terms \eqref{eq:QUiverDTNA} gives rise to two non-trivial conditions.
The first is given by
\begin{equation}
\sum_{i=1}^{5}\,  (|a_{i}|^{2}-|x_{i}|^{2})+|c_{1}|^{2}+|c_{2}|^{2}+|c_{3}|^{2}-|y_{1}|^{2}-|y_{2}|^{2}-|y_{3}|^{2}-|y_{4}|^{2}=0 
\end{equation}
which can be solved by choosing
\begin{align}\label{eq:CondNonAbD3LM2} 
|a_{1}|^{2}&=\sum_{i=1}^{4}\, |x_{i}|^{2}+|y_{1}|^{2}+|y_{2}|^{2}+|y_{3}|^{2}+|y_{4}|^{2}\, ,\nn\\
 |x_{5}|^{2}&=\sum_{i=2}^{5}\, |a_{i}|^{2}+|c_{1}|^{2}+|c_{2}|^{2}+|c_{3}|^{2}\, .
\end{align}
Furthermore,we find the constraint
\begin{equation}
\sum_{i=1}^{4}\,  (|b_{i}|^{2}-|z_{i}|^{2})+|d_{1}|^{2}+|d_{2}|^{2}+|d_{3}|^{2}-|z_{5}|^{2}-|w_{1}|^{2}-|w_{2}|^{2}-|w_{3}|^{2}=0
\end{equation}
which can be solved by choosing
\begin{align}\label{eq:CondNonAbD4LM2} 
|b_{1}|^{2}&=\sum_{i=1}^{4}\, |z_{i}|^{2}+|w_{1}|^{2}+|w_{2}|^{2}+|w_{3}|^{2}\, ,\nn\\
|z_{5}|^{2}&=\sum_{i=2}^{4}\, |b_{i}|^{2}+|d_{1}|^{2}+|d_{2}|^{2}+|d_{3}|^{2}\, .
\end{align}
We have $4$ real conditions on a total of $32$ complex parameters leaving us with $30$ complex parameters.

For the abelian $D$-terms, we obtain from \eqref{eq:QUiverDTA}
\begin{align}
&D_{1}=D_{2}=D_{5}=D_{6}=0\, ,\nn\\
&D_{8}=-D_{3}\kom D_{7}=-D_{8} \, ,\\
&D_{3}=-2 \left (\sum_{i=1}^{5}\,  (|a_{i}|^{2}-|x_{i}|^{2})+|c_{1}|^{2}+|c_{2}|^{2}+|c_{3}|^{2}-|y_{1}|^{2}-|y_{2}|^{2}-|y_{3}|^{2}-|y_{4}|^{2}\right )\, ,\nn \\
&D_{4}=-2 \left (\sum_{i=1}^{4}\,  (|b_{i}|^{2}-|z_{i}|^{2})+|d_{1}|^{2}+|d_{2}|^{2}+|d_{3}|^{2}-|z_{5}|^{2}-|w_{1}|^{2}-|w_{2}|^{2}-|w_{3}|^{2}\right )\nn \, .
\end{align}
Again, all abelian $D$-terms vanish upon utilising \eqref{eq:CondNonAbD3LM2} and \eqref{eq:CondNonAbD4LM2}.
The conclusion are therefore the same as for the small version of model I discussed in Sect.~\ref{sec:QuiverDFFlatness}.

\subsection{Higgsing Model II}\label{sec:HiggsinMod2} 

Let us briefly discuss the Higgsing procedure for Model II defined by the orientifold involution \eqref{eq:OAModel2}.
We consider the following bound states \cite{Wijnholt:2007vn} 
\begin{align}
\text{ch}(F_{a})&=\sum_{i=1}^{8}\, N_{i}^{(a)}\text{ch}(F_{i})=\left (3,2H-2E_{1}+E_{2}-E_{3}+E_{4}+E_{5},1\right )\nn\\
\mathbf{n}^{(a)}&=\left (4,1,2,2,1,1,2,2\right )\, ,\nn\\
\text{ch}(F_{b})&=\sum_{i=1}^{8}\, N_{i}^{(b)}\text{ch}(F_{i})=-\left (3,-E_{1}-2E_{2}+3E_{3}+3E_{4}+3E_{5},6\right )\nn\\
\mathbf{n}^{(b)}&=\left (2,1,3,3,0,6,3,3\right )\, ,\nn\\
\text{ch}(F_{b^{\prime}})&=\sum_{i=1}^{8}\, N_{i}^{(b^{\prime})}\text{ch}(F_{i})=\left (3,2H-2E_{1}+E_{2}-E_{3}+E_{4}+E_{5},1\right )\nn\\
\mathbf{n}^{(b^{\prime})}&=\left (2,1,3,3,6,0,3,3\right )\, .
\end{align}
For the bound state $2F_{a}$, we choose
\begin{align}
&\langle X_{13}\rangle=\left (\begin{array}{cccc}
a_{1} & 0 & 0 & 0 \\ 
0 & a_{2} & 0& 0  \\ 
0 &0&0&0\\
0&0&0&0\\
0 &0&a_{1}&0\\
0&0&0&a_{2}\\
0 &0&0&0\\
0&0&0&0
\end{array} \right )\kom \langle X_{14}\rangle=\left (\begin{array}{cccc}
0 &0&0&0\\
0&0&0&0\\
a_{2} &0&0&0\\
0&a_{1}&0&0\\
0 &0&0&0\\
0&0&0&0\\
0 &0&a_{2}&0\\
0&0&0&a_{1}
\end{array} \right )\nn\\
&\langle X_{23}\rangle=\left (\begin{array}{cccc}
a_{3} & 0 & 0 & 0\\ 
0 & 0 & a_{3} & 0
\end{array} \right )\kom \langle X_{24}\rangle=\left (\begin{array}{cccc}
0 & a_{3} & 0& 0 \\ 
0 & 0 & 0 & a_{3}
\end{array} \right )\nn\\
&\langle X_{36}\rangle=\left (\begin{array}{cc}
0 & 0 \\ 
0 & a_{5}  \\ 
0 & 0\\
a_{5} & 0
\end{array} \right )\kom \langle X_{46}\rangle=\left (\begin{array}{cc}
a_{4} & 0 \\ 
0 & 0  \\ 
0 & a_{4}\\
 0 & 0
\end{array} \right )\nn\\
&\langle X_{35}\rangle=\left (\begin{array}{cc}
a_{4} & 0 \\ 
0 &0  \\ 
0 & a_{4}\\
0 & 0
\end{array} \right )\kom \langle X_{45}\rangle=\left (\begin{array}{cc}
0 & 0 \\ 
a_{5} & 0  \\ 
0 & 0\\
0 & a_{5}
\end{array} \right )
\end{align}
For $F_{b}+F_{b^{\prime}}$ we use
\begin{align}
&\langle X_{13}\rangle=\langle X_{14}\rangle=\left (\begin{array}{cccccc}
e_{1} & 0 & 0 & 0 &0&0\\ 
0 & e_{1} & 0& 0&0&0  \\ 
0 &0&0&e_{1}&0&0\\
0&0&0&0&e_{1}&0
\end{array} \right ) \nn\\
&\langle X_{23}\rangle=\langle X_{24}\rangle=\left (\begin{array}{cccccc}
0&0&e_{1} & 0 & 0 & 0\\ 
0&0&0 & 0 & 0 & e_{1}
\end{array} \right ) \nn\\
&\langle X_{35}\rangle=\left (\begin{array}{cccccc}
e_{1} & 0 & 0 & 0 &0&0\\ 
0 & e_{1} & 0& 0&0&0  \\ 
0 &0&e_{1}&0&0&0\\
0&0&0&0&0&0\\
0&0&0&0&0&0\\
0&0&0&0&0&0
\end{array} \right )\kom \langle X_{45}\rangle=\left (\begin{array}{cccccc}
0 & 0 & 0 & e_{1} &0&0\\ 
0 & 0 & 0& 0&e_{1}&0  \\ 
0 &0&0&0&0&e_{1}\\
0&0&0&0&0&0\\
0&0&0&0&0&0\\
0&0&0&0&0&0
\end{array} \right )\nn\\
&\langle X_{35}\rangle=\left (\begin{array}{cccccc}
0 & 0 & 0 & 0 &0&0\\ 
0 & 0 & 0& 0&0&0  \\ 
0 &0&0&0&0&0\\
0&0&0&0&0&e_{1}\\
0&0&0&e_{1}&0&0\\
0&0&0&0&e_{1}&0
\end{array} \right )\kom \langle X_{45}\rangle=\left (\begin{array}{cccccc}
0 & 0 & 0 & 0 &0&0\\ 
0 & 0 & 0& 0&0&0  \\ 
0 &0&0&0&0&0\\
0&0&e_{1}&0&0&0\\
e_{1}&0&0&0&0&0\\
0&e_{1}&0&0&0&0
\end{array} \right )
\end{align}
We showed that both lead to the expected breaking patterns:
\begin{align}
2F_{a}&:\; \mathrm{U}(8)\times \mathrm{U}(2)^{3}\times \mathrm{U}(4)^{4}\raw \mathrm{U}(2)\, ,\\
F_{b}+F_{b^{\prime}}&:\, \mathrm{U}(4)\times \mathrm{U}(2)\times \mathrm{U}(6)^{6}\raw\mathrm{U}(1)\times \mathrm{U}(1)\, .
\end{align}

\begin{figure}[t!]
\centering
\includegraphics[scale=0.3]{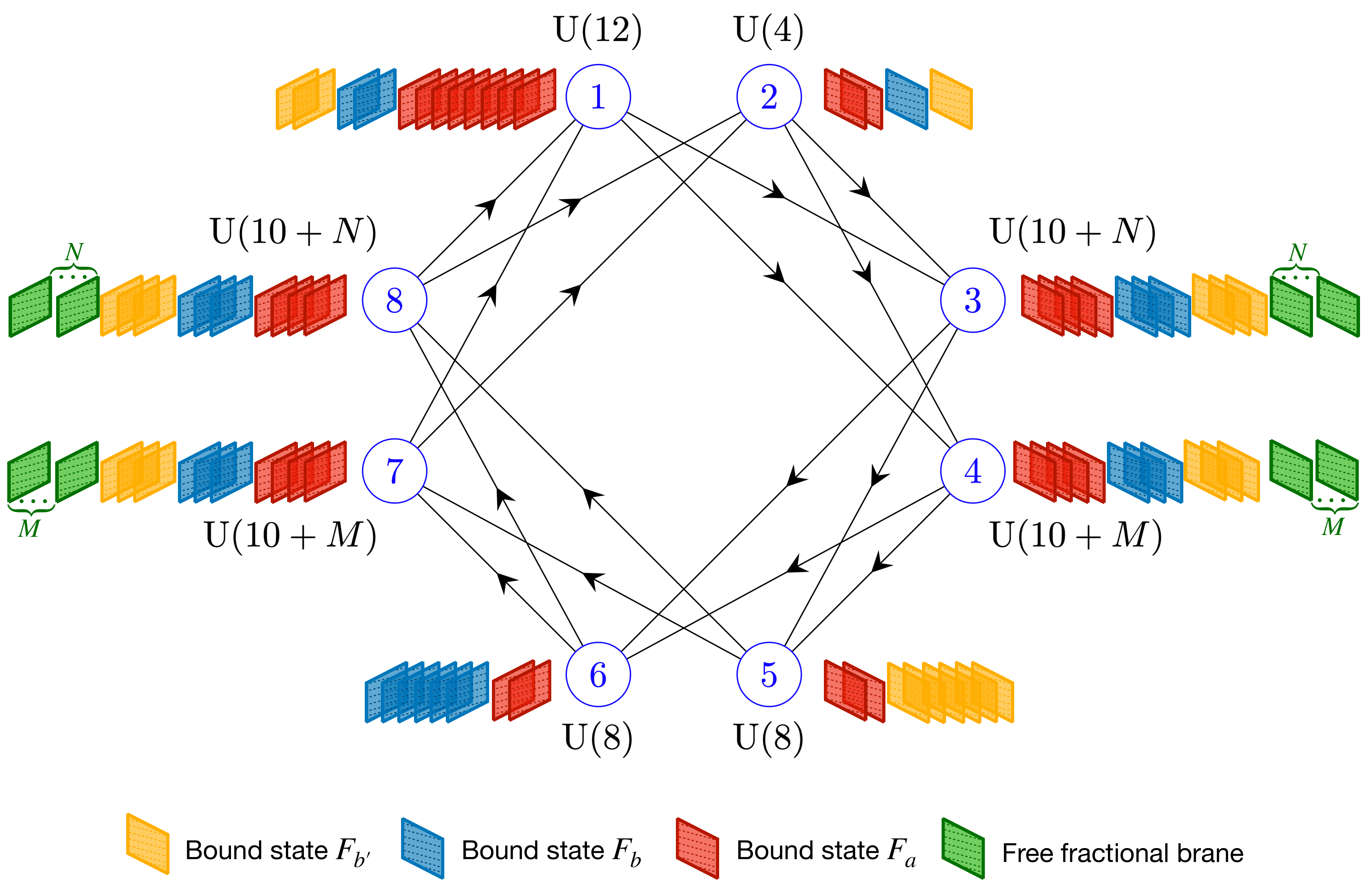}
\caption{Model II quiver diagram $\mathrm{dP}_{5}$ with the bound states $2F_{a}$ ({\color{red}red}), $F_{b}$ ({\color{blue}blue}) and $F_{b^{\prime}}$ ({\color{orange}orange}).
{\color{green}Green} fractional branes do not participate in any bound state.}\label{fig:QuivdP5Mod2} 
\end{figure}

\begin{figure}
\centering
\begin{tikzpicture}[scale=0.9]
\draw (-0.5,6.5) node[above=0pt] {\large\bf Higgsed quiver};
\draw (-0.5,5) node(1)[anchor=south,circle,draw,blue]{$a$}node[above=20pt] {$\mathrm{U}(2)$};
\draw (-2.5,4) node(8)[anchor=east,circle,draw,blue]{8}node[left=20pt] {$\mathrm{U}(N)$};
\draw (-2.5,3) node(7)[anchor=east,circle,draw,blue]{7}node[left=20pt] {$\mathrm{U}(M)$};
\draw (1.5,4) node(3)[anchor=west,circle,draw,blue]{3}node[right=20pt] {$\mathrm{U}(N)$};
\draw (1.5,3) node(4)[anchor=west,circle,draw,blue]{4}node[right=20pt] {$\mathrm{U}(M)$};
\draw (-0.5,2) node(5)[anchor=north,circle,draw,blue]{\small $b+b^{\prime}$}node[below=40pt] {$\mathrm{U}(1)\times \mathrm{U}(1)$};
\draw[directedThree,black] (1) -- (3);
\draw[directed,black] (1) -- (5);
\draw[directed,black] (5) -- (1);
\draw[directedThree,black] (1) -- (4);
\draw[directedThree,black] (3) -- (5);
\draw[directedThree,black] (4) -- (5);
\draw[directedThree,black] (5) -- (7);
\draw[directedThree,black] (5) -- (8);
\draw[directedThree,black] (7) -- (1);
\draw[directedThree,black] (8) -- (1);
\draw (8,6.5) node[above=0pt] {\large\bf Higgsed and orientifolded quiver};
\draw (8,5) node(1)[anchor=south,circle,draw,blue]{$a$}node[above=20pt] {$\mathrm{SU}(2)$};
\draw (6,3.5) node(8)[anchor=east,circle,draw,blue]{3}node[left=20pt] {$\mathrm{U}(3)$};
\draw (10,3.5) node(4)[anchor=west,circle,draw,blue]{4}node[right=20pt] {$\mathrm{U}(1)$};
\draw (8,2) node(5)[anchor=north,circle,draw,blue]{\small $b+b^{\prime}$}node[below=40pt] {$\mathrm{U}(1)$};
\draw[directed,black] (1) to[bend left=15]node[right=10pt] {$H_{d}$} (5);
\draw[directed,black] (5) to[bend left=15]node[left=10pt] {$H_{u}$} (1);
\draw[directedThree,black] (4) --node[right=12pt] {$L$} (1);
\draw[directedThreeBF,black] (5) to[bend left=20]node[left=10pt]{$d_{R}$} (8);
\draw[directedThreeBF,black] (5) to[bend right=20]node[right=10pt]{$e_{R}$} (4);
\draw[directedThreeFF,black] (5) --node[left=1pt] {$\nu_{R}$} (4);
\draw[directedThreeFF,black] (5) --node[right=1pt] {$u_{R}$} (8);
\draw[directedThree,black] (8) --node[left=12pt] {$Q$} (1);
\end{tikzpicture}
\caption{Left: Higgsed $\mathrm{dP}_{5}$ quiver diagram for Model II obtained from the quiver in Fig.~\ref{fig:QuivdP5Mod2}.
Right: Orientifolded quiver diagram for $N=3$ and $M=1$ via the involution \eqref{eq:OAModel2}.}\label{fig:QuivdP5Mod2Higgs} 
\end{figure}
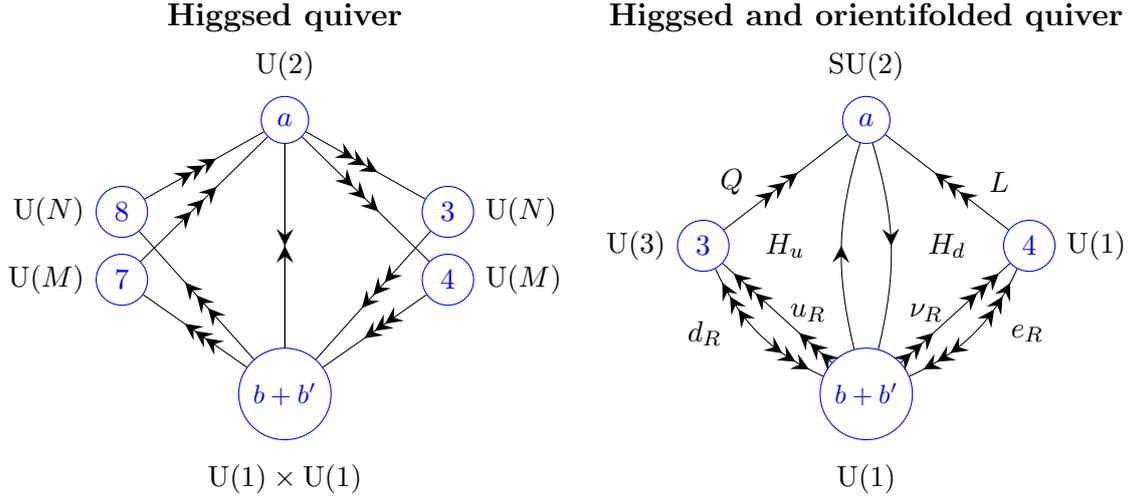

Altogether, we construct the Higgsed quiver $(2F_{a},NF_{3},MF_{4},F_{b}+F_{b^{\prime}},MF_{7},NF_{8})$ 
which is obtained from $(12F_{1},4F_{2},13F_{3},11F_{4},8F_{5},8F_{6},11F_{7},13F_{8})$, see Fig.~\ref{fig:QuivdP5Mod2}. Here, the bound state $2F_{a}$ is depicted in red, $F_{b}$ in blue and $F_{b^{\prime}}$ in green, while black dots represent fractional branes not participating in any bound state.
The Higgsed quiver is shown in Fig.~\ref{fig:QuivdP5Mod2Higgs}.
After performing the orientifolding, one ends up with the quiver on the right of Fig.~\ref{fig:QuivdP5Mod2Higgs} for $N=3$ and $M=1$.
The $\mathrm{U}(1)$ at the bottom node comes from identifying two different nodes in the covering quiver. 
The field content is computed from
\begin{align}
G&=\mathrm{U}(12)\times\mathrm{U}(4)\times( \mathrm{U}(10+N) \times \mathrm{U}(10+M) \times \mathrm{U}(8))^{2} \, ,\nn\\
\mathrm{dim}(G)&=688+40(N+M)+2(N^{2}+M^{2}) \, ,\nn\\
H&=\mathrm{U}(2)\times (\mathrm{U}(1)\times \mathrm{U}(N)\times \mathrm{U}(M))^{2} \, ,\nn\\
\mathrm{dim}(H)&=6+2(N^{2}+M^{2}) \, ,\nn\\
\dim(G/H)&=682+40(N+M)\, ,\nn\\
N_{\text{chiral}}&=1280 + 64 (N+M)\, ,\nn\\
N_{\text{chiral}}^{\text{Higgsed}}&=24(N+M)\, ,\nn\\
N_{\text{chiral}}-\dim(G/H)&=598+N_{\text{chiral}}^{\text{Higgsed}}\, .
\end{align}
We find $598$ complex scalars in bi-fundamentals between $\mathrm{U}(2)$ and one of the $\mathrm{U}(1)$'s
or between the two $\mathrm{U}(1)$'s.
There is no non-chiral matter between either $\mathrm{U}(N)$ or $\mathrm{U}(M)$ and $\mathrm{U}(2)$ ($\mathrm{U}(1)$).

\section{Line Orientifold of the Complex Cone over $\mathrm{dP}_5$}\label{App:InvolLocal}

\subsection*{Geometry}

\begin{figure}[ht]
  \centering
 \includegraphics[height=3cm]{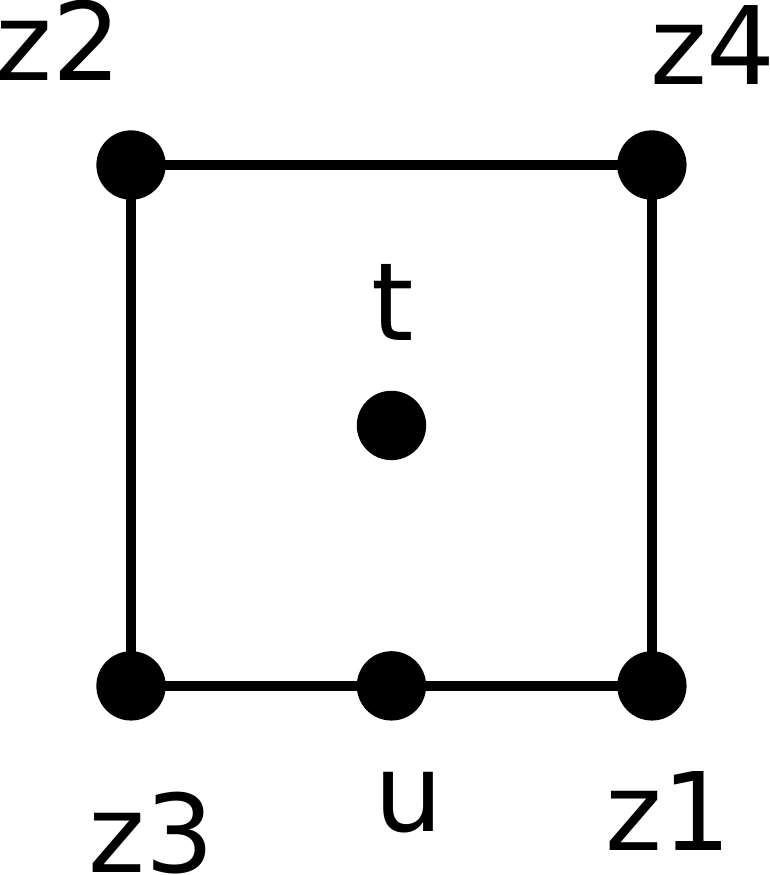}
  \caption{Toric data for the toric $\mathrm{dP}_5$.}
  \label{fig:toricData}
\end{figure}

This is a quick review of some of the properties of the complex cone
over $\mathrm{dP}_5$. At some specific loci in moduli space it can be described
as a $\mathbb{Z}_2\times\mathbb{Z}_2$ orbifold of the conifold, which can be
described as a particular phase of the following GLSM
\begin{equation}
  \begin{array}{c|cccccc}
    & z_1 & z_2 & z_3 & z_4 & u & t \\
    \hline
   \mathbb{C}^*_1 & 1 & 1 & 0 & 0 & 0 & -2 \\
   \mathbb{C}^*_2 & 0 & 0 & 1 & 1 & 0 & -2 \\
   \mathbb{C}^*_3 & 1 & 0 & 1 & 0 & -2 & 0
  \end{array}
\end{equation}
The $\mathbb{Z}_2\times\mathbb{Z}_2$ action can be seen easily by going to a phase
with $\xi_1=\xi_2<0$ and $\xi_3<0$, and gauge fixing $t=u=1$. This
leaves a $\mathbb{Z}_2\times\mathbb{Z}_2$ acting on the conifold coordinates $z_i$ as
\begin{subequations}
  \label{eq:glsm-orbifold}
  \begin{align}
    \sigma_1& \colon (z_1, z_2, z_3, z_4) \to (z_1, z_2, -z_3, -z_4)\, \\
    \sigma_2& \colon (z_1, z_2, z_3, z_4) \to (-z_1, z_2, -z_3, z_4)\, .
  \end{align}
\end{subequations}
The toric diagram is shown in Fig.~\ref{fig:toricData}.

The monomial ring is generated by the gauge invariants
\begin{equation}
  X = (z_1z_3)^2 t u^2\quad ; \quad Y = (z_2z_4)^2 t \quad ; \quad
  C = z_1z_2z_3z_4 tu \quad ; \quad Z = (z_1z_4)^2 tu \quad ; \quad
  W = (z_2z_3)^2 tu
\end{equation}
which satisfy
\begin{equation}
  \label{eq:complete-intersection}
  XY = ZW = C^2
\end{equation}
which is manifestly a double cover of the conifold branched over the
Cartier divisors $X=0$, $Y=0$, $Z=0$ and
$W=0$.

\subsection*{Orientifold}

\begin{figure}
  \centering
  \includegraphics[width=0.4\textwidth]{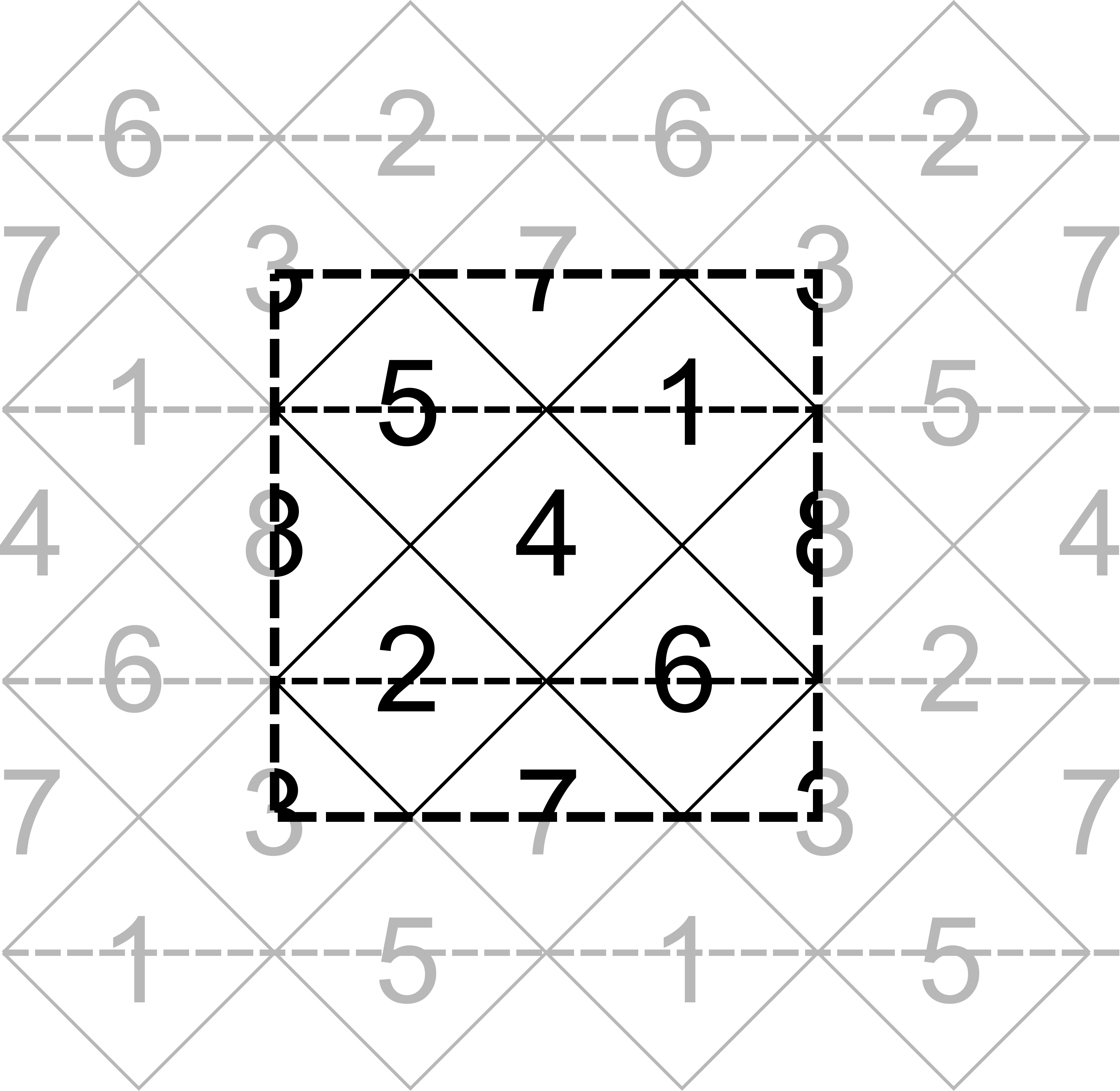}
  \caption{Line orientifold of the dimer model for the
    $\mathbb{Z}_2\times\mathbb{Z}_2$ orbifold of the conifold described in 
    text. We have highlighted a fundamental cell, and denoted 
    dashed line the involution in which we are interested.}
  \label{fig:dP5-orientifold}
\end{figure}

We are interested in the line orientifold shown in
figure~\ref{fig:dP5-orientifold}. This is in fact closely related to
the quotient studied in detail in
\cite{Franco:2007ii,Garcia-Etxebarria:2015lif,Collinucci:2016hgh},
so we can easily read the results from there. Using the
techniques in \cite{Franco:2007ii} we find that the
orientifold action on the fundamental mesons, identified with the
geometric fields introduced above, is given by
\begin{equation}
  \label{eq:meson-action}
  (X,Y,Z,W,C) \to (\epsilon X, \epsilon Y, W, Z, C)
\end{equation}
with $\epsilon$ the product of the signs of the fixed lines. Namely,
if the two fixed lines have the same sign we have that $\epsilon=+1$,
and the fixed locus is at $Z=W$, giving rise to a non-compact O7
plane. (As described in \cite{Collinucci:2016hgh} the sign of the
non-compact plane is determined by the projection, with $\Sp$
projection corresponding to the O7$^-$ case.) If the two fixed lines
have opposite sign, $\epsilon=-1$, corresponding to opposite choices
of Chan-Paton factor on the two fixed lines, the fixed locus is at
$X=Y=Z=W=C=0$.

\begin{figure}[htp]
  \centering
  \includegraphics[height=3cm]{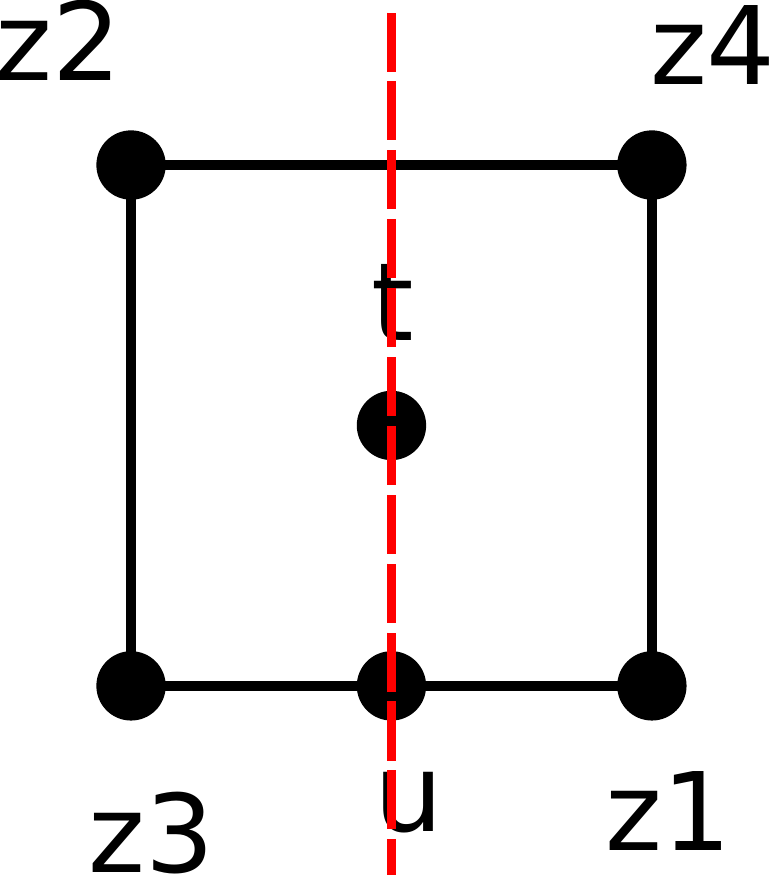}\hspace*{2.5cm}\includegraphics[height=3cm]{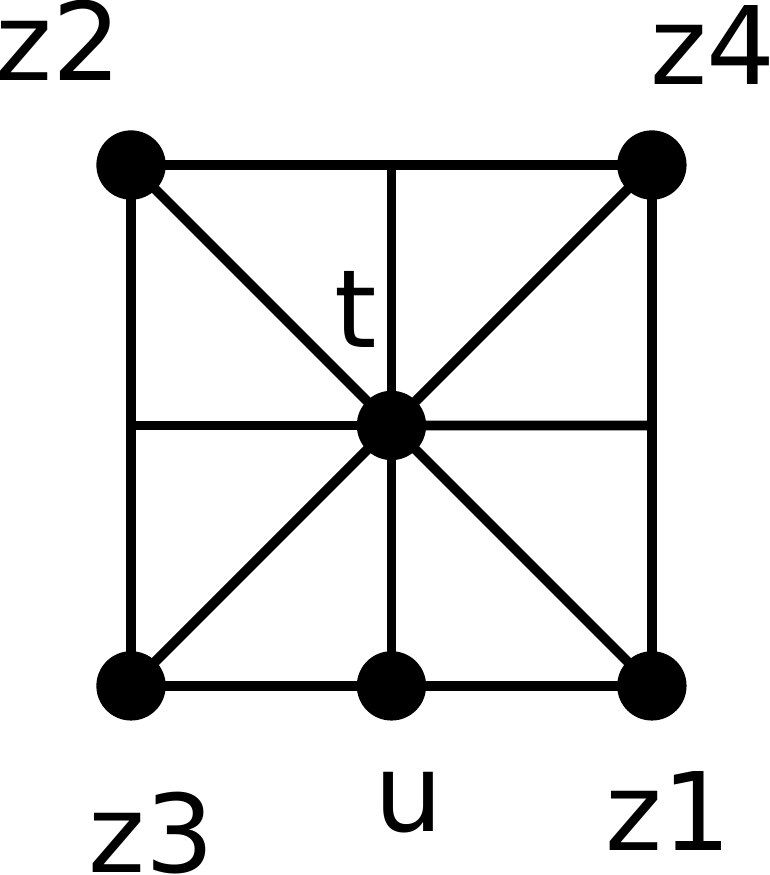}
  \caption{\emph{Left}: The involution acts on the toric diagram by reflecting
    along the dashed red line.
    \emph{Right}: A resolution of the toric singularity compatible with the
    involution, at least at the level of the toric diagram.}
  \label{fig:toricInvolution}
\end{figure}

The two projections of interest to us can be described in terms of the
toric coordinates by
\begin{equation}
  (z_1, z_2, z_3, z_4, u, t) \to (-z_3, z_4, z_1, -z_2, \epsilon u,
  \epsilon t)\, .
\end{equation}
The action on the
toric diagram is shown on the left of Fig.~\ref{fig:toricInvolution}.

We see that there are some resolutions of the dP$_5$ singularity that are
compatible with the involution, an example is on the right of
figure~\ref{fig:toricInvolution}.

\newpage

\bibliographystyle{JHEP}
\bibliography{reference}

\end{document}